\pdfoutput=1
\documentclass[PAPER, atlasdraft=false, UKenglish,cernpreprint, texmf, orcidlogo]{atlasdoc}
\usepackage{atlaspackage}
\usepackage{adjustbox}
\usepackage{multirow}
\usepackage{atlasbiblatex}

\usepackage{atlasphysics}
\usepackage{atlasbiblatex}
\usepackage{sansmath,amsmath,latexsym}
\usepackage{tabularht}
\newcolumntype{P}[1]{>{\centering\arraybackslash}p{#1}}

\graphicspath{{logos/}{figures/}}

\usepackage{ANA-STDM-2023-32-PAPER-defs}

\AtlasTitle{Effective field theory interpretation of ATLAS measurements involving the Higgs boson, electroweak bosons and the top quark}

\AtlasAbstract{%
Wilson coefficients in dimension-six effective field theory are constrained in a combined fit to several ATLAS measurements. These inputs probe Higgs-boson processes across multiple production and decay modes, di-Higgs signatures in the $b\bar{b}\gamma\gamma$ and $b\bar{b}\tau\tau$ final states, $WW$ and $WZ$ diboson signatures, electroweak $Zjj$ final states, high-mass Drell--Yan interactions, and $t \bar t$ events in both resolved and boosted topologies. Precision electroweak observables from LEP, SLD, and ATLAS are also included. A total of 48 parameters, including individual Wilson coefficients in the Warsaw basis and linear combinations of Wilson coefficients, are constrained simultaneously. Constraints on two-Higgs-doublet models and heavy-vector-boson models are also obtained by matching a relevant sub-set of the results with their parameters.  This combined fit provides the most comprehensive effective field theory interpretation of experimental data by the ATLAS Collaboration to date. No significant deviations from the Standard Model are observed.}

\AtlasRefCode{STDM-2023-32}

\PreprintIdNumber{CERN-EP-2026-113}

\AtlasJournal{Journal of High Energy Physics}
\AtlasCoverEgroupAnalysisTeam{atlas-stdm-2023-32-analysis-team@cern.ch}

\hypersetup{pdftitle={ATLAS document},pdfauthor={The ATLAS Collaboration}}

\begin{document}

\maketitle

\tableofcontents

\clearpage
\section{Introduction}
\label{sec:intro}

Precision measurements at colliders serve as crucial tests of the Standard Model (SM), with any observed deviations from its predictions offering indirect evidence for new physics.
The Standard Model Effective Field Theory (SMEFT)~\cite{Brivio:2017vri} offers a consistent and nearly model-independent framework to study the effects of Beyond-Standard Model (BSM) physics across different measurements, and a coherent description of the impact of heavy new-physics states at a mass scale $\Lambda$ well above the electroweak scale.

In SMEFT, predictions for experimental observables are expressed as an expansion in $E/\Lambda$, where $E$ denotes the typical energy exchanged in the process. This expansion is formulated in terms of higher-dimensional operators ${\mathcal{O}_i^{(d)}}$, constructed from gauge-invariant combinations of SM fields with energy dimension $d>4$:
\begin{equation}
\label{eq:smeft}
\mathcal{L}_\textrm{SMEFT} = \mathcal{L}_\textrm{SM} + \sum\limits_i \frac{c^{(5)}_i}{\Lambda} \mathcal{O}_i^{(5)} +  \sum\limits_i \frac{c^{(6)}_i}{\Lambda^2} \mathcal{O}_i^{(6)} + \dots,
\end{equation}
where $\mathcal{L}_\textrm{SM}$ denotes the SM Lagrangian and $i$ runs over all the operators that can be constructed at a given dimension.

Measurements of observables sensitive to the effect of SMEFT operators can be used to constrain $c_i^{(d)}/\Lambda^{d-4}$, where $c_i^{(d)}$ are the Wilson coefficients associated with the operators ${\mathcal{O}_i^{(d)}}$. In this paper, the interpretation is restricted to dimension-six operators, as contributions from higher-dimensional terms are suppressed by greater powers of $\Lambda$. The dimension-five operator (and its conjugate) is not considered here, as it induces lepton-number violation to which the measurements included in this combination are not sensitive.

When several operators contribute simultaneously to a given observable, correlations and potential cancellations can result in blind or weakly constrained directions in the parameter space, especially if constraints are obtained from individual analyses that are sensitive to the effects of only a subset of Wilson coefficients. Global fits, combining information from multiple processes, are therefore essential to fully exploit the constraining power of precision data and to obtain robust, model-independent bounds on SMEFT parameters.

Global fits in the SMEFT framework have been carried out both within and outside the Large Hadron Collider (LHC) collaborations. Within ATLAS and CMS, several analyses have provided constraints on dimension-six Wilson coefficients from Higgs boson, electroweak, top-quark, and diboson processes, and dedicated studies have addressed the challenges of performing global SMEFT fits at the LHC~\cite{HIGG-2022-17,CMS:2025eft}.

Outside the collaborations, several global analyses have been performed, providing comprehensive interpretations of Higgs boson, electroweak, and top-quark measurements within the SMEFT framework ~\cite{Giani:2023smefit,Falkowski:2015ModelIndependent,Ellis:2018UpdatedGlobalSMEFT,Falkowski:2020FlavourfulLikelihood,Dawson:2020PuttingSMEFTFitsToWork,Ethier:2021CombinedSMEFT}.  While these studies have achieved impressive sensitivity, they are limited by the detail level of publicly available experimental information, such as approximations to experimental likelihoods, or incomplete knowledge of correlated systematic uncertainties across measurements.
This paper builds upon and extends earlier ATLAS SMEFT combination efforts based primarily on Higgs boson measurements~\cite{HIGG-2022-17} by incorporating a substantially enlarged and more diverse set of Run-2 measurements. It also delivers, as an explicit output, the correlation matrix for the measured signal strengths, which allows the construction of simplified likelihood models for reinterpretations.

The measurements that are combined to extract constraints on the Wilson coefficients are:
\begin{itemize}
\item \textbf{ATLAS single-Higgs-boson data}~\cite{HIGG-2020-16,HIGG-2018-28,HIGG-2019-09,HIGG-2021-20,HIGG-2020-20,HIGG-2019-04,HIGG-2020-23,HIGG-2021-08,HIGG-2018-42,HIGG-2019-14,HIGG-2022-17}: a combined measurement of Higgs boson production and decay in exclusive kinematic regions of the production phase space, defined within the Simplified Template Cross-Section (STXS) framework, is used. This combined measurement provides sensitivity to a wide variety of SMEFT operators given the set of production modes and decay channels included. Compared to the previous ATLAS global combination, new channels are included, such as the boosted $H\to b\bar{b}$ channel, focusing on fully hadronic $H\to b\bar{b}$ events in the regime of large Higgs boson transverse momentum, $H\rightarrow Z \gamma$ and $H\rightarrow \mu\mu$.
\item \textbf{ATLAS di-Higgs channels}~\cite{HDBS-2021-10,HDBS-2019-27}:  the measurements of the Higgs boson pair (di-Higgs) production in the $b\bar{b}\gamma \gamma$ and $b\bar{b}\tau\tau$ channels are also included, exploiting the sensitivity of the di-Higgs analyses to the trilinear Higgs boson self-coupling.
\item \textbf{ATLAS electroweak data}~\cite{STDM-2017-24,STDM-2018-03,STDM-2017-27}: differential cross-section measurements for $WW$ production, $WZ$ production, and $Z$-boson production via vector boson fusion (VBF) are included. The electroweak analyses target triple gauge couplings and four-fermion operators.
\item \textbf{ATLAS High Mass Drell--Yan (HMDY) cross-sections}~\cite{EXOT-2022-42,STDM-2018-41}: neutral ($Z\rightarrow \tau^+ \tau^-$) and charged current Drell--Yan cross-sections at high mass are included, extending the set of channels used in the previous ATLAS global combination. The HMDY analyses provide sensitivity to four-fermion operators.
\item \textbf{ATLAS top-quark distributions}~\cite{TOPQ-2018-26,TOPQ-2019-23}: differential cross-sections for dilepton top-quark pair ($t\bar{t}$) events and $t\bar{t}$ events with a high $p_\mathrm{T}$ top quark are incorporated in the global combination, representing a new addition relative to the earlier ATLAS results. The top-quark analyses are sensitive to heavy-flavour four-fermion operators.
\item \textbf{Electroweak precision data}~\cite{ALEPH:2005ab,STDM-2012-20,TOPQ-2018-29,TOPQ-2023-28,STDM-2019-24}: combined measurements of electroweak precision observables (EWPO) on the $Z$ and $W$ resonances performed at LEP and SLD ($Z$ pole observables, $W$ branching ratios) as well as at ATLAS (lepton flavour universality ratios and $W$-boson width) are also included. The EWPO are sensitive to couplings between weak bosons and fermions.
\end{itemize}
In addition to presenting constraints on SMEFT Wilson coefficients, this work also shows how the resulting limits can be matched to selected ultraviolet (UV)-complete models.

This paper is organised as follows. Section~\ref{sec:input} provides a brief description of the different measurements used to derive the results. Studies of overlapping data regions among ATLAS analyses are also summarised, together with the strategy adopted to deal with these overlaps. The theoretical model, the simulation of SMEFT and SM samples, and the methodology used to perform the interpretation, are described in Section~\ref{sec:theory}, together with the impact of new physics on each observable. Section~\ref{sec:stat} describes the statistical model and the correlation scheme used in the combination. Section~\ref{sec:result} summarises the studies performed to determine the Wilson coefficients to which the combination is sensitive, and presents the final results, including both full-likelihood and simplified-likelihood results. The reinterpretation of the SMEFT constraints to UV complete models such as two-Higgs-doublet models (2HDM) ~\cite{Branco:2011iw} and heavy-vector-boson ($Z'$) models~\cite{Langacker:2008yv} is also presented. Finally, Section~\ref{sec:conclusion} presents the conclusions.


%

%
%
%
%
%

%

%
\section{Data and input measurements}
\label{sec:input}

%
Given the diversity of the measurements entering the global SMEFT interpretation, the level of detail reported in this section is necessarily heterogeneous. For some measurements, specific observables, their precise definitions, and the associated phase-space selections are explicitly discussed, particularly when these aspects are central to the SMEFT sensitivity. Other analyses are summarised more concisely, with details provided through references to the original publications.

An overview of all the input analyses entering the combination is given in Table~\ref{tab:summary-analyses-no-ref}, together with the corresponding references, size of the dataset and the targeted SMEFT operators.

\begin{table}[!htbp]
\centering
\caption{Summary of analyses entering the combined SMEFT interpretation, together with the corresponding references, and the sensitivity to SMEFT effects.}
\label{tab:summary-analyses-no-ref}
\renewcommand{\arraystretch}{1.1}
\setlength{\tabcolsep}{3pt}
\resizebox{\textwidth}{!}{
\begin{tabular}{@{}lcc@{}}
\toprule
Process & Dataset [fb$^{-1}$] & SMEFT sensitivity \\
\midrule

\multicolumn{3}{l}{\textbf{Higgs boson measurements}~\cite{HIGG-2020-16,HIGG-2018-28,HIGG-2019-09,HIGG-2021-20,HIGG-2020-20,HIGG-2019-04,HIGG-2020-23,HIGG-2021-08,HIGG-2018-42,HIGG-2019-14}} \\
$pp\to H\to \gamma\gamma,\ ZZ^*\!\to 4\ell,\ WW^*\!\to \ell\nu\ell\nu$
& 140& Higgs boson operators \\
$pp\to H\to Z\gamma,\ \mu^+\mu^-$
& 140& Higgs boson operators \\
$pp\to H\to b\bar b,\ \tau^+\tau^-$
& 126--140& Higgs boson operators \\

\multicolumn{3}{l}{\textbf{Electroweak measurements}~\cite{STDM-2017-24,STDM-2018-03,STDM-2017-27}} \\
$pp\to W^+W^-\to e^\pm\nu\,\mu^\mp\nu$
& 36 &  Four-fermion operators and triple gauge couplings \\
$pp\to W^\pm Z\to \ell^\pm\nu\,\ell^+\ell^-$
& 36 &  Four-fermion operators and triple gauge couplings \\
$pp\to Zjj\to \ell^+\ell^- jj$
& 140 &  Four-fermion operators and triple gauge couplings \\

\multicolumn{3}{l}{\textbf{HMDY measurements}~\cite{EXOT-2022-42,STDM-2018-41}} \\
$pp\to Z/\gamma^*\to \tau^+\tau^-$
& 140 &  Four-fermion operators involving leptons \\
$pp\to W^\pm\to \ell^\pm\nu$
& 140 & Four-fermion operators involving leptons \\

\multicolumn{3}{l}{\textbf{Top-quark measurements}~\cite{TOPQ-2018-26,TOPQ-2019-23}} \\
$pp\to t\bar t\to Wb\,Wb\to e^\pm\mu^\mp \nu \nu b b$
& 140 & Four-fermion (heavy-flavour) operators \\
$pp\to t\bar t\to Wb\,Wb \to qq'b\,\ell\nu b$
& 140 & Four-fermion (heavy-flavour) operators \\

\multicolumn{3}{l}{\textbf{Di-Higgs measurements}~\cite{HDBS-2021-10,HDBS-2019-27}} \\
$pp\to HH\to b\bar b\gamma\gamma,\ b\bar b\tau^+\tau^-$
& 140 & Trilinear Higgs boson self-coupling \\

\multicolumn{3}{l}{\textbf{Electroweak precision observables}~\cite{ALEPH:2005ab,STDM-2012-20,TOPQ-2018-29,TOPQ-2023-28,STDM-2019-24}} \\
LEP, SLD, ATLAS EWPO
& -- & Weak boson--fermion couplings \\

\bottomrule
\end{tabular}
}
\end{table}

\subsection{Higgs boson measurements}

Higgs boson measurements enter the SMEFT interpretation through combined measurements of Higgs boson production and decay performed in exclusive kinematic regions of the production phase-space defined within the simplified template cross-section (STXS) framework~\cite{deFlorian:2016spz,LesHouches,Berger:2019wnu,Amoroso:2020lgh}.
The STXS framework categorises Higgs boson production cross-sections into mutually exclusive bins defined by the Higgs boson production mode and by key kinematic observables, such as jet multiplicities and the transverse momentum, $\pT$, of the Higgs boson or of the associated system of final-state particles. This categorisation is designed to minimise theoretical uncertainties, reduce model dependence, and enhance sensitivity to physics beyond the SM in different regions of phase-space. These measurements, based on proton--proton collision data collected by the ATLAS experiment in 2015-2018 at $\sqrt{s}=13$~\TeV, provide sensitivity to modifications of Higgs boson couplings to fermions and vector bosons, as well as to loop-induced interactions.

The full set of input analyses, together with the targeted Higgs boson decay channels, production modes, and integrated luminosities, is summarised in Table~\ref{tab:lumi}. Further details can be found in the references listed therein.
For each Higgs boson decay mode, the branching fraction used corresponds to theoretical calculations at the highest available order~\cite{deFlorian:2016spz}. The product of the STXS production cross-sections and the corresponding Higgs boson branching fractions is reparameterised in terms of the SMEFT Wilson coefficients.
\begin{table}[!htbp]
\caption{For each input analysis of the combined Higgs-boson STXS measurements the decay channels, targeted production modes, integrated luminosity and references to individual analyses are reported. Gluon--gluon fusion and vector boson fusion production modes are referred to as ggF and VBF, respectively. }
\label{tab:lumi}
\vspace{3pt}

\centering
\begin{tabular}{llccccc}
\toprule

Decay channel & Production mode                 & $\mathcal{L}$   \([\textrm{fb}^{-1}]\) & Ref.                      &    \\
\midrule
\hyy &  \tggF, \tVBF, \tWH, \tZH, \tttH, \ttH         & $140$  & \cite{HIGG-2020-16}  \\
\hzz  & \tggF, \tVBF, \tWH\, + \tZH, \tttH\, + \ttH  & $140$  & \cite{HIGG-2018-28}   \\

\htt        &  \tggF, \tVBF, \tWH \, + \tZH, \tttH\, + \ttH         & $140$                                              & \cite{HIGG-2019-09} \\
\hww        &  \tggF, \tVBF                  & $140$                                              & \cite{HIGG-2021-20}     \\
\hbb        &   \tWH, \tZH                  & $140$                                              & \cite{HIGG-2020-20}   \\
&  \tVBF                        & $126$                                              & \cite{HIGG-2019-04}  \\

&  \tttH\, + \ttH                        & $140$                                              & \cite{HIGG-2020-23}    \\
&   inclusive      & $140$                                              & \cite{HIGG-2021-08}   \\
\hzgam       &   inclusive         & $140$                                              & \cite{HIGG-2018-42}      \\
\hmm       &   \tggF \,+ \tttH \,+ \ttH, \tVBF \, + \tWH \, + \tZH         & $140$                                              & \cite{HIGG-2019-14}     \\
\bottomrule
\end{tabular}
%


\end{table}

\subsection{Differential cross-section measurements for electroweak processes}

Measurements of differential cross-sections unfolded at particle level of weak boson production and decay, based on either partial or full Run 2 data, are used in this interpretation.
These comprise \WW production in the $e^\pm\nu\,\mu^\mp\nu$ final state~\cite{STDM-2017-24}, \WZ production in the \lllv final state~\cite{STDM-2018-03} (where $\ell=e,\mu$), and VBF $Z$ boson production (EW $Zjj$ ) with subsequent decay of the $Z$ boson in $e^+e^-$ or $\mu^+\mu^-$ pairs~\cite{STDM-2017-27}.

For the \WW\ measurement, the fiducial region is defined by the presence of exactly one electron and one muon of opposite charge with $m_{\ell\ell}>55\,\GeV$, and by vetoing events containing jets with $p^\text{jet}_\mathrm{T}>35\,\GeV$ in order to suppress top-quark backgrounds.
For the \WZ\ measurement, the fiducial phase-space requires three charged leptons with the same-flavour opposite-sign lepton pair satisfying $66<m_{\ell\ell}<116\,\GeV$, and a minimum transverse mass\footnote{The transverse mass is defined as
$m_T^W = \sqrt{ 2 \,  \pT^\nu \cdot   \pT^\ell \cdot \left[ 1 - \cos \Delta \phi (\ell, \nu) \right] },$ where $\Delta\phi(\ell, \nu)$ is the angle between the charged lepton and the neutrino in the transverse plane, and $ \pT^\ell$ and $ \pT^\nu$ are the transverse momenta of the charged lepton from the $W$ boson decay and of the neutrino, respectively.} of the $W$ boson of $m_\mathrm{T}^W>30\,\GeV$.
For electroweak $Z$-boson production in association with two jets, the fiducial region targets a VBF topology with a dilepton pair consistent with a $Z$-boson decay and two jets with a large invariant mass, $m_{jj}>1000\,\GeV$, enhancing the purely electroweak contribution from the other sources of $Zjj$ production.

For each of the three electroweak processes, a single differential distribution is used as input for the SMEFT interpretation. The differential cross-section as a function of the leading lepton $\pT^{\ell}$ and the transverse mass, $m_{\mathrm{T}}^{WZ}$, are employed for the \WW measurement and \WZ measurement, respectively. The differential cross-section as a function of the signed azimuthal angle between the two jets, $\Delta\phi_{jj}$\footnote{The signed azimuthal angle between the two jets is defined as $\Delta\phi_{jj} = \phi_{f} - \phi_{b}$, where the two highest transverse momentum jets are ordered by their rapidity $y$ such that $y_{f} > y_{b}$}, is employed for EW $Zjj$  production, as this variable is especially sensitive to SMEFT operators modifying triple gauge couplings.
The measurements and key phase-space requirements are summarised in Table~\ref{tab:ew-measurements}.
The unfolded fiducial cross-sections from these measurements are directly compared to particle-level SMEFT predictions, parameterised in each measurement bin as a function of the Wilson coefficients.
\begin{table} [!htbp]
\caption{ATLAS electroweak measurements entering the combined SMEFT interpretation, together with the most important phase-space requirements in each measurement, the observables used in the interpretation, and the integrated luminosity of the analysed data.}
\centering
{  \renewcommand{\arraystretch}{1.05}%
\begin{tabular}{@{}lllll@{}}
\toprule
Process & Main phase-space requirements & Observable & $\mathcal{L}$ [fb$^{-1}$] & Ref. \\
\midrule
$pp\to W^+W^-\to e^\pm\nu\,\mu^\mp\nu$  & $\mll>55\, \GeV$, $p^\text{jet}_\textrm{T}>35\,\GeV$& $\pT^{\ell}$ & 36 & \cite{STDM-2017-24}\\
$pp\to W^\pm Z\to \ell^\pm\nu\,\ell^+\ell^-$  & $\mll\in(66,116)\, \GeV$, $m_\mathrm{T}^W>30\,\GeV$ & \mtwz & 36 & \cite{STDM-2018-03}\\
$pp\to Zjj\to \ell^+\ell^- jj$  & $m_{jj}>1000\, \GeV$, $\mll\in(81,101)\, \GeV$ & $\Delta\phi_{jj}$ & 140 & \cite{STDM-2017-27}\\
\bottomrule
\end{tabular}
\renewcommand{\arraystretch}{1.0}%
}
\label{tab:ew-measurements}
\end{table}

\subsection{Differential cross-section measurements for neutral- and charged-current HMDY processes}

Measurements of neutral- and charged-current Drell--Yan processes at high invariant masses are included in the global combination, using the full Run 2 data.
For neutral-current HMDY, the differential cross-section is measured in the $\tau^+\tau^-$ final state. Following the fiducial phase space requirements, events are required to have two reconstructed hadronically decaying $\tau$-leptons with $p_\mathrm{T}(\tau) > 65$~GeV, $|\eta(\tau)| < 2.5$, and visible invariant mass $m_{\tau\tau}^{vis} > 100$~GeV. Cross-sections are measured differentially in $m_{\tau\tau}^{vis}$ and the measured fiducial cross-sections are unfolded to particle level.
For charged-current Drell--Yan, differential cross-sections for $W$-boson production are measured in the high transverse-mass  ($\mtw$) region. Results are obtained separately for both $W$-boson charges and lepton flavour ($e$ and $\mu$). Single- and double-differential Born-level cross-sections are measured up to $\mtw=5$~TeV and $|\eta(\ell)|=2.4$, within a fiducial region requiring $\pt^\ell > 65$~GeV and $\pt^\nu > 85$~GeV.

The HMDY measurements entering the global interpretation are summarised in Table~\ref{tab:dy-measurements}. The unfolded fiducial cross-sections for both analyses are directly compared to particle-level SMEFT predictions, parameterised in each bin as functions of the Wilson coefficients.

\begin{table}  [!htbp]
\caption{The neutral-current and charged-current HMDY analyses entering the combined SMEFT interpretation, together with the most important phase-space requirements in each measurement, the observables used in the interpretation, and the integrated luminosity of the data analysed in each measurement.}
\centering
{  \renewcommand{\arraystretch}{1.0}%
\begin{tabular}{@{}lllll@{}}
\toprule
Process & Important phase-space requirements & Observable & $\mathcal{L}$ [fb$^{-1}$] & Ref. \\
\midrule
$pp\rightarrow Z/\gamma^*\rightarrow \tau^{\pm}\tau^{\mp}$ & $m_{\tau\tau}^{vis}>100$~GeV &  $m_{\tau\tau}^{vis}$ & 140 & \cite{EXOT-2022-42} \\
$pp\rightarrow W \rightarrow \ell^{\pm} \nu$ (with $\ell = e,\mu$)& $200<\mtw<5000$~GeV & $\mtw$ & 140 & \cite{STDM-2018-41}\\
\bottomrule
\end{tabular}
}
\label{tab:dy-measurements}
\end{table}

\subsection{Differential $t\bar{t}$ cross-section measurements}

The top-quark data included in the global SMEFT interpretation comprise the measurements of unfolded differential cross-sections for $\ttbar$ production in the dilepton and in the boosted top-quark final states, using Run 2 data.

In the dilepton channel~\cite{TOPQ-2018-26}, the $\ttbar$ pair decays via $\ttbar \to W^+W^-\bbbar$, with both $W$ bosons decaying leptonically. Events with one electron and one muon of opposite charge are selected. Absolute and normalised differential cross-sections are measured for eight leptonic observables at particle level, in a fiducial region defined by the lepton transverse momentum, $\pT^\ell>27\ (25)$~GeV for the leading (subleading) lepton and pseudorapidity $|\eta(\ell)|<2.5$.
The distribution of the transverse momentum of the leading lepton $\pT^\ell$ is used for the SMEFT interpretation.

In the boosted channel~\cite{TOPQ-2018-26}, events with a leptonically decaying top quark and a hadronically decaying high-$\pT$ top quark are selected. Differential cross-sections are unfolded to particle level, with small-$R$ jets ($R=0.4$) clustered from stable particles with $|\eta|<4.5$, and large-$R$ jets ($R=1.0$) built from selected small-$R$ jets. Large-$R$ jets are required to satisfy $\pt>355$~GeV, $|\eta|<2.0$, invariant mass $120 < m_J < 220$~GeV, and contain at least one $b$-jet. The reconstructed highest-$\pT^J$ large-$R$ jet is identified as the hadronic top-quark candidate.
The differential cross-section as a function of the transverse momentum of the hadronically decaying top quark $\pT^{t,h}$ is used for the SMEFT interpretation, as this variable is the most sensitive to SMEFT effects.
The measurements are summarised in Table~\ref{tab:top-measurements}.

\begin{table}  [!htbp]
\caption{The top-quark processes entering the combined SMEFT interpretation, together with the most important phase-space requirements in each measurement, the observables used in the interpretation, and the integrated luminosity analysed in the measurement.}
\centering
{  \renewcommand{\arraystretch}{1.0}%
\begin{tabular}{@{}lllll@{}}
\toprule
Process & Important phase-space requirements & Observable & $\mathcal{L}$ [fb$^{-1}$] & Ref. \\
\midrule
$pp\to t\bar t\to Wb\,Wb\to e^\pm\mu^\mp \nu \nu b b$  & $\pT^{\ell}$ >27 GeV, $|\eta(\ell)|<2.5$ & $\pT^{\ell}$& 140& \cite{TOPQ-2018-26}\\
$pp\to t\bar t\to Wb\,Wb \to qq'b\,\ell \nu b$ & $\pT^J$ > 355 GeV & $\pT^{t,h}$  & 140& \cite{TOPQ-2019-23}\\
\bottomrule
\end{tabular}
}
\label{tab:top-measurements}
\end{table}

\subsection{Di-Higgs measurements}
Measurements of di-Higgs production at reconstruction level with decays into $b\bar{b}\gamma\gamma$ and $b\bar{b}\tau\tau$ are included in the global SMEFT combination. The analyses probe SMEFT effects in the gluon--gluon fusion (ggF) production mode, which dominates di-Higgs production, while SMEFT effects on VBF di-Higgs production are expected to be negligible. The observables entering the SMEFT interpretation are the measured signal yields in the analysis categories, together with the reconstructed di-Higgs invariant mass spectrum, $m_{HH}$, which provides the dominant sensitivity to modifications of the Higgs boson self-coupling.

The $HH\to b\bar{b}\gamma\gamma$ analysis~\cite{HDBS-2021-10} uses the full Run 2 data and defines a single signal region targeting events with exactly two high-\pT isolated photons and exactly two $b$-tagged jets. The photons are required to satisfy $\ET/m_{\gamma\gamma} > 0.35$ and $0.25$ for the leading and subleading photon, respectively. For the $HH \to b\bar{b}\tau\tau$ analysis~\cite{HDBS-2019-27}, events are required to contain exactly two $b$-tagged jets with $p_\mathrm{T}>45\,\GeV$ and $20\,\GeV$ respectively and exactly two $\tau$-lepton candidates selected in one of three signal regions: one hadronic region targeting events with two oppositely charged hadronically decaying $\tau$-leptons, and two lepton--hadron regions targeting events with one leptonically decaying $\tau$-lepton and one hadronically decaying $\tau$-leptons, with electron and muon channels combined.

The measurements entering the interpretation are summarised in Table~\ref{tab:HH-measurements}. SMEFT effects are included by accounting for modifications of the ggF total cross-section and the $m_{HH}$ spectrum, as well as changes to the total cross-section of single-Higgs boson backgrounds. In addition, both the total and partial widths of the Higgs boson are parameterised as functions of the Wilson coefficients.

\begin{table}[!htbp]
\caption{The di-Higgs measurements entering the combined SMEFT interpretation, together with the targeted production mode and integrated luminosity of the analysed data.  }
\centering
\renewcommand{\arraystretch}{1.0}
\begin{tabular}{@{}lllrl@{}}
\toprule
Process & Production mode & $\mathcal{L}$ [fb$^{-1}$] & Ref. \\
\midrule
$HH \to b\bar{b}\gamma \gamma$ & ggF & 140 & \cite{HDBS-2021-10}\\
$HH \to b\bar{b}\tau\tau$ & ggF & 140 & \cite{HDBS-2019-27}\\
\bottomrule
\end{tabular}
\label{tab:HH-measurements}
\end{table}

\subsection{Electroweak precision observables}

The EWPO considered in this analysis probe the couplings of the $Z$ and $W$ bosons to charged leptons, neutrinos, and heavy quarks. Key observables measured at LEP and SLD~\cite{ALEPH:2005ab} include the total and partial $Z$ widths, $\Gamma_Z$, the hadronic cross-section at the $Z$ pole (defined as the total cross-section for Z decays into hadrons), $\sigma_\mathrm{had}$, the ratio of hadronic to leptonic $Z$-boson decays for charged leptons, $\Rl$, and the fractions of hadronic $Z$ decays into charm and bottom quarks, $\Rc$ and $\Rb$, respectively. Forward--backward asymmetries $A_\mathrm{FB}^{0,f}$, lepton polarisation and asymmetry measurements, $W$-boson branching ratios are also included.

Additionally, ATLAS measurements of lepton-flavour-universality ratios~\cite{STDM-2012-20,TOPQ-2018-29,TOPQ-2023-28}, $R_W^{\tau/\mu}$ and $R_{WZ}^{\mu/e}$, and the $W$-boson total width \cite{STDM-2019-24} are considered. These observables provide sensitivity to deviations from the SM through their dependence on left- and right-handed electroweak couplings.

Table~\ref{tab:EWPO_list} summarises the 26 EWPO included in the combined SMEFT fit.

\begin{table}[!htbp]
\centering
\caption{Electroweak precision observables~\cite{ALEPH:2005ab,STDM-2012-20,TOPQ-2018-29,TOPQ-2023-28,STDM-2019-24}, included in the analysis. The second column corresponds to the experimental value, the third to the theory prediction in the $\{{m}_W, {m}_Z,{G}_F\}$ scheme, and the fourth is the ratio of the two values. The correlation between the observables is shown in Table~\ref{tab:EWPO-correlation}.}
\label{tab:EWPO_list}
\renewcommand{\arraystretch}{1.1}
\begin{tabular}{@{}l*3{r@{ $\pm$ }l}@{}}
\toprule
Observable  & \multicolumn{2}{c}{Measurement} & \multicolumn{2}{c}{Prediction} & \multicolumn{2}{c}{Ratio}\\
\midrule
$\Deltaalpha$	&$	0.05903	$&$	0.00009	$&$	0.05911	$&$	0.00098	$&$	0.999	$&$	0.017	$\\
$\GammaZ$\,[GeV]	&$	2.4955	$&$	0.0023	$&$	2.4945	$&$	0.0010	$&$	1.0004	$&$	0.0010	$\\
$\Rel$	&$	20.804	$&$	0.050	$&$	20.751	$&$	0.010	$&$	1.0025	$&$	0.0024	$\\
$\Rmu$	&$	20.784	$&$	0.034	$&$	20.751	$&$	0.010	$&$	1.0016	$&$	0.0017	$\\
$\Rtau$	&$	20.764	$&$	0.045	$&$	20.799	$&$	0.010	$&$	0.9983	$&$	0.0022	$\\
$\Rc$	&$	0.1721	$&$	0.0030	$&$	0.1722	$&$	0.0001	$&$	0.999	$&$	0.017	$\\
$\Rb$	&$	0.21629	$&$	0.00066	$&$	0.21587	$&$	0.00010	$&$	1.0019	$&$	0.0031	$\\
\sigmahad\,[pb]	&$	41481	$&$	33	$&$	41489	$&$	8	$&$	0.9998	$&$	0.0008	$\\
$\AeSLD$	&$	0.1516	$&$	0.0021	$&$	0.1470	$&$	0.0025	$&$	1.031	$&$	0.022	$\\
$\AeLEP$	&$	0.1498	$&$	0.0049	$&$	0.1470	$&$	0.0025	$&$	1.019	$&$	0.037	$\\
$\AmuSLD$	&$	0.142	$&$	0.015	$&$	0.147	$&$	0.003	$&$	0.97	$&$	0.11	$\\
$\AtauSLD$	&$	0.136	$&$	0.015	$&$	0.147	$&$	0.003	$&$	0.92	$&$	0.11	$\\
$\AtauLEP$	&$	0.1439	$&$	0.0043	$&$	0.1470	$&$	0.0025	$&$	0.979	$&$	0.035	$\\
$\AFBe$	&$	0.0145	$&$	0.0025	$&$	0.0162	$&$	0.0006	$&$	0.89	$&$	0.18	$\\
$\AFBmu$	&$	0.0169	$&$	0.0013	$&$	0.0162	$&$	0.0006	$&$	1.042	$&$	0.084	$\\
$\AFBtau$	&$	0.0188	$&$	0.0017	$&$	0.0162	$&$	0.0006	$&$	1.159	$&$	0.095	$\\
$\AFBb$	&$	0.0992	$&$	0.0016	$&$	0.1031	$&$	0.0018	$&$	0.962	$&$	0.024	$\\
$\AFBc$	&$	0.0707	$&$	0.0035	$&$	0.0737	$&$	0.0014	$&$	0.959	$&$	0.053	$\\
$\Ab$	&$	0.923	$&$	0.020	$&$	0.935	$&$	<0.001	$&$	0.987	$&$	0.022	$\\
$\Ac$	&$	0.670	$&$	0.027	$&$	0.668	$&$	0.001	$&$	1.003	$&$	0.040	$\\
$\GammaW$\,[GeV]	&$	2.198	$&$	0.049	$&$	2.090	$&$	0.001	$&$	1.052	$&$	0.022	$\\
$\BrWe$	&$	0.1071	$&$	0.0016	$&$	0.1082	$&$	<0.0001	$&$	0.990	$&$	0.015	$\\
$\BrWmu$	&$	0.1063	$&$	0.0015	$&$	0.1082	$&$	<0.0001	$&$	0.983	$&$	0.014	$\\
$\BrWtau$	&$	0.1138	$&$	0.0021	$&$	0.1082	$&$	<0.0001	$&$	1.052	$&$	0.018	$\\
$R_{WZ}^{\mu / e}$	&$	0.9990	$&$	0.0042	$&$	1.0000	$&$	<0.0001	$&$	0.999	$&$	0.004	$\\
$R_{W}^{\tau / \mu}$	&$	0.992	$&$	0.013	$&$	1.000	$&$	<0.001	$&$	0.992	$&$	0.013	$\\
\bottomrule
\end{tabular}
\renewcommand{\arraystretch}{1.0}
\end{table}

\subsection{Event overlap between input analyses}

Potential overlaps of selected events among the Higgs boson, electroweak, HMDY, top-quark, and di-Higgs measurements entering the combination are assessed using three complementary approaches:
\begin{itemize}
\item using run number and event number, which provide a unique identifier for each recorded data event, where information is available;
\item performing bootstrap-based tests of statistical correlations, where information is available and the size of the event overlap is not negligible;
\item comparing selection criteria.
\end{itemize}

Run number and event number information is available for the Higgs-boson, HMDY and di-Higgs input measurements. Pairwise comparisons between these analyses show that the number of overlapping events is typically small, corresponding to relative event fractions well below 1\% of the events in their respective signal regions. A few exceptions are described here, together with their treatment.

The overlap for the $H\rightarrow \tau\tau$ and high-mass $\tau\tau$ Drell--Yan analyses is concentrated in the lowest-mass region of the latter, and the first bin of the HMDY $\tau\tau$ analysis is removed accordingly. In the $H\to WW^*$ measurement, the 0-jet $WW$ control region overlaps with the dedicated $WW$ analysis. The 0-jet $WW$ control region is therefore excluded from the combination and the associated normalisation parameter is treated as correlated with that of the $WW$ measurement. Overlaps also occur in shared top-quark enriched control regions. However, since these regions are used to measure freely floating background normalisation factors, the overlaps do not bias the combined result.

Sizeable overlaps not addressed by these exclusions are investigated using bootstrap techniques~\cite{ATL-PHYS-PUB-2021-011}. A subset of the analyses are re-evaluated on Poisson-fluctuated replicas of the nominal dataset, and correlations between the signal strengths of the individual measurements are extracted from the resulting distributions. The tested combinations, including $t\bar{t}$ boosted, $t\bar{t}(H\to b\bar{b})$, $V(H\to b\bar{b})$, $t\bar{t}$ dilepton, and $H\to WW^*$, do not exhibit significant cross-analysis correlations.

For analyses without run number and event number information, comparison of selection criteria indicates that any potential overlaps arise only in regions governed by floating background normalisations, and are therefore expected to be negligible. Based on these studies, the remaining overlaps among the input analyses are not expected to bias the combined interpretation.


\section{Theoretical predictions}
\label{sec:theory}

\subsection{SMEFT model}

In SMEFT models with dimension-six operators, the measured cross-sections are related to the corresponding scattering amplitudes $\mathcal{M}_{\text{SMEFT}}$ through the following relation:
\begin{equation}
\label{eqn:SMEFT_amplitude}
|\mathcal{M}_{\text{SMEFT}}|^2
=
|\mathcal{M}_{\text{SM}}|^2
+
2 \sum_i \frac{c_i}{\Lambda^2}
\mathrm{Re}\!\left(
\mathcal{M}_{\text{SM}}^{*}\,\mathcal{M}_i
\right)
+
\sum_{i,j} \frac{c_i c_j}{\Lambda^4}
\left(
\mathcal{M}_i^{*}\,\mathcal{M}_j
\right) ,
\end{equation}
where $i$ and $j$ run over all possible dimension-six operators. The first term is the SM contribution, while the second term represents the SM-BSM interference, linear in the BSM coupling and suppressed by the square of the new physics scale ($1/\Lambda^2$). The third term is the pure BSM contribution, quadratic in the BSM coupling and suppressed by the fourth power of the new physics scale ($1/\Lambda^4$).\newline
While the linear terms generally provide the dominant contribution, quadratic terms can become relevant for high $\pT$ regions, where their omission may lead to unphysical negative cross-sections. Interference effects with dimension-8 operators also enter at order $\mathcal{O}(\Lambda^{-4})$, but these are neglected in the present analysis. As a consequence, the difference observed between the purely linear model and the linear-plus-quadratic model can be considered a qualitative estimate of the uncertainty arising from the missing $\mathcal{O}(\Lambda^{-4})$ contributions.

Throughout this study, $\Lambda$ is set to  $1~\TeV$, which is a conventional reference scale used to express dimension-six Wilson coefficients. This choice does not correspond to a physical new-physics mass scale, since only the combinations $c_i/\Lambda^2$ are directly constrained by experiment. Fixing $\Lambda$ provides a useful normalisation for reporting and comparing limits on different coefficients~\cite{Brivio:2201.04974}.

An observable $O_b$, e.g. the cross-section of a given process in a measurement bin $b$, can be expressed as
\begin{equation}
O_b = O^\text{SM}_b\left(1 + \sum_i A_{bi}c_i + \sum_{i} B_{bi}c_i^2 + \sum_{i<j} C_{bij}c_ic_j\right),
\label{eq:smeft-para}
\end{equation}
where $O^\text{SM}_b$ is its SM value and $A_{bi}$, $B_{bi}$, and $C_{bij}$ correspond to the linear, quadratic, and cross term values, and are obtained from simulation based on matrix elements from Eq. \ref{eqn:SMEFT_amplitude}.

Higgs boson production and decay are factorised based on the narrow width approximation with the expression of Equation~\ref{eq:smeft-para} applied separately to the production cross-section $\sigma_p$ (in each STXS bin), partial decay width $\Gamma_d$, as well as to the total Higgs boson width $\Gamma_\text{tot}$ and combined as $\sigma_p \times \Gamma_d / \Gamma_\text{tot}$.
A full Taylor expansion of this ratio is avoided for single-Higgs and di-Higgs parameterisations, since it breaks down for operators affecting dominant Higgs boson decays or with large uncertainties (see Ref.~\cite{HIGG-2022-17}). In the single-Higgs case, the linear model instead uses the ratio of SMEFT-modified partial and total widths, which inherently introduces non-linear terms in the Wilson coefficients. For the di-Higgs case, the branching-ratio and production cross-section modifications are multiplied rather than added. For electroweak bosons, whose widths are tightly constrained, the full expansion is used.

\subsection{SMEFT parameterisation}

\subsubsection{Input parameters and symmetry assumptions}
The $\{m_W,m_Z,G_\mathrm{F}\}$ electroweak input parameter scheme is used for the theoretical prediction of observables at the LHC and LEP.
The measurements of the $Z$ boson mass $m_Z$ at LEP~\cite{ALEPH:2005ab} and the $W$ boson mass $m_W$ at ATLAS~\cite{STDM-2019-24} are sufficiently model-independent for these inputs to remain valid in a more general SMEFT scenario.
The value of the Fermi constant $G_\mathrm{F}$ is determined from muon decays, which are affected by the Wilson coefficients \cHlthree and \cllone, such that the true value of $G_\mathrm{F}$ is a function of these parameters.
Quarks from the first two generations and leptons from all three generations are assumed to be massless, and the four-flavour scheme is used for parton distribution functions. In this interpretation, the top-flavour symmetry scheme is adopted, which assumes a $U(2)^3$ symmetry in the quark sector. Under this scheme, the quarks of the first two generations and the third generation are treated independently. In the lepton sector, flavour diagonality is assumed, so that the lepton-antilepton pair of each generation is described independently.
These relaxed assumptions allow for a dedicated treatment of heavy-flavour quarks and each lepton generation, enabling tests of lepton flavour universality violation. The assumptions introduced in the quark sector with respect to a general flavour symmetry scheme are justified by the set of measurements that are included in this work, which would not provide sufficient sensitivity to distinguish between operators involving light quarks.

Only operators that conserve charge-parity (CP) symmetry, i.e. CP-even operators, are considered and the Wilson coefficients are assumed to be real-valued. The list of operators and corresponding Wilson coefficients relevant to the processes studied in this paper is given in Tables~\ref{tab:relevant-operators-1} and \ref{tab:relevant-operators-2}. The Warsaw basis operators are grouped into purely bosonic operators, operators containing both boson and fermion fields, four-lepton operators, four-fermion operators containing both quark and lepton fields, and four-quark operators. Operators with negligible impact on the combined measurements within the sensitivity of the datasets considered are not included in these tables.

\begin{table}
\caption{Dimension-six operators affecting the analysed processes (excluding four-fermion operators with quark fields).
The Warsaw basis Wilson coefficients and the corresponding operators are listed in the first and second leftmost columns, respectively.
The remaining columns indicate the processes that are affected by the operator.
Operator groups are separated by horizontal lines.
First and second (third) generation left-handed quark fields are denoted as $q$ ($Q$).
First and second (third) generation right-handed quark fields are denoted as $u,d$ ($t,b$).
Left and right-handed lepton fields are denoted $l$ and $e$, respectively.
The lower-case roman letter flavour indices run over 1, 2 for quarks and 1, 2, 3 for leptons.}
\label{tab:relevant-operators-1}
\centering
\small
\renewcommand{\arraystretch}{1.15}
\begin{tabular}{@{}llP{0.08\linewidth}P{0.08\linewidth}P{0.08\linewidth}P{0.08\linewidth}P{0.08\linewidth}P{0.08\linewidth}}
\toprule
\multicolumn{2}{@{}l}{Wilson coefficient and operator}  & \multicolumn{5}{c}{Affected process group}\\
& &             & ATLAS  & ATLAS & ATLAS & ATLAS & ATLAS  \\
& & EWPO        &  Higgs &  electroweak &  $t \bar t$ &  HMDY &  di-Higgs \\
\midrule
\cHbox      &  \scalebox{0.8}{$(H^\dagger H)\square(H^\dagger H)$}                                                    &             &$\checkmark$ &              &              &                & $\checkmark$ \\
\cG         &  \scalebox{0.8}{\defG}                                                                                  &             &$\checkmark$ & $\checkmark$ & $\checkmark$ &                &                \\
\cW         &  \scalebox{0.8}{$\epsilon^{IJK} W_\mu^{I\nu} W_\nu^{J\rho} W_\rho^{K\mu}$}                              &             &$\checkmark$ & $\checkmark$ &              &                & $\checkmark$\\
\cHDD       &  \scalebox{0.8}{$\ \left(H^\dag D_\mu H\right)^* \left(H^\dag D_\mu H\right)$}                          &             &$\checkmark$ & $\checkmark$ &              &                & $\checkmark$\\
\cHG        &  \scalebox{0.8}{\defHG}                                                                                 &             &$\checkmark$ &              & $\checkmark$ &                & $\checkmark$  \\
\cHB        &  \scalebox{0.8}{\defHB}                                                                                 &             &$\checkmark$ &              &              &                & $\checkmark$\\
\cHW        &  \scalebox{0.8}{\defHW}                                                                                 &             &$\checkmark$ &              &              &                & $\checkmark$\\
\cHWB       &  \scalebox{0.8}{$ H^\dag \sigma^I H\, W^I_{\mu\nu} B^{\mu\nu}$}                                         &$\checkmark$ &$\checkmark$ & $\checkmark$ &              &                & $\checkmark$\\
\hline
\ceHtwo     &  ${(H^\dag H)(\bar l_p e_r H)}_{22}$                                                                    &             &$\checkmark$ &              &              &                & $\checkmark$ \\
\ceHthree   &  ${(H^\dag H)(\bar l_p e_r H)}_{33}$                                                                    &             &$\checkmark$ &              &              &                & $\checkmark$ \\
\ctH        &  \scalebox{0.8}{\deftH}                                                                                 &             &$\checkmark$ &              &               &               &$\checkmark$  \\
\cbH        &  \scalebox{0.8}{\defbH}                                                                                 &             &$\checkmark$ &              &              &                & $\checkmark$ \\
\hline
\cHloneone  &  \scalebox{0.8}{$(H^\dag i\overleftrightarrow{D}_\mu H)(\bar{\ell}_p \gamma^\mu \ell_r)_{11}$}                     &$\checkmark$ &$\checkmark$ &$\checkmark$ &                &               & \\
\cHlonetwo  &  \scalebox{0.8}{$(H^\dag i\overleftrightarrow{D}_\mu H)(\bar{\ell}_p \gamma^\mu  \ell_r)_{22}$}                     &$\checkmark$ &$\checkmark$ &$\checkmark$ &                &               & \\
\cHlonethree&  \scalebox{0.8}{$(H^\dag i\overleftrightarrow{D}_\mu H)(\bar{\ell}_p \gamma^\mu  \ell_r)_{33}$}                     &$\checkmark$ &$\checkmark$ &$\checkmark$ &                & $\checkmark$  & \\
\cHlthreeone   &  \scalebox{0.8}{$(H^\dag i\overleftrightarrow{D}^I_\mu H)(\bar{\ell}_p \sigma^I \gamma^\mu  \ell_r)_{11}$}       &$\checkmark$ &$\checkmark$ &$\checkmark$ &                & $\checkmark$  & $\checkmark$ \\
\cHlthreetwo   &  \scalebox{0.8}{$(H^\dag i\overleftrightarrow{D}^I_\mu H)(\bar{\ell}_p \sigma^I \gamma^\mu  \ell_r)_{22}$}       &$\checkmark$ &$\checkmark$ &$\checkmark$ &                & $\checkmark$  & $\checkmark$ \\
\cHlthreethree   &  \scalebox{0.8}{$(H^\dag i\overleftrightarrow{D}^I_\mu H)(\bar{\ell}_p \sigma^I \gamma^\mu  \ell_r)_{33}$}     &$\checkmark$ &$\checkmark$ &$\checkmark$ &                & $\checkmark$  & $\checkmark$ \\
\cHeone     &  \scalebox{0.8}{$(H^\dag i\overleftrightarrow{D}_\mu H)(\bar e \gamma^\mu e)_{11}$}                     &$\checkmark$ &$\checkmark$ &$\checkmark$ &                &               & \\
\cHetwo     &  \scalebox{0.8}{$(H^\dag i\overleftrightarrow{D}_\mu H)(\bar e \gamma^\mu e)_{22}$}                     &$\checkmark$ &$\checkmark$ &$\checkmark$ &                &               & \\
\cHethree   &  \scalebox{0.8}{$(H^\dag i\overleftrightarrow{D}_\mu H)(\bar e \gamma^\mu e)_{33}$}                     &$\checkmark$ &$\checkmark$ &             &                & $\checkmark$  & \\
\cHjone     &  \scalebox{0.8}{$(H^\dag i\overleftrightarrow{D}_\mu H)(\bar q \gamma^\mu q)$}                          &$\checkmark$ &$\checkmark$ &$\checkmark$ &                &               & \\
\cHjthree   &  \scalebox{0.8}{$(H^\dag i\overleftrightarrow{D}^I_\mu H)(\bar q \sigma^I \gamma^\mu q)$}               &$\checkmark$ &$\checkmark$ &$\checkmark$ &                & $\checkmark$ & $\checkmark$\\
\cHu        &  \scalebox{0.8}{$(H^\dag i\overleftrightarrow{D}_\mu H)(\bar u \gamma^\mu u)$}                          &$\checkmark$ &$\checkmark$ &$\checkmark$ &                &               & \\
\cHd        &  \scalebox{0.8}{$(H^\dag i\overleftrightarrow{D}_\mu H)(\bar d \gamma^\mu d)$}                          &$\checkmark$ &$\checkmark$ &$\checkmark$ &                &               & \\
\cHQone     &  \scalebox{0.8}{$(H^\dag i\overleftrightarrow{D}_\mu H)(\bar Q \gamma^\mu Q)$}                          &$\checkmark$ &$\checkmark$ & $\checkmark$&                &               & \\
\cHQthree   &  \scalebox{0.8}{$(H^\dag i\overleftrightarrow{D}^I_\mu H)(\bar Q \sigma^I \gamma^\mu Q)$}               &$\checkmark$ &$\checkmark$ & $\checkmark$ & $\checkmark$   &             & \\
\cHt        &  \scalebox{0.8}{$(H^\dag i\overleftrightarrow{D}_\mu H)(\bar t \gamma^\mu t)$}                          &             &$\checkmark$ &              &               &               & \\
\cHb        &  \scalebox{0.8}{$(H^\dag i\overleftrightarrow{D}_\mu H)(\bar b \gamma^\mu b)$}                          &$\checkmark$ &$\checkmark$ &              &               &               & \\
\hline
\ctG       &  \scalebox{0.8}{\deftG}                                                                                  &             &$\checkmark$ &              &              &                & $\checkmark$ \\
\ctW       &  \scalebox{0.8}{\deftW}                                                                                  &             &$\checkmark$ &              & $\checkmark$ &                & $\checkmark$\\
\ctB       &  \scalebox{0.8}{\deftB}                                                                                  &             &$\checkmark$ &              &              &                & $\checkmark$\\
\cH        &  \scalebox{0.8}{\defH}                                                                                   &             &$\checkmark$ &              &              &                & $\checkmark$ \\
\ceW       &  \scalebox{0.8}{\defeW}                                                                                  &             &$\checkmark$ &              &              & $\checkmark$   & \\
\ceB       &  \scalebox{0.8}{\defeB}                                                                                  &             &$\checkmark$ &              &              & $\checkmark$   & \\

\hline
\cllonetwoone    &  \scalebox{0.8}{$(\bar l \gamma_\mu l)(\bar l \gamma^\mu l)_{1221}$}                               &$\checkmark$ &$\checkmark$ &$\checkmark$ &              & $\checkmark$    & $\checkmark$\\
\bottomrule
\end{tabular}
\end{table}

\begin{table}
\caption{Dimension-six operators affecting the analysed processes (four-fermion operators with quark fields only).
The Warsaw basis Wilson coefficients and the corresponding operators are listed in the first and second columns, respectively.
The remaining columns indicate the processes that are affected by the operator.
First and second (third) generation left-handed quark fields are denoted as $q$ ($Q$).
First and second (third) generation right-handed quark fields are denoted as $u,d$ ($t,b$).
Left and right-handed lepton fields are denoted $l$ and $e$, respectively.
The lower-case roman letter flavour indices run over 1, 2 for quarks and 1, 2, 3 for leptons. EWPO are not affected by the operators listed.}
\label{tab:relevant-operators-2}
\captionsetup[table]{skip=1pt}
\centering
\small
\renewcommand{\arraystretch}{1.05}
\begin{tabular}{@{}llP{0.08\linewidth}P{0.08\linewidth}P{0.08\linewidth}P{0.08\linewidth}P{0.08\linewidth}}
\toprule
\multicolumn{2}{@{}l}{Wilson coefficient and operator}  & \multicolumn{5}{c}{Affected process group}\\
&    & ATLAS  & ATLAS & ATLAS & ATLAS & ATLAS  \\
&    &  Higgs &  electroweak &  $t \bar t$ &  HMDY &  di-Higgs \\
\midrule
$c_{lq,11^{(1)}}$ &   \scalebox{0.8}{$(\bar l \gamma_\mu l)(\bar q \gamma^\mu q)_{11}$}                         & &$\checkmark$ & & &\\
$c_{lq,22^{(1)}}$ &   \scalebox{0.8}{$(\bar l \gamma_\mu l)(\bar q \gamma^\mu q)_{22}$}                         & &$\checkmark$ & & &\\
$c_{lq,33^{(1)}}$ &   \scalebox{0.8}{$(\bar l \gamma_\mu l)(\bar q \gamma^\mu q)_{33}$}                         & & & & $\checkmark$ &\\
$c_{lq,11^{(3)}}$ &   \scalebox{0.8}{$(\bar l \gamma_\mu \sigma^I l)(\bar q \gamma^\mu \sigma^I q)_{11}$}       & &  & & $\checkmark$ &\\
$c_{lq,22^{(3)}}$ &   \scalebox{0.8}{$(\bar l \gamma_\mu \sigma^I l)(\bar q \gamma^\mu \sigma^I q)_{22}$}       & &$\checkmark$ & & $\checkmark$ &\\
$c_{lq,33^{(3)}}$ &   \scalebox{0.8}{$(\bar l \gamma_\mu \sigma^I l)(\bar q \gamma^\mu \sigma^I q)_{33}$}       & &  & & $\checkmark$ &\\
$c_{lu,11}$&  \scalebox{0.8}{$(\bar l \gamma_\mu l)(\bar u \gamma^\mu u)_{11}$}                            & &$\checkmark$ & & &\\
$c_{lu,22}$&  \scalebox{0.8}{$(\bar l \gamma_\mu l)(\bar u \gamma^\mu u)_{22}$}                            & &$\checkmark$ & & &\\
$c_{lu,33}$&  \scalebox{0.8}{$(\bar l \gamma_\mu l)(\bar u \gamma^\mu u)_{33}$}                            & & & & $\checkmark$&\\
$c_{ld,11}$ &  \scalebox{0.8}{$(\bar l \gamma_\mu l)(\bar d \gamma^\mu d)_{11}$}                           & &$\checkmark$ & & &\\
$c_{ld,22}$ &  \scalebox{0.8}{$(\bar l \gamma_\mu l)(\bar d \gamma^\mu d)_{22}$}                           & &$\checkmark$ & & &\\
$c_{ld,33}$ &  \scalebox{0.8}{$(\bar l \gamma_\mu l)(\bar d \gamma^\mu d)_{33}$}                           & & & & $\checkmark$&\\
$c_{eu,33}$ &  \scalebox{0.8}{$(\bar e \gamma_\mu e)(\bar u \gamma^\mu u)_{33}$}                           & & & & $\checkmark$&\\
$c_{ed,33}$ &  \scalebox{0.8}{$(\bar e \gamma_\mu e)(\bar d \gamma^\mu d)_{33}$}                           & & & & $\checkmark$&\\
$c_{ej,33}$ &  \scalebox{0.8}{$(\bar e \gamma_\mu e)(\bar q \gamma^\mu q)_{33}$}                           & & & & $\checkmark$&\\
\hline
\cjjoo &  \scalebox{0.8}{\defjjoo}                                                                         & &$\checkmark$ & & &\\
\cjjoe &  \scalebox{0.8}{\defjjoe}                                                                         & &$\checkmark$ & & &\\
\cjjto &  \scalebox{0.8}{\defjjto}                                                                         & &$\checkmark$ & & &\\
\cjjte &  \scalebox{0.8}{\defjjte}                                                                         & &$\checkmark$ & & &\\
\cuuo &  \scalebox{0.8}{\defuuo}                                                                           & &$\checkmark$ & & &\\
\cjuo &  \scalebox{0.8}{\defjuo}                                                                           & &$\checkmark$ & & &\\
\cjdo &  \scalebox{0.8}{\defjdo}                                                                           & &$\checkmark$ & & &\\\hline
\cQjoo & \scalebox{0.8}{\defQjoo}                                                                          & $\checkmark$ & & $\checkmark$& &  \\
\cQjoe & \scalebox{0.8}{\defQjoe}                                                                          & $\checkmark$ & & $\checkmark$& &  \\
\cQjto & \scalebox{0.8}{\defQjto}                                                                          & $\checkmark$ & & $\checkmark$& &  \\
\cQjte & \scalebox{0.8}{\defQjte}                                                                          & $\checkmark$ & & $\checkmark$& &  \\
\ctuo & \scalebox{0.8}{\deftuo}                                                                            & $\checkmark$ & & $\checkmark$& &  \\
\ctue & \scalebox{0.8}{\deftue}                                                                            & $\checkmark$ & &  $\checkmark$& &  \\
\ctdo & \scalebox{0.8}{\deftdo}                                                                            & $\checkmark$ & &  $\checkmark$& &  \\
\ctde & \scalebox{0.8}{\deftde}                                                                            & $\checkmark$ & &  $\checkmark$& &  \\
\cQuo & \scalebox{0.8}{\defQuo}                                                                            & $\checkmark$ & & $\checkmark$& &  \\
\cQue & \scalebox{0.8}{\defQue}                                                                            & $\checkmark$ & & $\checkmark$& &  \\
\cQdo & \scalebox{0.8}{\defQdo}                                                                            & $\checkmark$ & & $\checkmark$& &  \\
\cQde & \scalebox{0.8}{\defQde}                                                                            & $\checkmark$ & & $\checkmark$& &  \\
\ctjo & \scalebox{0.8}{\deftjo}                                                                            & $\checkmark$ & & $\checkmark$& &  \\
\ctje & \scalebox{0.8}{\deftje}                                                                            & $\checkmark$ & & $\checkmark$& &  \\ \hline
$c_{Qe,33}$ & \scalebox{0.8}{$(\bar Q \gamma_\mu Q)(\bar e \gamma^\mu e)_{33}$}                           & & & & $\checkmark$ & \\
$c_{Ql,33^{(1)}}$ & \scalebox{0.8}{$(\bar Q \gamma_\mu Q)(\bar l \gamma^\mu l)_{33}$}                          & & & & $\checkmark$ & \\
$c_{Ql,33^{(3)}}$ & \scalebox{0.8}{$(\bar Q \sigma^I \gamma_\mu Q)(\bar l \sigma^I \gamma^\mu l)_{33}$}        & & & & $\checkmark$ & \\
$c_{be,33}$ & \scalebox{0.8}{$(\bar b \gamma_\mu b)(\bar e \gamma^\mu e)_{33}$}                           & & & & $\checkmark$ & \\
$c_{bl,33}$ & \scalebox{0.8}{$(\bar b \gamma_\mu b)(\bar l \gamma^\mu l)_{33}$}                           & & & & $\checkmark$ & \\
\bottomrule
\end{tabular}
\end{table}

\subsubsection{Simulation of the effect of dimension-six operators on LHC processes}
\label{sec:dimsixsim}

Dedicated samples representing the SM, linear, and quadratic effects of dimension-six operators, as well as dimension-six cross terms, are generated at leading-order (LO) using \MGNLO~\cite{Alwall:2011uj} with the SMEFTsim 3.0 model~\cite{Brivio:2017btx,Brivio:2020onw}. Events are interfaced to \PYTHIA 8~\cite{Sjostrand:2014zea} for parton showering and hadronisation, while the decays of bottom and charm hadrons are simulated with \EVTGEN~\cite{EvtGen}.
For single-Higgs loop-induced processes ($gg\to H$, $gg \to ZH$, and $H\to gg$) as well as for $t\bar{t}$ analyses, calculations are performed with the SMEFTatNLO model~\cite{Degrande:2020evl}, which provides next-to-leading-order (NLO) QCD corrections. For the \(H \to \gamma\gamma\) and \(H\to Z\gamma\) decays, analytical calculations including one-loop electroweak SMEFT contributions are used~\cite{Actis:2008ug,Bredenstein:2006nk}.

Minor modifications to the single-Higgs parameterisation, relative to the one published in Ref.~\cite{HIGG-2022-17}, are implemented, to obtain a common STXS parameterisation between ATLAS and CMS. These consist of using fixed renormalisation and factorisation scales instead of dynamic ones in the ggH samples generated with \amc .

Dimension-six operators can also modify the mass and width of intermediate particles, which becomes an important correction in cases where intermediate particles are on shell. To account for these effects in the linearised SMEFT expansion, propagator corrections are computed with SMEFTsim 3.0 for the Higgs boson measurements using dummy fields, following the methodology of Refs.~\cite{Brivio:2020onw,Brooijmans:2018xbu}. In short, on-shell effects from dimension-six operators are incorporated consistently at linear order by treating mass and width shifts as propagator corrections. Quadratic propagator corrections are expected to be small and are neglected. The electroweak measurements ($WW$, $WZ$, and EW $Zjj$) are dominated by on-shell bosons. In these cases, instead of propagator corrections, a branching-ratio correction is applied. For each final state boson, a linear Wilson-coefficient-dependent correction of $-\frac{\delta\Gamma^\text{SMEFT}}{ \Gamma^\text{SM}}$ is included, using the predictions of the \ewpdlhc tool~\cite{Mildner:2024wbl}.
This approach is validated to yield results consistent with the propagator correction method, while avoiding the need for large-scale Monte Carlo generation and ensuring consistency with EWPO.

Higher-order QCD and electroweak corrections are included under the assumption that their relative effect does not change in the presence of dimension-six operators. The differential cross-section predictions in SMEFT are obtained as a product of the best available SM particle-level prediction in each bin and the ratio of the \MGNLOPY SMEFT over SM prediction. It is also assumed that the relative effect of theory uncertainties derived for the SM processes remains valid in the presence of dimension-six operators.

The effects of dimension-six operators on background processes are not taken into account, with two exceptions.
For the interpretation of the EW $Zjj$ measurement, their effects are simulated both for electroweak $Zjj$ production and for $Z$ + jets production via the strong interaction, as both contribute comparably to the signal region. The second exception is $WW$ production in the 0-jet $H\to WW^*$ signal region. In this case, the $WW$ background normalisation is modelled analogously to the signal parameterisation in the $WW$ measurement. In the di-Higgs channels $b\bar{b}\tau\tau$ and $b\bar{b}\gamma\gamma$, the single-Higgs background is parameterised explicitly.
Other background processes are either subdominant compared with the signal or are normalised to data in kinematically similar control regions.

An assumption made for the Higgs-boson cross-section measurements in the STXS categories is that the acceptance is constant within each bin, compared to the statistical uncertainty in the measurement of that bin's cross-section. Therefore, no impact of new physics, i.e. SMEFT operators, on the acceptance is taken into account for the STXS categories. SMEFT operators can however modify the kinematics of the Higgs boson decay products, potentially leading to acceptance differences between the SM and SMEFT. For most decays, in particular two-body decays, this effect is found to be negligible within the range of the coefficients probed by the analyses.

Dedicated studies are performed to investigate the effect of SMEFT operators on the acceptance for $H\to ZZ^* \to 4\ell$ and $H\to WW^* \to \ell\nu\ell\nu$ decays. Due to the specific selection targeting the signal phase-space, variations in the acceptance are observed and accounted for through dedicated corrections. The acceptance parameterisation is re-derived for the $H\to ZZ^* \to 4\ell$ and $H\to WW^* \to \ell\nu\ell\nu$ channels using SMEFTsim 3.0 for all the parameters that affect the Higgs boson decay.
For $H\to ZZ^* \to 4\ell$, a uniform acceptance parameterisation is applied across the STXS bins. For $H\to WW^* \to \ell\nu\ell\nu$, a uniform correction is applied across the ggF and VBF production modes.
Further details on how acceptance effects enter the SMEFT parameterisation can be found in Ref.~\cite{HIGG-2022-17}.

\subsubsection{Predictions for electroweak precision observables}
\label{sec:ewpd-pred}

SM and SMEFT predictions in the $\{m_W,m_Z,G_\mathrm{F}\}$ input scheme are obtained using the \ewpdlhc tool~\cite{Mildner:2024wbl}, which provides SMEFT parameterisations compatible with the symmetry assumptions commonly adopted in LHC interpretations. The SMEFT predictions include contributions quadratic in dimension-six Wilson coefficients, incorporated consistently with their treatment in LHC analyses.
The implementation is cross-checked against Refs.~\cite{Corbett:2021eux,Dawson:2019clf}.

Compared to previous approaches~\cite{PhysRevD.71.075009,JHEP012014151,Falkowski:2015ModelIndependent,JHEP072015018,JHEP022016069,deBlas:2018Wq,PhysRevD.99.033001,SciPostPhys.6.6.064,s10052-019-6977-z,Ellis:2018UpdatedGlobalSMEFT,Falkowski:2020FlavourfulLikelihood,Ellis:2020unq,Corbett:2021eux,JHEP052023208,JHEP092024091}, \ewpdlhc incorporates both additional experimental inputs and refined theoretical modelling: on the experimental side, it includes new observables such as SLD asymmetries, LEP $\tau$ polarisation measurements, ATLAS lepton-flavour-universality tests, and updated determinations of the $W$ boson mass and width. On the theoretical side, it accounts for lepton-flavour-universality violation, uses updated Particle Data Group (PDG) input values~\cite{pdg2024}, and implements an improved treatment of SM theory uncertainties. These include both parametric uncertainties in the SM prediction, which stem from uncertainties in input parameters, as well as theory uncertainties, which arise for example due to missing higher-order corrections in theoretical calculations.

The values of the electroweak input parameters $m_Z$ and $G_\mathrm{F}$ are taken from the PDG~\cite{ALEPH:2005ab,pdg2024}. The $W$ boson mass and width are taken from the most recent ATLAS measurement~\cite{STDM-2019-24}. The combined extraction of the $W$ boson mass and width from this measurement is used as input to the fit, with their correlation taken into account. Predictions for EWPO also depend on $\alpha_s$, which is set to the combined value provided by the Flavour Lattice Averaging Group (FLAG)~\cite{FlavourLatticeAveragingGroupFLAG:2021npn}. At loop level, the predictions also depend weakly on the top quark mass $m_t$, and the Higgs boson mass $m_H$, which are set to their PDG values~\cite{pdg2024}.

\subsubsection{Expected impact of SMEFT terms on analysis observables}

\begin{figure*}[]
\centering
\captionsetup{skip=-13pt}
\includegraphics[angle=90,width=0.84\linewidth]{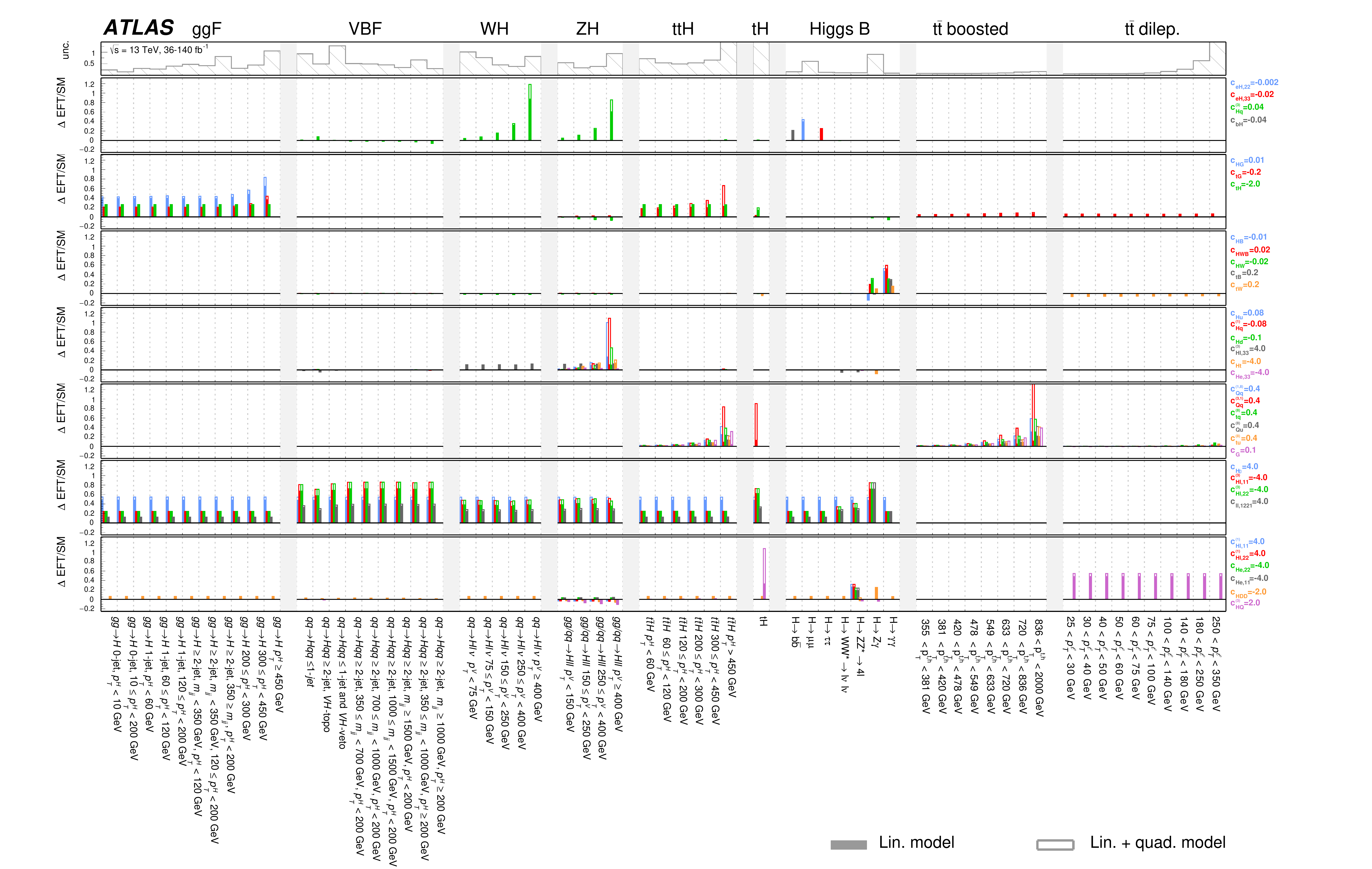}
\caption{Expected impact of the most relevant SMEFT operators on Higgs boson and top-quark observables of the most relevant SMEFT operators relative to the SM predictions for the linearised SMEFT model (shaded histogram) and the SMEFT model including quadratic terms (open histogram). The coefficient values are arbitrary and chosen for visualisation.  The impact, $\Delta \ \text{EFT/SM}$, is defined as the difference between SMEFT and SM predictions, normalised to the SM value. To judge the experimental sensitivity to constrain the operators from the data, the total uncertainty in the measurement of each observable is shown in the top panel. }
\label{fig:impact_1}
\end{figure*}

\begin{figure*}[]
\centering
\captionsetup{skip=-12pt}
\includegraphics[angle=90,width=0.95\linewidth]{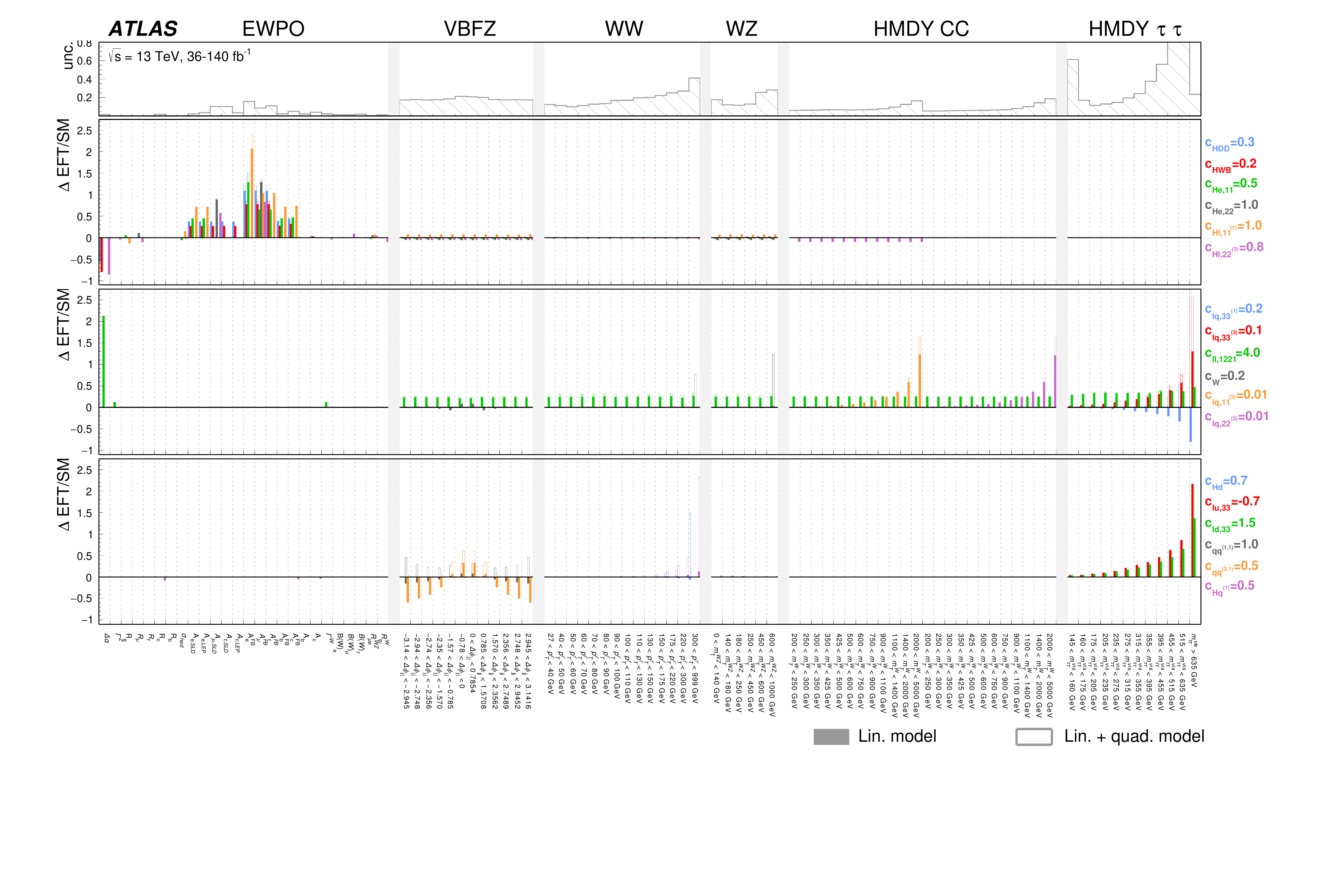}
\caption{Expected impact of the most relevant SMEFT operators on electroweak, HMDY and EWPO observables. The theoretical predictions and measurement uncertainties are presented in the same way as for Figure \ref{fig:impact_1}.}
\label{fig:impact_2}
\end{figure*}

The linear and linear-plus-quadratic effects of operators on the different observables are summarised in Figures~\ref{fig:impact_1} and ~\ref{fig:impact_2}, which can be used to identify the measurement bins that are the most sensitive to a given operator, and compare their sensitivity with the remaining measurements.

\subsection{Simulation of SM processes}
The SM predictions are taken from each published input analysis and are briefly described below.
\subsubsection{Higgs boson processes}

All analyses except the \vbfbb\ measurement share a common set of signal samples, which are described in the following paragraphs. The samples used for \vbfbb\ are described separately at the end of this section.
The signal modelling is summarised here; further technical details are provided in the corresponding analysis publications.

Higgs boson production via gluon--gluon fusion is simulated using the \powhegbox~\cite{Nason:2004rx,Frixione:2007vw,Alioli:2010xd,Alioli:2008tz} NNLOPS implementation~\cite{Hamilton:2013fea,Hamilton:2015nsa} and normalised to the inclusive next-to-next-to-next-to-leading-order QCD calculation with NLO electroweak corrections applied~\cite{deFlorian:2016spz,Anastasiou:2015ema,Anastasiou:2016cez,Actis:2008ug,Anastasiou:2008tj}. The \tVBF\ and \tVH\ production processes are generated at NLO accuracy in QCD using the \powhegbox~\cite{Nason:2009ai,Luisoni:2013cuh} generator and normalised to the highest available QCD and electroweak precision. Higgs boson production in association with a $t\bar{t}$ pair is simulated at NLO accuracy in QCD using either \powhegbox~\cite{Hartanto:2015uka} or \amc, depending on the decay channel. Parton showering, hadronisation and multiple parton interactions are modelled using \PYTHIA[8.2].

In the \vbfbb\ analysis, the \powhegbox\ generator with the NNPDF3.0~\cite{Ball:2014uwa} set of Parton Distribution Functions (PDFs) is used to simulate the \tggF\ and \tVBF\ production processes, interfaced to \PYTHIA\ for parton showering. NLO electroweak corrections are calculated using
MadGraph and applied as a function of the generated Higgs boson transverse momentum. Contributions from \tVH\ and \tttH\ production are generated using \powhegbox\ with the \nnpdf\ set of PDFs.

\subsubsection{Electroweak processes}

The \WW\ and \WZ\ production processes are modelled using \SHERPA~\cite{Bothmann:2019yzt} generator at NLO QCD accuracy for up to one additional parton and LO for up to three additional partons, using NNPDF3.0nnlo PDFs~\cite{Ball:2014uwa} and MEPS@NLO matching~\cite{Hoeche:2011fd,Hoeche:2012yf} with virtual QCD corrections from \OPENLOOPS~\cite{Cascioli:2011va,Denner:2016kdg}. The gluon-initiated  \evmv production processes are simulated using \SHERPA at LO precision for up to one additional parton emission, with the parton-shower modelling being the same as in the quark-initiated samples. Electroweak $Zjj$ production is simulated at NLO QCD with \HERWIG{7.1.5}~\cite{Bahr:2008pv,Bellm:2015jjp} interfaced to \VBFNLO{3.0.0}~\cite{Arnold:2011wj}, using MMHT2014 PDFs~\cite{Harland-Lang:2014zoa}. The default \HERWIG parameters are applied for showering, hadronisation, and the underlying event, while \EVTGEN is used for bottom- and charm-hadron decays.

\subsubsection{Neutral- and charged-current HMDY processes}

For neutral-current HMDY, the Drell--Yan production of $\tau$-lepton pairs is modelled using \SHERPA~\cite{Bothmann:2019yzt} with matrix-element calculations at NLO in QCD for up to two additional partons, and LO accuracy for up to five additional partons using the \textsc{Comix}~\cite{Gleisberg:2008fv} and \OPENLOOPS~\cite{Buccioni:2019sur,Cascioli:2011va,Denner:2016kdg} libraries. These calculations are matched to the \SHERPA parton shower using the MEPS@NLO prescription~\cite{Hoeche:2011fd,Hoeche:2012yf,Catani:2001cc,Hoeche:2009rj}, based on Catani--Seymour dipole factorisation. The \NNPDF[3.0nnlo] PDF set is used.

For charged-current HMDY, the Drell--Yan process is modelled at NLO with \powhegbox~\cite{Nason:2004rx,Frixione:2007vw,Alioli:2008gx,Alioli:2010xd} and the CT10 PDF~\cite{Lai:2010vv}, interfaced to \PYTHIA{8} for parton showering, hadronisation, and underlying-event modelling.
The predictions are corrected to approximate next-to-next-to-leading-order (NNLO) precision in perturbative QCD using invariant-mass-dependent $k$-factors computed with VRAP v0.9~\cite{Anastasiou:2003ds} and weighted using the CT14NNLO PDF set~\cite{Dulat:2015mca}. These corrections are combined with NLO electroweak effects computed with MCSANC~\cite{mcsanc}, excluding QED final-state radiation.

\subsubsection{$t\bar{t}$ processes}

In the dilepton channel, the nominal \ttbar\ sample is simulated with \POWHEGBOX\ ~\cite{Frixione:2007nw,Nason:2004rx, Frixione:2007vw, Alioli:2010xd} at NLO using NNPDF3.0 PDFs~\cite{Ball:2014uwa}, interfaced to \PYTHIA with the A14 tune~\cite{ATL-PHYS-PUB-2014-021} and NNPDF2.3 PDFs~\cite{Ball:2012cx} for parton showering, hadronisation, and the underlying event. The $h_\text{damp}$ parameter is set to $h_\text{damp}=1.5 m_t$, with $m_t$ denoting the top-quark mass, and both renormalisation and factorisation scales are set to the top-quark transverse mass. In the boosted channel, the nominal \ttbar\ sample is generated with \POWHEGBOX\ ~\cite{Frixione:2007nw,Nason:2004rx, Frixione:2007vw, Alioli:2010xd} at NLO using NNPDF3.0 PDFs~\cite{Ball:2014uwa}, with $h_\text{damp}=1.5 m_t$ and renormalisation and factorisation scales set to $\sqrt{m_t^2 + \pT^2}$. \PYTHIA with the A14 tune models parton showering, hadronisation, and the underlying event.

\subsubsection{Di-Higgs processes}

In both $HH\to b\bar{b}\gamma\gamma$ and $HH \to b\bar{b}\tau\tau$ analyses, di-Higgs ggF production is simulated at NLO including finite top-mass effects using \POWHEGBOX~\cite{POWHEG_1,POWHEG_2,POWHEG_3,Campbell:2014kua,Heinrich_2017,Heinrich_2019} with the PDF4LHC15~\cite{Butterworth:2015oua} set of PDFs. Parton showering and hadronisation are modelled with \PYTHIA{8} using NNPDF2.3LO and the A14 tune. Samples are normalised to the best available SM cross-section predictions.

%

%

%

%

%
%
%
%
%
%
%
%
%
%
%
%
%
%
%
%


%

\section{Statistical model}
\label{sec:stat}


This section describes the statistical models used to interpret the measurements.  The statistical model used for the Higgs boson inputs differs from the ones available for the remaining measurements, therefore they are described separately.

For all Higgs boson measurements, the likelihood function for each signal region $k$, with one or more bins~$r$, is modelled as:
\begin{equation}
\label{eq:Lbasic}
L(\vec{N_k} | \vec{\mu}, \vec{\theta}) = \prod_{r} \text{Poisson}\left( N_{k,r} | s_k(\vec{\mu},\vec{\theta}) \cdot f_s^{k,r}(\vec{\theta})~+~ b_{k,r}(\vec{\theta}) \right),
\end{equation}

\noindent where $N_{k,r}$ is the observed event count of bin $r$ in region $k$, $s_k$ is the expected signal count in region $k$, $f_s^{k,r}$ is the expected fraction of the signal in region $k$ that is
contained in bin $r$, and $b_{k,r}$ represents the expected event count from background processes. The ensemble of Parameters of Interest (POIs) $\vec{\mu}$ describes the Higgs boson signal normalisation, while $\vec{\theta}$ represents the set of \textit{nuisance parameters} taking into account the systematic uncertainties that originate from theoretical and experimental sources, as well as additional degrees of freedom without prior constraints such as background yields.
The full likelihood function is then the product of the likelihood functions for each signal region $k$ and the Gaussian or log-normal probability density functions that constrain the nuisance parameters.

The signal yield for region $k$ is modelled with the scale factors $ \mu^{i,k^\prime,X}_k$ applied to the SM Higgs boson production cross-section times branching ratio, for each Higgs boson production process $i$ and decay $X$, in a fiducial region $k^\prime$ defined at the particle level, i.e. using stable particles from the simulation before detector effects. The expression of the signal yield $s_k$ depends on $\mu^{i,k^\prime,X}_k$ and the nuisance parameters $\vec\theta$ as follows:
\begin{equation}
s_k^\textrm{STXS}\left(\vec{\mu_k},\vec{\theta}\right)  =  \mathcal{L} \times  \sum\limits_{i,k^\prime,X}   \mu^{i,k^\prime,X}_k \times (\sigma \times \BR)_\text{SM, MC}^{i,k^\prime,X}(\vec{\theta}) \times \epsilon^{i,k^\prime,X}_{\textrm{STXS},k}(\vec{\theta}),
\end{equation}
\noindent where $\mathcal{L}$ is the integrated luminosity, $(\sigma \times \BR)_\text{SM, MC}^{i,k^\prime,X}$ is the calculation, at the highest available order, of the SM Higgs boson cross-section for the production process $i$ in particle-level region $k^\prime$ multiplied by the SM Higgs boson branching ratio to the final state $X$. The factors $\epsilon^{i,k^\prime,X}_{\textrm{STXS},k}$ represent the products of acceptance times efficiency of the reconstruction-level region $k$ for the particle-level fiducial phase space region $k^\prime$ and Higgs boson decay $X$.
The original signal parameters $\vec{\mu}$ are replaced with expressions that parameterise the SMEFT predictions, i.e., $\vec{\mu} \rightarrow \vec{\mu}(\vec{c})$, so that the likelihood of Eq.~\eqref{eq:Lbasic} is directly expressed in terms of the parameters $\vec{c}$, the SMEFT Wilson coefficients.

Di-Higgs measurements are also interpreted with a Poissonian likelihood, with the difference that SMEFT effects are included at reconstruction level, i.e. SMEFT modifications are obtained from the reweighting of fully simulated di-Higgs samples and no reparameterisation step is needed.

The remaining measurements (EW, HMDY and top-quark measurements) use instead a Gaussian likelihood constructed from published results. The predicted and measured cross-sections, $x^\text{pred}_b$ and $x^\text{meas}_b$, in each bin $b$ of the unfolded distributions of the differential measurements are expressed as:
\begin{align}
\label{eq:smww-signal-para}
x^\text{pred}_b\left(\boldsymbol{c},\boldsymbol \theta_\text{theo}\right)&=x^{\text{SM}}_b\left(1 + \sum_i A_{bi}c_i + \sum_{i} B_{bi}c_i^2 + \sum_{i<j} B_{bij}c_ic_j\right)\times\prod_{k}^{n_\text{theo}}\left(1+\theta_{k}u_{b,k}\right), \\ \nonumber
x^\text{meas}_b\left(\boldsymbol \theta_\text{exp}\right)&=x_b \times \prod^{n_\text{exp}}_k\left(1+\theta_{\text{exp},k}v_{b,k}\right),
\end{align}
where $x^\text{SM}_b$ and $x_b$ are the nominal SM cross-section predictions and measurements, $i$ and $j$ run over all dimension-six Wilson coefficients $\boldsymbol{c}$, $A_i$ and $B_{ij}$ are the linear and quadratic SMEFT effects, respectively,  $\boldsymbol \theta$ represents the nuisance parameters, $v_{b,k}$ and $u_{b,k}$ are the relative sizes of the theory and experimental uncertainty $k$ on the prediction in bin $b$.

The likelihood $L\left(\boldsymbol x | \boldsymbol{c},\boldsymbol\theta\right)$ for an individual measurement is modelled as a multivariate Gaussian:
\begin{equation}
\begin{aligned}
L\left(\boldsymbol x \middle| \boldsymbol{c},\boldsymbol\theta\right)=&\frac{1}{\sqrt{\left(2\pi\right)^{n_\text{bins}}\mathrm{det}\left(V\right)}}\exp\left(-\frac{1}{2}\Delta{\boldsymbol x}^{\intercal}\left(\boldsymbol{c},\boldsymbol\theta\right) V^{-1}\Delta{\boldsymbol x}\left(\boldsymbol{c},\boldsymbol\theta\right)\right)\times\prod^{n_\text{theo}}_if_i\left(\theta_{\text{theo},i}\right) \times\prod^{n_\text{exp}}_if_i\left(\theta_{\text{exp},i}\right).
\end{aligned}
\label{eq:likelihood}
\end{equation}
Here, $\boldsymbol\theta$ represents all nuisance parameters, while $\theta_{\text{theo},i}$ and $\theta_{\text{exp},i}$ are the nuisance parameters corresponding to the theoretical and experimental uncertainties, respectively. The terms $f_i$ represent the Gaussian constraints on nuisance parameters, $V$ is the covariance matrix, and the vector $\Delta{\boldsymbol x}=(\Delta x_1,\dots, \Delta x_{n_\text{bins}})$ is the difference between measurement and prediction, which in a given measurement bin $b$ is defined as
\begin{equation}
\Delta x_b\left(\boldsymbol{c},\boldsymbol \theta\right)= x^\text{meas}_b(\boldsymbol{\theta})-x^\text{pred}_b\left(\boldsymbol{c},\boldsymbol \theta\right).
\end{equation}

Finally, the multivariate Gaussian model introduced above is also used for the interpretation of EWPO. In this case, the model contains no nuisance parameters and both theoretical and experimental uncertainties are included in the covariance matrix. The matrix of the Pearson correlation coefficients for the electroweak precision observables, obtained as the sum of the theoretical and experimental contributions, is shown in Table~\ref{tab:EWPO-correlation}.

The combined likelihood is obtained as the product of the measurement likelihoods, where the nuisance parameters that describe the same effect in different measurements are equated, and only one constraint for each independent nuisance parameter is included.

\begin{table}
\caption{Correlation matrix for the electroweak precision observables.}\label{tab:EWPO-correlation}
\centering
\begin{adjustbox}{scale=0.45}
\begin{tabular}{@{}lrrrrrrrrrrrrrrrrrrrrrrrrrrr@{}}
\toprule
&$\Deltaalpha$ & $\GammaZ$ & $\Rel$ & $\Rmu$ & $\Rtau$ & $\Rc$ & $\Rb$ & $\sigmahad$ & $\AeSLD$ & $\AeLEP$ & $\AmuSLD$ & $\AtauSLD$ & $\AtauLEP$ & $\AFBe$ & $\AFBmu$ & $\AFBtau$ & $\AFBb$ & $\AFBc$ & $\Ab$ & $\Ac$ & $\GammaW$ &  $\BrWe$ & $\BrWmu$ & $\BrWtau$ & $R_{WZ}^{\mu / e}$ & $R_{W}^{\tau / \mu}$ \\
\midrule
$\Deltaalpha$ & 1.000 & & & & &  & & & &  & & & &  & & & &  & & & &  &\\
$\GammaZ$ &  $-$0.307  &  1.000& &  & &  & & & &  & & & &  && & &  & & & & & & & &  &\\
$\Rel$ & $-$0.107 & 0.043 & 1.000 & & &  & & & &  & & & &  & & & &  & & & &  &\\
$\Rmu$ & 0.043 & 0.084 & 0.098 & 1.000 & &  & & & &  & & & &  & & & &  & & & &  &\\
$\Rtau$ &$-$0.118  & 0.064 & 0.069 & $-$0.029  &1.000&  & & & &  & & & &  & & & &  & & & &  &\\
$\Rc$ & $-$0.006 & 0.003 & 0.001& 0.002 & 0.001& 1.000& & & &  & & & &  & & & &  & & & &  & & & &  &\\
$\Rb$ & 0.010 & $-$0.007 & $-$0.002 & $-$0.004& $-$0.003 &$-$0.178 & 1.000& & &  & & & &  & & & &  & & & &  & & & &  &\\
$\sigmahad$	& 0.035  & $-$0.324 & 0.093& 0.107 & 0.076 & $-$0.001 & 0.002& 1.000& &  & & & &  & & & &  & & & &  & & & &  &\\
$\AeSLD$	 & $-$0.711  & 0.253 & 0.086& 0.124  & 0.095 & 0.005 & $-$0.009&$-$0.029& 1.000&  & & & &  & & & &  & & & &  & & & &  &  \\
$\AeLEP$ & $-$0.423 & 0.150 & 0.051& 0.074  & 0.056& 0.003 &$-$0.006 & $-$0.017& 0.350& 1.000 & & & &  & & & &  & & & &  & & & &  &\\
$\AmuSLD$ & $-$0.153  & 0.054 & 0.018& 0.027  & 0.020 & 0.001 & $-$0.002& $-$0.006&0.127& 0.075 &1.000 & & &  & & & &  & & & &  & & & &  &\\
$\AtauSLD$ & $-$0.153 & 0.054 & 0.018 & 0.027& 0.020 & 0.001& $-$0.002& $-$0.006& 0.127 & 0.075& 0.027& 1.000& &  & & & &  & & & &  & & & &  &\\
$\AtauLEP$ & $-$0.467  & 0.166 & 0.056  & 0.082& 0.062 & 0.003& $-$0.006&$-$0.019 &  0.387 & 0.239& 0.083 & 0.083& 1.000&  & & & &  & & & &  & & & &  &\\
$\AFBe$ & $-$0.200 & 0.078 & $-$0.331&   0.036& 0.029 & 0.001& $-$0.003& $-$0.007& 0.166 &0.099 & 0.036&0.036 &0.109  &1.000 & & & &  & & & &  & & & &  &\\
$\AFBmu$ & $-$0.363  & 0.131 & 0.062&  0.074 & 0.049&  0.002&$-$0.005 &$-$0.012 & 0.301 & 0.179 & 0.065&0.065 &0.198 & 0.063 & 1.000 & & &  & & & &  & & & &  &\\
$\AFBtau$ & $-$0.287  & 0.103 & 0.047&  0.048 & 0.047&0.002  & $-$0.004& $-$0.010& 0.238& 0.141 &0.051 &0.051 &0.156 &0.048  & 0.162 &1.000  & &  & & & &  & & & &  &\\
$\AFBb$ & $-$0.690  & 0.245 & 0.083& 0.121  & 0.092&  0.031& $-$0.075& $-$0.028& 0.551 &0.328 &0.119 & 0.119& 0.362&0.155  &0.282 & 0.223&1.000 &  & & & &  & & & &  &\\
$\AFBc$ & $-$0.340  & 0.121 & 0.041& 0.060  &0.045 & $-$0.054& 0.060&$-$0.014 & 0.272& 0.162 &0.059 & 0.059&0.179 &0.077  &0.139 &0.110 &0.366 &1.000  & & & &  & & & &  &\\
$\Ab$ & $-$0.010  & 0.003 & 0.001&  0.002 & 0.001& 0.040 & $-$0.079&  - & 0.008& 0.005 & 0.002& 0.002&0.005 &0.002  &0.004 &0.003 &0.047 &  $-$0.015 & 1.000 & & &  & & & &  &\\
$\Ac$ & $-$0.038  & 0.014 & 0.005&  0.007 &0.005 & $-$0.060 &0.039 & $-$0.002& 0.032& 0.019 & 0.007& 0.007& 0.021&0.009  & 0.016& 0.013& 0.036 & 0.052 & 0.110& 1.000 & &  & & & &  &\\
$\GammaW$& $-$0.023  & 0.010 & 0.003& 0.005 &0.004 & - & $-$0.001&  $-$0.002& 0.019& 0.011 &0.004 &0.004 &0.013 & 0.005 &0.010& 0.008& 0.019& 0.009 & - & 0.001& 1.000& && &&  &\\
$\BrWe$ & -  & - & -& -  & -& - & - & - & - &-  &- &- & -&  -& -& -&- &-  & -&- & - &1.000 & & &  &\\
$\BrWmu$ & -  & - & -& -  & -&  -&  -& - & - &-  &- &- &- & - &- & -& -& - &-  &- & - & 0.136 & 1.000& &  &\\
$\BrWtau$ & -  & - & -& -  & - & - & - & - &- &-  & -& -&- & - &-& -&- &-  &-  &- & - & $-$0.201& $-$0.122 &1.000 &  &\\
$R_{WZ}^{\mu / e}$ & -  & - & -& -  & - &  - &- & -&- &-  &- & -&- &-  &- & -& - & -&- &- & - &- &- & -& 1.000 &\\
$R_{W}^{\tau / \mu}$ & -  & - & -& -  & - &  -&- & -&- &  -& -& -&- &  -& - & -& - &- &- & -&  - & - & - & - & - &1.000\\
\bottomrule
\end{tabular}
\end{adjustbox}
\end{table}

\subsection{Treatment of systematic uncertainties}
\label{sec:combination}
Using the full experimental likelihood ensures a consistent treatment of systematic uncertainties across the different analyses, which is crucial for an accurate estimation of the total uncertainty.

Systematic uncertainties are correlated when they describe the same effect and are defined consistently across analyses. For example, uncertainties related to electron and muon reconstruction, identification, trigger efficiencies, and energy scale and resolution are typically correlated, with exceptions arising when different calibration or estimation schemes are used, such as in the $t\bar{t}$ dilepton channel. Similarly, jet energy scale and resolution uncertainties are also typically correlated, except in cases where alternative uncertainty schemes are employed.
The luminosity uncertainty is correlated across all analyses, following the prescription in Ref. \cite{newlumi}. For single-Higgs and di-Higgs measurements, the same correlation scheme as in Ref. \cite{HDBS-2022-03} has been applied. The impact of approximations in the correlation scheme is tested and found to be negligible.

A pruning procedure of the nuisance parameters has been applied by removing a set of nuisance parameters that does not modify the uncertainty obtained on any POI by more than 1$\%$. A summary of the systematic uncertainty correlation is provided in Table~\ref{tab:NP}.
\begin{table}[]
\caption{Summary of the correlation of systematic uncertainties across the different analyses. For each systematic uncertainty, the symbols indicate whether the corresponding nuisance parameters are assumed to be correlated between the analyses listed in the columns: the crosses denote uncorrelated nuisance parameters, checkmarks denote fully correlated ones and the $\approx$ symbol indicates that only a subset of the corresponding nuisance parameters is correlated. Where no symbol is shown, the uncertainty is negligible or not considered in that analysis.}

\resizebox{\textwidth}{!}{
\begin{tabular}{l|l|l|l|l|l|l|l|l}
\multicolumn{2}{l|}{Systematic uncertainties}                 & Single Higgs &$t\bar{t}$ dilepton &$t\bar{t}$ boosted & HMDY $\tau\tau$              & HMDY CC       & WW,WZ, EW $Zjj$     &   di-Higgs              \\ \hline
\multicolumn{2}{l|}{Muons}                                          & $\checkmark$ & $\approx$          & $\approx$      &  $\checkmark$       & $\checkmark$           & $\times$        & $\approx$      \\ \hline
\multirow{3}{*}{Electron/Photon}        & Electron efficiency                   & $\checkmark$ & $\approx$          & $\checkmark$      &                     & $\checkmark$           & $\times$        &   $\approx$        \\
& Scale and resolution & $\checkmark$&                    &  $\approx$        & $\checkmark$         &   $\approx$           & $\times$        & $\checkmark$       \\
& Photon efficiency                          & $\checkmark$ &                    &                   &                     &                        &                 & $\checkmark$      \\ \hline
\multirow{3}{*}{Jets}       & Jet energy resolution                & $\checkmark$ &                    &                   &   $\checkmark$      & $\times$               & $\checkmark$    &  $\approx$        \\
& Jet Energy scale                     & $\checkmark$ & $\approx$          & $\approx$         &   $\checkmark$      & $\checkmark$           & $\approx$    & $\approx$          \\
& JVT                                  & $\checkmark$ &                    &                   &                     &                        & $\checkmark$    & $\checkmark$      \\ \hline
\multicolumn{2}{l|}{Missing transverse energy}                      & $\checkmark$ & $\checkmark$       & $\times$          &                     & $\checkmark$           &                 & $\times$          \\ \hline
\multirow{7}{*}{Theoretical}& $t \bar{t} $            & $\approx$  & $\approx$            &$\approx$          &                     & $\approx$              &$\checkmark$    & $\checkmark$         \\
& Diboson: cross-section               &              & $\times$           & $\times$          &                     &  $\checkmark$          &                 &                   \\
& Single top               & $\checkmark$ &  $\checkmark$       &    $\checkmark$        &                     & $\checkmark$           & $\checkmark$    &                   \\
& Z+jets modelling                     &              &                    & $\times$          &                     & $\times$               & $\times$    &                   \\
& WW modelling                         & $\checkmark$ &                    &                   &                     &                        &   $\checkmark$  &                   \\
& Higgs boson BR                             & $\checkmark$ &                    &                   &                     &                        &                 &  $\checkmark$     \\ \hline
\multirow{4}{*}{Other}                  & Luminosity                           & $\checkmark$ & $\checkmark$       & $\checkmark$      &  $\checkmark$       & $\checkmark$           &   $\checkmark$  & $\checkmark$      \\
& Pile-up modelling                     & $\checkmark$ &                    & $\checkmark$      &                     & $\checkmark$           &   $\checkmark$  & $\times$          \\
& Flavour tagging                      & $\times$     & $\times$           & $\times$          &                     &     $\times$           &   $\times$      &                   \\
& $\tau$-lepton               & $\checkmark$ &                    &                   &  $\checkmark$       &                        &                 &     $\checkmark$  \\

\bottomrule
\end{tabular}
}


\label{tab:NP}
\end{table}

\subsection{Estimation of confidence intervals}
The profile likelihood ratio test statistic is constructed from the likelihood as follows:
\begin{equation}
q\left(c_i\right)=-2\log \frac{L\left(\boldsymbol x \middle|c_i,\hat{\hat{\boldsymbol{\theta}}}\right)}{L\left(\boldsymbol x \middle|\hat{c_i},\hat{\boldsymbol{\theta}}\right)},
\label{eq:teststat}
\end{equation}
where $\hat{\hat{\boldsymbol{\theta}}}$ is the maximum likelihood estimate of the nuisance parameters for a fixed value of $c_i$, while $\hat{c_i}$ and $\hat{\boldsymbol{\theta}}$ are the unconditional maximum likelihood estimates of $c_i$ and the nuisance parameters, respectively.

Confidence intervals are derived using Wilks' theorem, assuming that $q(c_i)$ follows a $\chi^2$ distribution. Following the conclusion of Ref. \cite{wilkspaper}, the validity of the assumptions of the Wilks' theorem have been investigated and are reported in the appendix.


%
\section{Results}
\label{sec:result}

This section presents the results of the analysis. It starts with the constraints obtained for individual Wilson coefficients in the Warsaw basis. The methodology used to obtain the full results is then described, showing how correlations between coefficients are taken into account to identify the most and least constrained directions in parameter space. Results are presented for both linear and linear-plus-quadratic models, and the section concludes with a discussion of how the SMEFT limits can be interpreted in terms of specific UV-complete models.

\subsection{Expected constraints on individual Warsaw basis coefficients}
\begin{figure}
\centering
\includegraphics[width=1\linewidth]{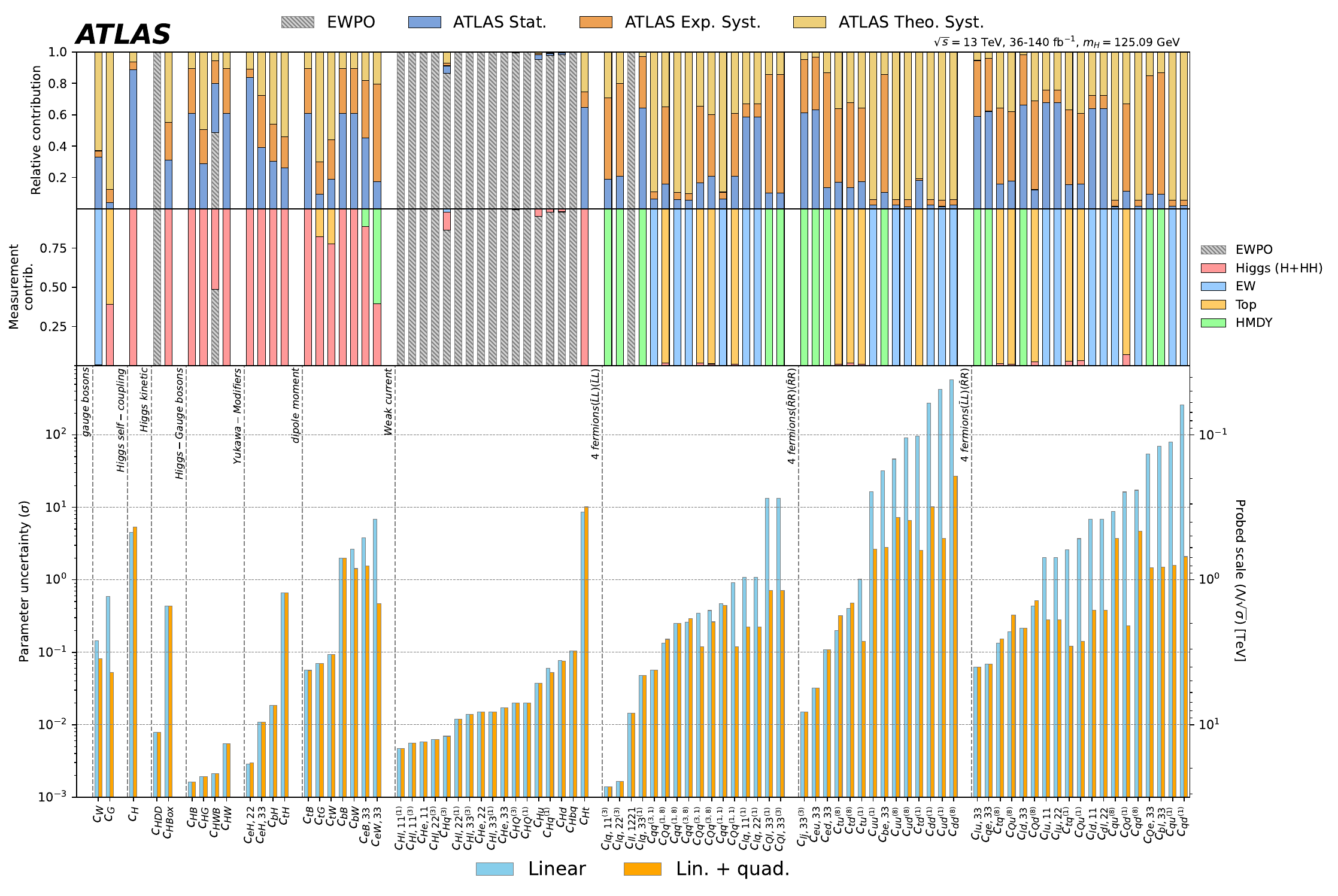}
\caption{Expected results from the fits of individual Warsaw basis coefficients. Top: uncertainties split by statistical, and experimental and theoretical systematic components for each analysis, with the exception of EWPO where the information is not available. Middle: expected contribution of each analysis to the sensitivity to each operator. Bottom: the symmetrised 68$\%$ CL uncertainty $\sigma$ and the corresponding probed scale of new physics $\Lambda/\sqrt\sigma$ for the linear and linear-plus-quadratic models.}
\label{fig:summary_oaat}
\end{figure}

Figure~\ref{fig:summary_oaat} shows the 68\% Confidence Level (CL) limits for the linear and linear-plus-quadratic models obtained on Asimov data~\cite{Cowan2011} for each coefficient of the Warsaw basis shown in Tables~\ref{tab:relevant-operators-1} and ~\ref{tab:relevant-operators-2}. The constraints are obtained by fixing the remaining coefficients to zero. The parameters are grouped following the definitions of Ref. \cite{Brivio_2021} and ordered by sensitivity in the linear model. The lower panel reports the absolute uncertainties, the middle panel matches the constraints to the input measurements, while the upper panel shows the breakdown of the uncertainties into statistical and systematic components. The systematic component is further divided into experimental and theoretical components, where the latter includes uncertainties on the signal and background modelling. The breakdown of the uncertainties is obtained with simplified Gaussian models.

The total number of operators that can be constrained individually using the included measurements is 86. Constraints on four-fermion operators are dominated by EW, top-quark and HMDY measurements, while Yukawa-coupling modifiers, Higgs-to-gauge-bosons coupling and dipole operators constraints are mostly dominated by measurements in the Higgs sector. The $W$-boson self-coupling $\cW$ is constrained by EW measurements, while the gluon self-coupling $\cG$ is constrained by a mix of top-quark and Higgs boson measurements. The trilinear Higgs boson self-coupling is constrained by Higgs boson measurements, with the strongest contribution coming from di-Higgs measurements. The weak current operators are constrained almost entirely by EWPO, with the exception of the weakly measured $\cHt$ coefficient, which is constrained by Higgs boson measurements. The best constrained coefficients are $c_{lq,11^{(3)}}$ and $c_{lq,22^{(3)}}$, which are constrained by the high-mass tails of the charged-current Drell--Yan measurements.

\subsection{Identification of the most sensitive directions and fit basis definition}

The 86 Wilson coefficients $\vec c$ included in the combination cannot all be simultaneously constrained. The available input measurements are not sufficient to resolve the degeneracies of the SMEFT effects on the observables included. Because of that, the best measured directions in the SMEFT parameter space are identified and used to define a new basis of fit parameters $\vec c^\prime$ that can be simultaneously measured. This set of parameters is found considering:
\begin{equation}
V^{-1}_{\textrm{SMEFT}}=P^T V^{-1}_{\mu} P,
\end{equation}
where $V_\textrm{SMEFT}$ and $V_{\mu}$ are the covariance matrices of the Wilson coefficients and of the $\mu$ measurement, respectively, and $P$ is the linearised SMEFT response matrix.
In the limit of Gaussian measurements, the matrix $V^{-1}_{\textrm{SMEFT}}$ represents the Fisher information matrix of its linearised SMEFT model. A Principal Component Analysis (PCA) is performed on $V^{-1}_{\textrm{SMEFT}}$ in sub-groups of parameters, and the eigenvectors with the largest eigenvalues are selected in each group as the directions that can be measured. The parameters are grouped to ensure both fit stability and fit-parameter interpretability. A threshold of 10 on the uncertainty of the considered parameter, corresponding to an eigenvalue of 0.01, is applied to select the directions to measure. The remaining directions are fixed to their SM values.

\begin{figure}
\centering
\includegraphics[width=0.9\textwidth]{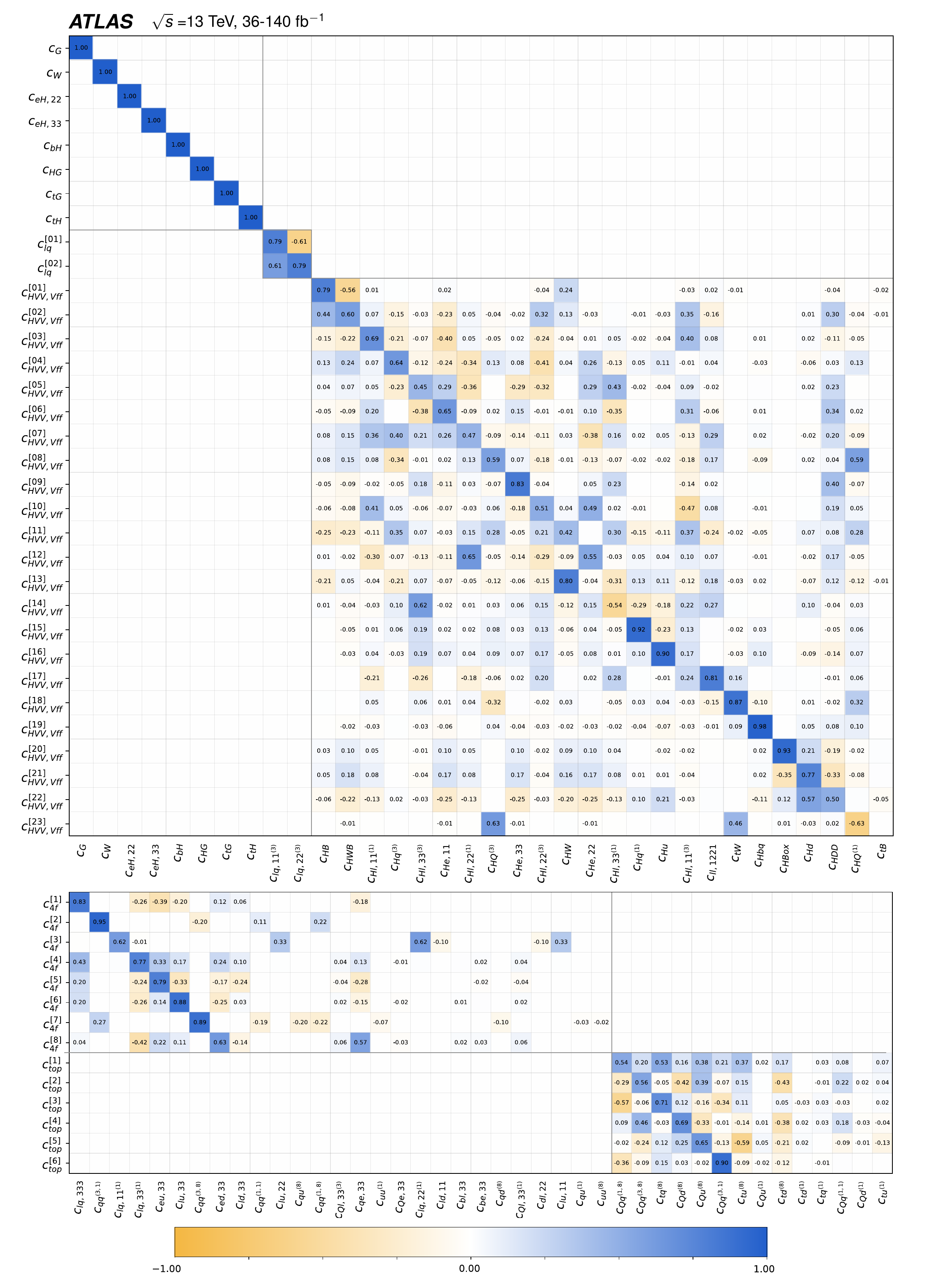}
\caption{Definition of the 47 directions composing the fit basis for the global SMEFT fit. Each row corresponds to a fitted linear combination of Warsaw basis vectors. Entries are rounded to the second
decimal place and their size is also indicated by colour code. Contributions to the fit basis directions with an absolute value below 0.01 are omitted. }\label{fig:fitbasis_def_GEFT}
\end{figure}

The PCA identifies 47 directions as fit basis, and their definition in terms of the Warsaw basis coefficients are shown in Figure~\ref{fig:fitbasis_def_GEFT}. The following Warsaw basis coefficients can be individually constrained:
\begin{itemize}
\item $\cW$ and $\cG$ affecting the $W$ boson and gluon self-coupling.
\item the Yukawa-coupling modifiers $\ceHtwo$, $\ceHthree$ and $\cbH$, for muons, $\tau$-leptons and $b$-quarks, respectively.
\item $\ctG$, $\ctH$ and $\cHG$: operators modifying the top--gluon, Higgs--top and Higgs--gluon couplings,  mostly constrained by ggF and $t\bar{t}H$ processes.
\end{itemize}

Separate principal component analyses are performed on the remaining coefficients, grouped as follows:
\begin{itemize}
\item \textbf{$c_{lq}$}: mixing of the $c_{lq,11^{(3)}}$ and $c_{lq,22^{(3)}}$, measured by the HMDY charged-current analysis.
\item \textbf{$\ctop$}: four-fermion operators including heavy quarks, mostly measured by $t\bar{t}$ measurements (6 out of 14 directions measured).
\item \textbf{$\cfour$}: operators involving four light quarks or two quarks and two leptons (8 out of 33 directions measured).
\item \textbf{$\cHVVEW$}: since the operators affecting EWPO are strongly correlated, a large number of operators are fitted together, alongside operators affecting $H \rightarrow \gamma \gamma$, the $Z$-boson--lepton couplings ($\cHe$, $\cHlone$), the high-energy tails of $VH$ production ($\cuH$,$\cHd$ and $\cHQone$) and the electroweak couplings ($\cHlthree$, $\cllonetwoone$) (23 out of 30 directions measured).
\end{itemize}
The following scheme shows the translation from the Wilson coefficients $\vec{c}$ to the fit basis $\vec{c}^\prime$. The single Wilson coefficients are shown on the left side and the POIs of the fit basis on the right side. The parameters that can be constrained individually like $\ceHtwo$ are listed on both sides.
{\scriptsize
\begin{align*}
\vec{c}  =  ~&  \{  \ceHtwo \} ~\cup                                                                &  \vec{c}^\prime =~   &    \{ \ceHtwo \} ~ \cup  \\
&  \{  \ceHthree \} ~\cup                                                                          &                        & \{  \ceHthree \}~ \cup \\
&  \{  \cG \} ~\cup                                                                                &                        & \{ \cG \}  ~\cup \\
&  \{  \cHG \} ~\cup                                                                               &                        & \{ \cHG \}  ~\cup \\
&  \{  \ctG \} ~\cup                                                                               &                        & \{ \ctG \}  ~\cup \\
&  \{  \ctH \} ~\cup                                                                               &                        & \{ \ctH \}  ~\cup \\
&  \{  \cbH \} ~\cup                                                                               &                        & \{ \cbH \}  ~\cup \\
&  \{  \cW  \} ~\cup                                                                               &                        & \{  \cW \}  ~\cup \\
&  \{   c_{lq,11^{(3)}}, c_{lq,22^{(3)}}   \}     ~ \cup                                                           &                        & \{  c_{lq}^{[1]}, c_{lq}^{[2]} \}  ~\cup \\
&  \{ \ctdo , \cQdo , \cQde , \cQjoo , \cQjoe , \cQjto , \cQjte , \cQuo , \cQue , \ctde , \ctjo ,                                        &                        &    \{ \ctopone, \ctoptwo, \ctopthree, \\
&   \ctje , \ctuo , \ctue , \cjjoo , \cjjoe , \cjjto , \cjjte , \cuuo , \cjue , \cjde   \} ~\cup                                        &                        &  \ctopfour, \ctopfive, \ctopsix, \}  ~\cup \\
&  \{ c_{Qe,22} , c_{Ql,22^{(1)}} , c_{Ql,22^{(3)}} , c_{be,22} , c_{bl,22} , c_{ed,22} , c_{eu,22} , c_{ej,22} , c_{ld,22} ,                                &                        &    \{ \cfourfone,  \cfourftwo,  \cfourfthree,  \\
& c_{lu,22} , c_{lq,22^{(3)}} , c_{Qe,33} ,   c_{Ql,33^{(3)}} , c_{Ql,33^{(1)}} , c_{be,11} , c_{bl,11} , c_{ed,11} ,                            &                        &    \{  \cfourffour,  \cfourffive,  \cfourfsix , \\
&   c_{eu,11} , c_{ej,11} , c_{lq,11^{(1)}} , c_{lq,22^{(1)}} , c_{lq,11^{(3)}} , c_{lu,11} , c_{ld,11}    \}  ~\cup                            &                        &       \cfourfseven,  \cfourfeight \} ~\cup \\
&  \{ \cHDD , \cHQone , \cHQthree , \cHb , \cHd , \cHeone ,                                   &                        &    \{ \cHVVEWone, \cHVVEWtwo, \cHVVEWthree, \cHVVEWfour , \\
& \cHjthree , \cHloneone , \cHlonetwo ,\cHlonethree ,\cHlthreeone, \cHlthreetwo,\cHlthreethree ,                             &                       &   \cHVVEWfive, \cHVVEWsix, \cHVVEWseven, \cHVVEWeight , \\
&   \cHB,\cHW , \cHWB , \cHbox  ,                                 &                        &     \cHVVEWnine, \cHVVEWten, \cHVVEWeleven, \cHVVEWtwelve, \\
&c_{bB} ,  c_{bW}  , \cHetwo,  \cHethree , \cHjone ,                                &                        &    \cHVVEWthirteen, \cHVVEWfourteen, \cHVVEWfifteen, \cHVVEWsixteen , \\
&  c_{eW}^{33} ,   c_{eB}^{33}  , \ctB  ,                                  &                        &     \cHVVEWseventeen, \cHVVEWeighteen, \cHVVEWnineteen, \cHVVEWtwenty,  \\
&  \cHlthreethree , \cHu , \cllonetwoone  \}                              &                        &      \cHVVEWtwentyone, \cHVVEWtwentytwo, \cHVVEWtwentythree \}
\end{align*}}%
The results of the analysis are expressed in terms of constraints placed on the SMEFT coefficients of the fit basis, first for the linear model and then for the linear-plus-quadratic model. The results are shown in terms of 68$\%$ and 95$\%$ CL limits on the set of coefficients composing the fit basis, where all numbers should be considered normalised by the energy scale $\Lambda = 1$ TeV.

\subsection{Linear results}
Two sets of results are shown for the linear model: one without di-Higgs and one including it, both using the full experimental likelihood. This is due to the fact that the di-Higgs analyses are found to not be well described with simplified Gaussian models as the SMEFT effects are included at the level of reconstructed detector observables. The first set of results is meant to be the one reproducible using simplified likelihood models, while the second is meant to show the impact of di-Higgs measurements, and provide a constraint of the Higgs boson self-coupling in SMEFT.

\subsubsection{Results without di-Higgs measurements}

\begin{figure}{}
\centering
\includegraphics[width=0.97\textwidth]{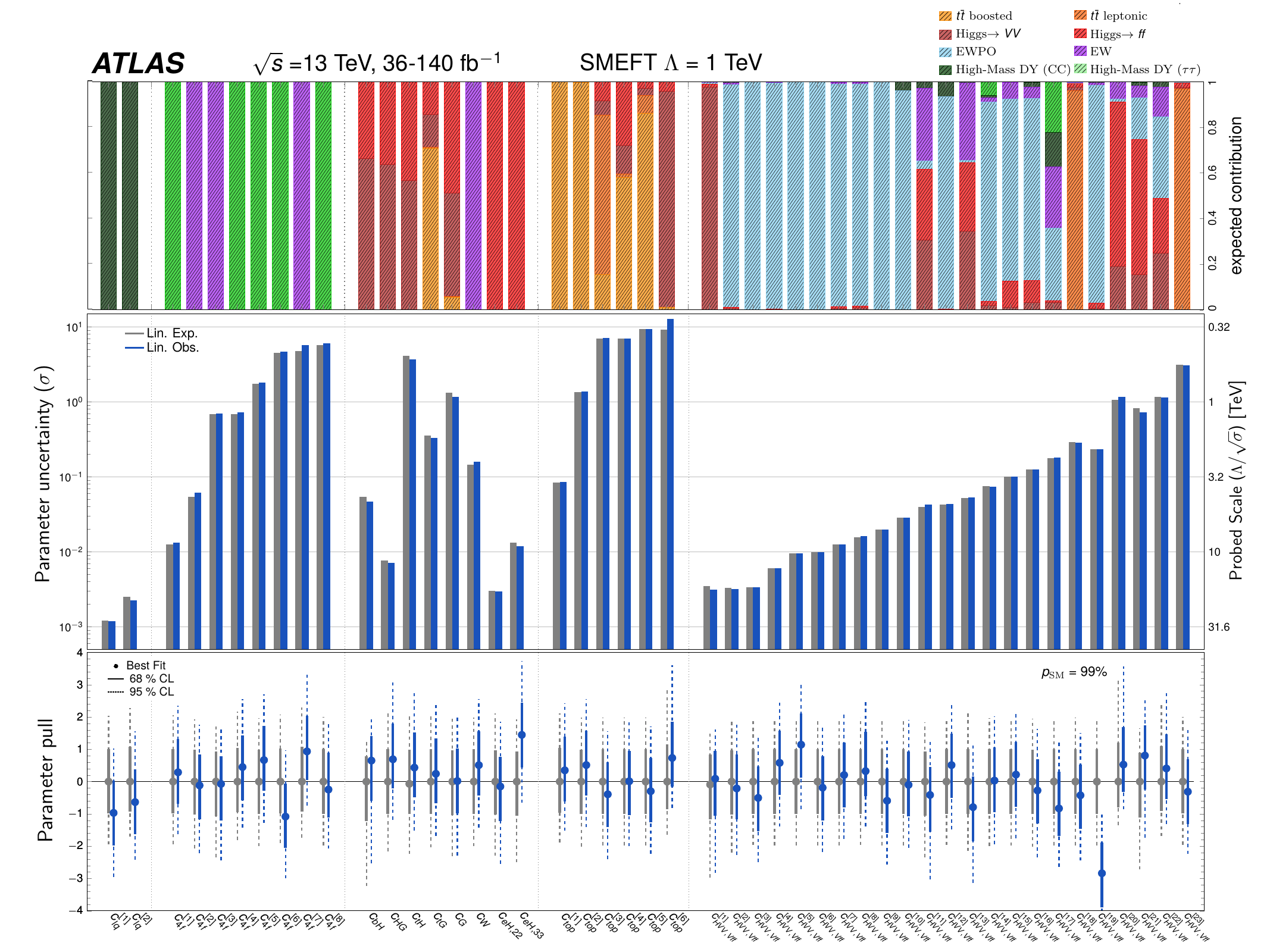}
\caption{Comparison between the expected (grey) and observed (blue) parameters of the fit basis $\vec c^\prime$ for the linear model, where all other coefficients and nuisance parameters are profiled. The upper panel shows the expected breakdown of contributions to the sensitivity of each measurement from the various channels included. The middle panel shows the symmetrised 68 $\%$ CL uncertainty $\sigma$ on a logarithmic scale, together with the corresponding probed energy scale of new physics $\Lambda/\sqrt\sigma$. The lower panel shows, for each operator, the measured value and the 68$\%$ (solid) and 95$\%$ (dashed) CL intervals, divided by the symmetrised uncertainty shown in the middle panel.
}\label{fig:lin_exp_vs_obs}
\end{figure}

Figure~\ref{fig:lin_exp_vs_obs} shows the constraints on the 47 Wilson coefficients of the fit basis for the linear model without the inclusion of the di-Higgs channels. The remaining 39 directions are fixed to their SM value of zero. The lower panel shows the measured value normalised by the symmetrised uncertainty, which is shown in the middle panel on a logarithmic scale. The upper panel shows the expected breakdown of contributions to the sensitivity of each measurement from the various channels included, calculated from the information matrix of the measurements, removing one at a time each set of measurements considered. The contribution $z_i(c)$ is calculated for each sub-set of measurements $i$ and each coefficient $c$ as:
\begin{equation}
z_i(c) = \frac{1}{\sigma_i(c)^2\sum_j \frac{1}{\sigma_j(c)^2}},
\end{equation}
where $\sigma_i(c)$ is the uncertainty on the coefficient $c$ when considering only the sub-set of measurements $i$.

No significant deviation from the SM is observed: all coefficients are compatible with the SM expectation within 2$\sigma$. The only exception is $\cHVVEWnineteen$, constrained by EWPO and pulled by the well-known discrepancy between the $A_{FB}^{0,b}$ and $A_{FB}^{0,c}$ measurements and the SM expectations~\cite{ALEPH:2013dgf}. Good agreement between the observed data and the SM expectations is found, corresponding to a $p$-value of 99$\%$.

HMDY measurements show the strongest constraints on the fit basis for the $c_{lq}^{[1]}$ and $c_{lq}^{[2]}$ coefficients, excluding new physics up to a scale of around 30 TeV. The \textbf{$\cfour$} group is dominated by the EW and the HMDY $\tau\tau$ measurements, while the \textbf{$\ctop$} group is dominated by the $t\bar t$ dilepton and boosted measurements, with the exception of the last direction, which is instead constrained by $t\bar{t}H$ Higgs boson production. As expected, many directions in the \textbf{$\cHVVEW$} group are dominated by EWPO. However, the most constrained direction ($\cHVVEWone$) is probed through the $H \rightarrow \gamma\gamma$ decay.

\begin{figure}
\centering
\includegraphics[width=1\textwidth]{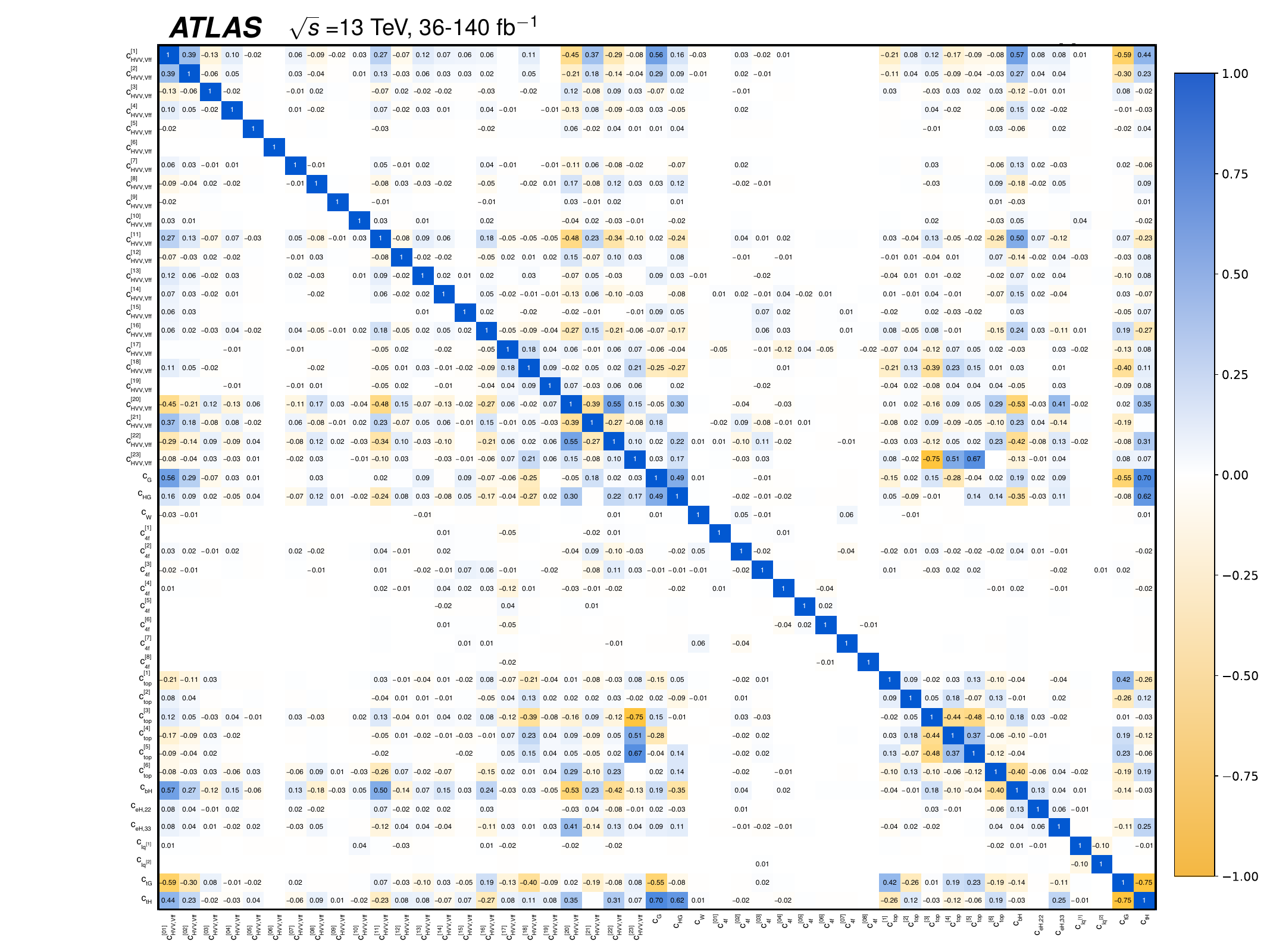}
\caption{Observed correlation matrix for the linear model without di-Higgs measurements.}\label{fig:corrmat_obs_lin}
\end{figure}

The observed linear correlation matrix is shown in Figure~\ref{fig:corrmat_obs_lin}, showing low but non-negligible correlation between the fit basis parameters, ranging from a few percent to $75\%$. The largest correlations are observed between $\cHVVEWtwentythree$ and $\ctopthree$, both constrained by top-quark measurements and between $\ctG$ and $\ctH$, both constrained by Higgs boson measurements.

\subsubsection{Results including di-Higgs measurements}

\begin{figure}{}
\centering
\includegraphics[width=0.95\textwidth]{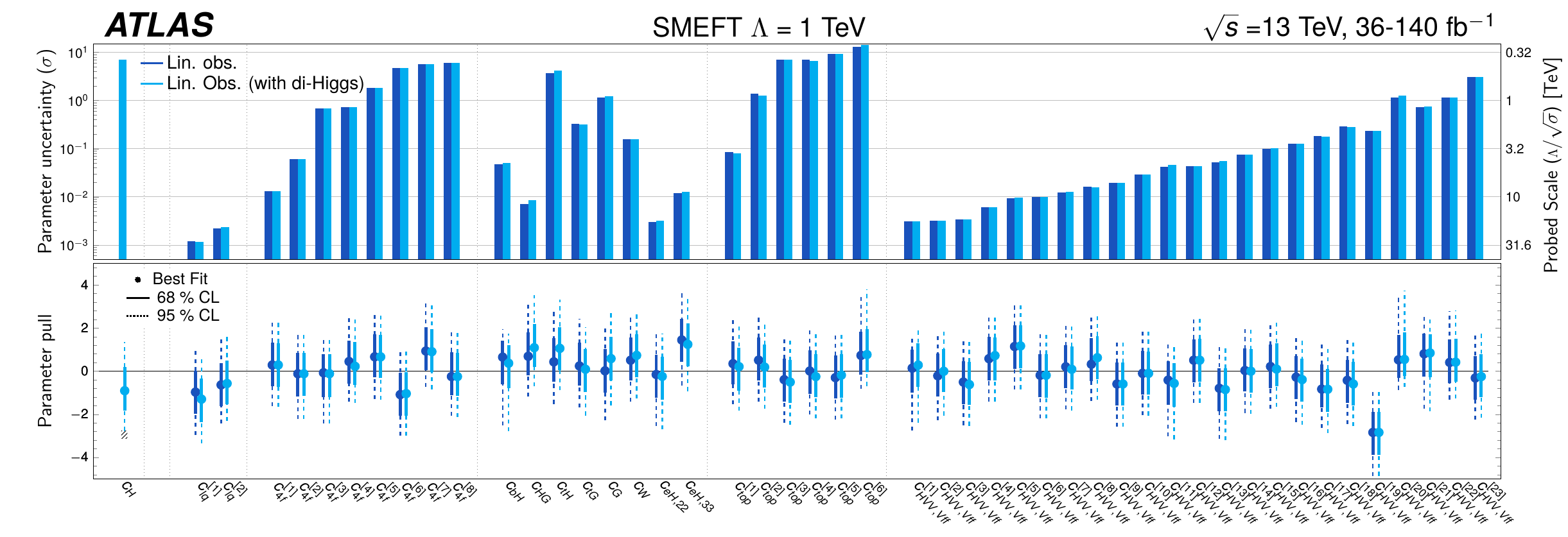}
\caption{Observed constraints on the fit basis SMEFT coefficients before (dark blue) and after (light blue) including the di-Higgs input measurements in the combined interpretation, using the linear SMEFT model. The symmetrised uncertainty $\sigma$ is shown in the upper panel, together with the corresponding probed energy scale of new physics $\Lambda/\sqrt\sigma$. The lower panel shows, for each operator, the measured value and the 68$\%$ (solid) and 95$\%$ (dashed) CL intervals, divided by the symmetrised uncertainty shown in the upper panel. The hashed box indicates where the statistical model becomes undefined due to predicted negative yields in an analysis category caused by the value of the Wilson coefficient scanned.}\label{fig:HHvsnoHH}
\end{figure}

Figure~\ref{fig:HHvsnoHH} shows the comparison between the observed fit results when including and not including the di-Higgs measurements, for a linear SMEFT model. The inclusion of di-Higgs measurements allows the constraint on the Higgs boson self-coupling $\cH$, while all other coefficients are not significantly affected. The result represents the first simultaneous extraction of the Higgs boson self-coupling coefficient $\cH$  with other coefficients affecting the Higgs boson couplings to other SM particles.

\subsection{Linear-plus-quadratic results}\label{sec:quad}
\begin{figure}
\centering
\includegraphics[width=1\textwidth]{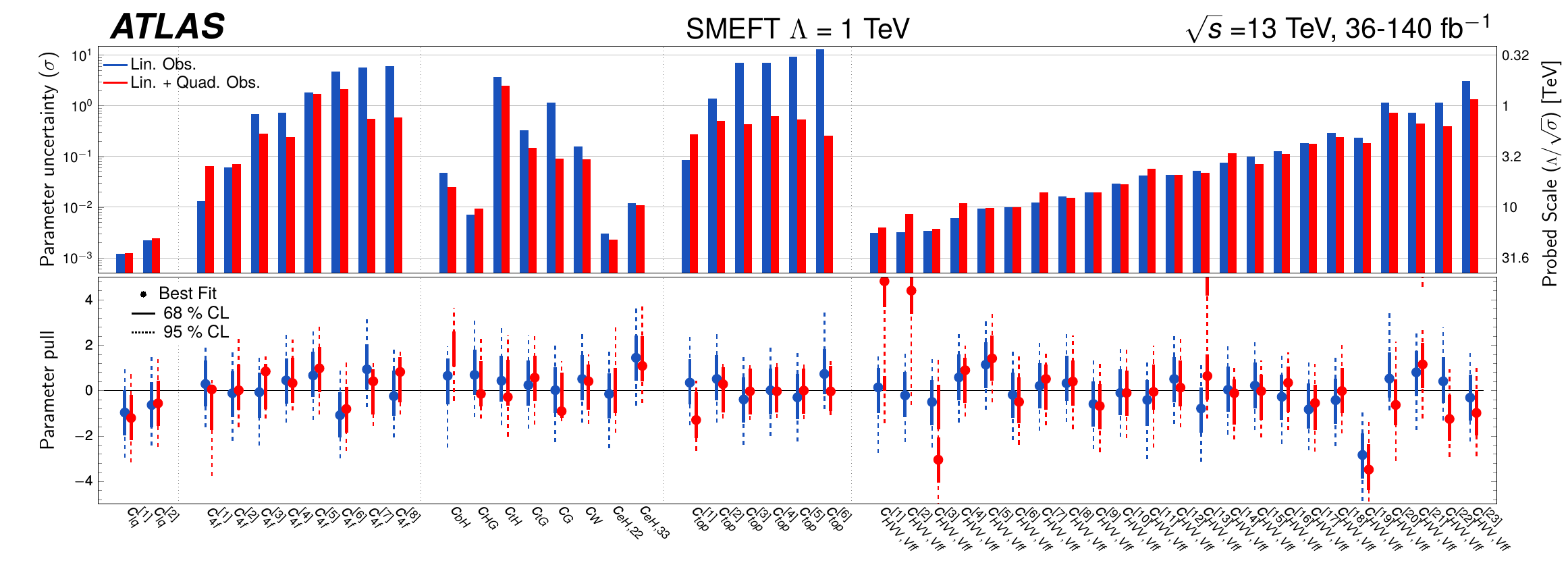}
\caption{Comparison of the constraints on the fit basis $\vec c^\prime$ SMEFT coefficients for the linear (blue) and the linear-plus-quadratic (red) SMEFT models, where all other coefficients and nuisance parameters are profiled. The symmetrised uncertainty $\sigma$ is shown in the upper panel, together with the corresponding probed energy scale of new physics $\Lambda/\sqrt\sigma$. The lower panel shows, for each operator, the measured value and the 68$\%$ (solid) and 95$\%$ (dashed) CL intervals, divided by the symmetrised uncertainty shown in the upper panel.  }\label{fig:quad}
\end{figure}

Figure~\ref{fig:quad} shows the observed 68$\%$ and 95 $\%$ CL limits on the 47 Wilson coefficients of the fit basis for the linear and linear-plus-quadratic model, without including di-Higgs measurements.

The complex model with 47 simultaneously floating parameters and quadratic terms gives rise to a non-trivial likelihood landscape with multiple local minima. To prevent finding local instead of glocal minima, many likelihood scans are performed, starting the minimisations at different points in the parameter space. The starting points are often chosen close to the minima of conditional fits of a different parameter. The resulting confidence intervals, obtained from the merging of the likelihood shapes around multiple minima, are often disjoint, such as for the $\cHVVEWthree$ and $\cHVVEWthirteen$ coefficients. The $\cbH$ and $\ceHtwo$ coefficients have two well-separated minima. In both cases the lower minimum lies furthest from the SM prediction. Nonetheless, the SM value is well inside the 95$\%$~CL interval.

Constraints in the linear-plus-quadratic model are stronger than in the linear model for the four-fermions and the top-quark groups, while they are weaker for some of the directions in the \textbf{$\cHVVEW$} group. This is due to the presence of different minima in the likelihood scans that merge together, an effect that is larger for expected results where minima are often degenerate.

Since quadratic terms are the same order in the SMEFT expansion as dimension-eight operators, the results obtained with the linear-plus-quadratic model should be interpreted as an indicator of the possible impact of higher-order terms in the SMEFT expansion.

\subsection{Simplified likelihood models}
An alternative likelihood function, based on a multivariate Gaussian model approximation of the input measurements, can be constructed from the covariance matrix of the measurements at the best-fit values, which is available in the auxiliary materials. An identical SMEFT re-parameterisation to the case of the full likelihood model is then performed to obtain a simplified likelihood SMEFT model.

\begin{figure}
\centering
\includegraphics[width=1\textwidth]{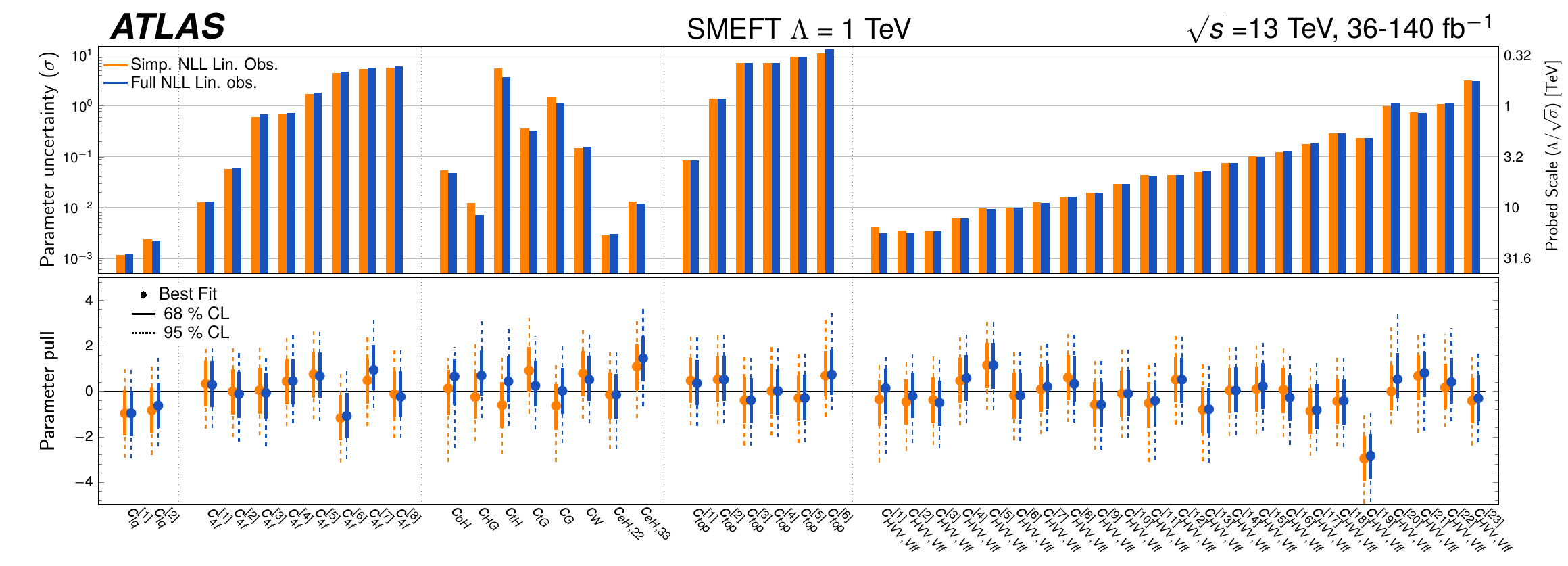}
\includegraphics[width=1\textwidth]{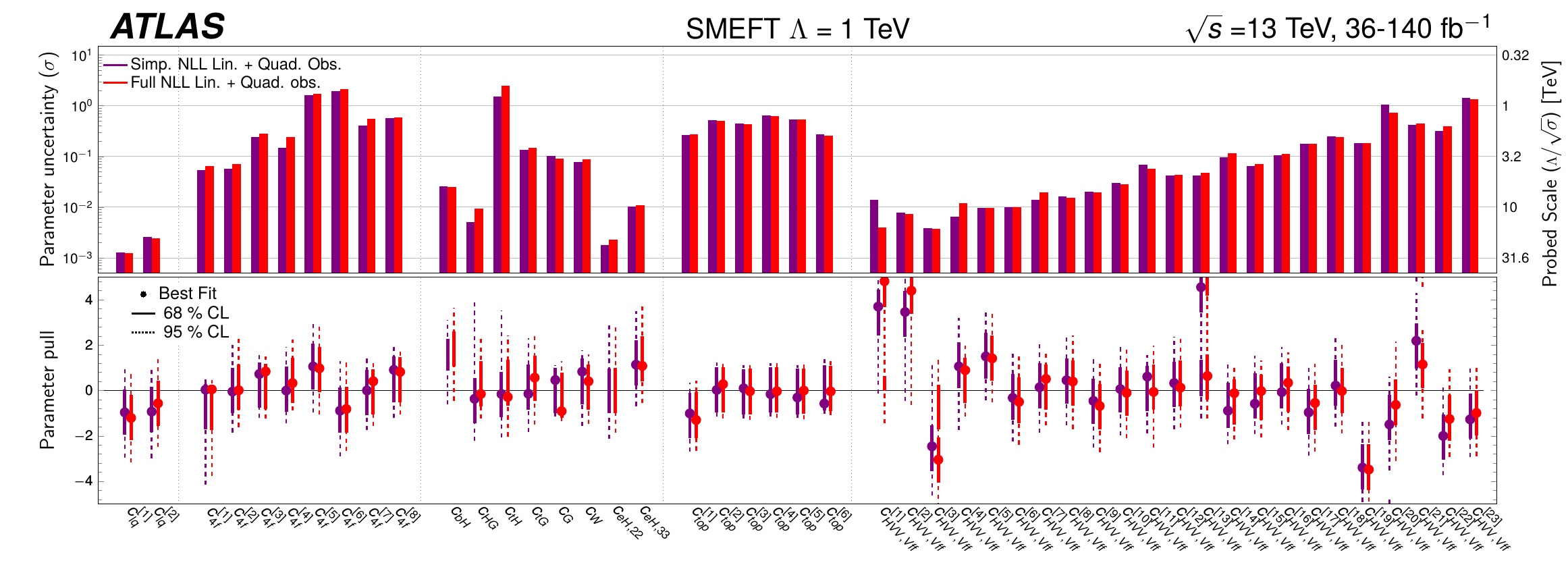}
\caption{Comparison between the full (blue and red) and simplified (orange and purple) likelihood constraints on the SMEFT coefficients composing the fit basis $\vec c^\prime$, for the linear (top) and linear-plus-quadratic (bottom) model and observed data, where all other coefficients and nuisance parameters are profiled. The symmetrised uncertainty $\sigma$ is shown in the upper panel, together with the corresponding probed energy scale of new physics $\Lambda/\sqrt\sigma$. The lower panel shows, for each operator, the measured value and the 68$\%$ (solid) and 95$\%$ (dashed) CL intervals, divided by the symmetrised uncertainty shown in the upper panel.}\label{fig:MVG_quad_obs}
\end{figure}

Figure~\ref{fig:MVG_quad_obs}~(top) shows the comparison of the observed constraints obtained with full and simplified likelihoods for the linear SMEFT model without di-Higgs measurements. Figure~\ref{fig:MVG_quad_obs} (bottom) shows the corresponding comparison for the linear-plus-quadratic model using observed data without di-Higgs measurements. The simplified models are found to reproduce well the results obtained with the full likelihood. The largest differences are observed for the Yukawa-coupling modifiers $\cbH$ and $\ctH$, the top-gluon coupling $\ctG$, and the Higgs-gluon coupling $\cHG$. In the few instances where the simplified multivariate-Gaussian likelihood exhibits substantial deviation from the full likelihood results, the differences can be attributed either to non-Gaussian features arising in the tails of the underlying measurements or to strong parameter correlations.

\subsection{EFT to UV matching}
\label{sec:results-matching}
SMEFT provides a framework to connect precision collider measurements to UV-complete theories through matching procedures. It allows patterns of effective operators that deviate from the SM expectations to be identified and, in the absence of significant deviations, the resulting constraints on Wilson coefficients to be interpreted in terms of specific SM extensions.
Assuming that new physics will appear at a mass scale well above the electroweak scale, the parameters of a given UV-complete model can be mapped onto SMEFT Wilson coefficients by integrating out the heavy states~\cite{Henning:2014wua}.

In this work, Two-Higgs-Doublet Models (2HDM)~\cite{Branco:2011iw} and models with an additional neutral gauge boson~($Z'$)~\cite{Langacker:2008yv} are considered, since their parameters are constrained by measurements from different sectors. Constraints on 2HDM scenarios based on single Higgs boson data were previously published by ATLAS~\cite{HIGG-2022-17}, while the present analysis extends the interpretation to a global combination including di-Higgs measurements. Heavy-vector-boson models can be matched to both four-fermion and Higgs-sector operators, and can be constrained by HMDY measurements and electroweak precision observables.

\subsubsection{Matching to two-Higgs-doublet models}
In two-Higgs-doublet models an additional doublet of scalar complex fields $\Phi_2$ is added to the SM Higgs doublet $\Phi_1$, leading, after electroweak symmetry breaking, to five physical Higgs bosons: the SM CP-even boson $h$, the heavier CP-even scalar $H$, the pseudo scalar $A$ and the charged Higgs bosons $H^{\pm}$. The model considered here assumes CP conservation and a (softly broken) $Z_2$ discrete symmetry that forbids quartic terms of the scalar field potential V($\Phi_1$, $\Phi_2$) that contain odd powers of either $\Phi_1$ or $\Phi_2$.

The models are defined by seven free parameters: the four boson masses ($m_h$, $m_H$, $M_A$, $m_{H^{\pm}}$), two mixing angles $\alpha$ and $\beta$, defined as the mixing angle between the two CP-even Higgs fields and the ratio of the vacuum expectation values of the two doublets, respectively, and the parameter $m^2_{12}$, which softly breaks the
$Z_2$ symmetry of the potential. Different types of 2HDM can be defined depending on the couplings of SM particles to the doublets. The 2HDM Type-I model is considered here, where all fermions couple to the same Higgs doublet.

Taking the 2HDM Lagrangian, expanding all of its terms and integrating out the heavy fields, it is possible to obtain an expression that can be matched to terms of dimension six and dimension eight in the SMEFT framework. In particular, the SMEFT coefficients that can be matched at dimension six are:
\begin{itemize}
\item The Yukawa coupling modifiers for top quarks, bottom quarks, $\tau$-leptons, and muons, respectively: $\ctH$, $\cbH$, $\ensuremath{c_{\tau H}}$, $\ensuremath{c_{\mu H}}$. The impact of other Yukawa coupling modifiers is negligible.
\item The Higgs boson self-interaction modifier: $\cH$.
\end{itemize}
The SMEFT operators are matched to the 2HDM Type-I parameters through the expressions of Ref.~\cite{PhysRevD.102.055012}:
\begin{align}
\frac{v^2\cH}{\Lambda^2} &= \frac{\cos^2 (\beta - \alpha) M_A^2}{v^2}, \\
\frac{v^2\cbH}{\Lambda^2} &= -y_b  \frac{\cos(\beta - \alpha)}{\tan\beta}, \\
\frac{v^2\ctH}{\Lambda^2} &= -y_t  \frac{\cos(\beta - \alpha)}{\tan\beta}, \\
\frac{v^2c_{\tau H}}{\Lambda^2} &= -y_{\tau}  \frac{\cos(\beta - \alpha)}{\tan\beta}, \\
\frac{v^2c_{\mu H}}{\Lambda^2} &= -y_{\mu}  \frac{\cos(\beta - \alpha)}{\tan\beta},
\end{align}
where $y_i = \sqrt{2} m_i / v$ for $i=b,t,\tau,\mu$, and $v$ is the SM vacuum expectation value. The 2HDM constraints are obtained by placing the above equations for Wilson coefficients in the likelihood fit, while setting all other coefficients to zero.

For the expansion of the 2HDM in SMEFT to be valid, the heavy degrees of freedom of the model must be decoupled, and the masses of the additional bosons must be much higher than the weak scale, corresponding to the so-called alignment limit ($\cos(\beta - \alpha) \rightarrow 0$). The parameter $m_h$ is assumed to be 125.09~GeV, and the other masses are assumed to be degenerate and are set to 1~TeV ($M_A=m_H=m_{H^{\pm}}=1$ TeV) . The matching is done for dimension-six operators only (i.e. at order $\Lambda^{-2}$), although the role of dimension-eight operators (and quadratic effects at dimension six) is known to be non-negligible~\cite{Dawson_2022}.

In the SMEFT approach, operators corresponding to the coupling strength modifiers between the Higgs boson and the weak bosons appear only at dimension eight. For the 2HDM Type-I model, neglecting these contributions results in weaker constraints in the high-$\tan \beta$ region.

\begin{figure}
\centering
\includegraphics[width=0.5\linewidth]{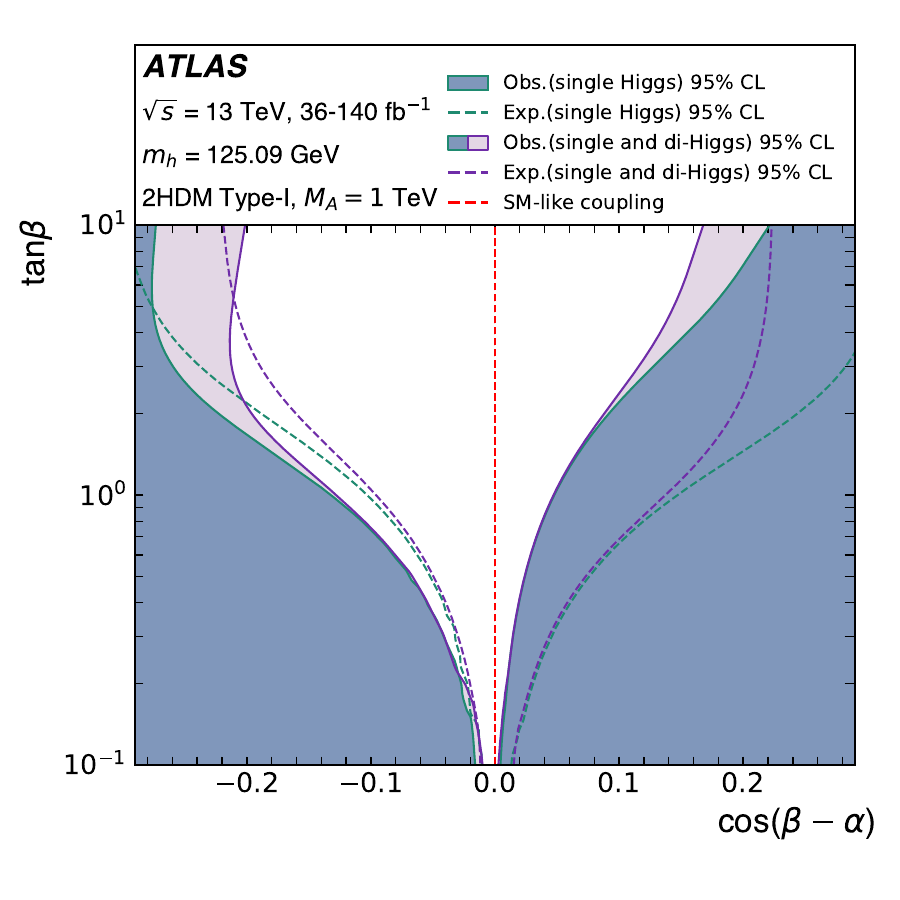}
\caption{Regions of the 2HDM ($\tan \beta$, $\cos(\beta - \alpha)$) parameter plane excluded at 95$\%$ CL in the SMEFT-based approach by the measured rates of single Higgs boson production and decays (blue filled region) and by the combination of single Higgs and di-Higgs analyses (violet filled region, with the overlap with the single-Higgs exclusion also shown in blue) in the Type-I model. The dashed lines show the borders of the corresponding expected exclusion regions for the SM hypothesis. Results are derived assuming that $|\cos(\beta - \alpha)| \ll $1, near the alignment limit represented by the red dashed line, and that the masses of the non-SM Higgs bosons are large compared to the SM $v$.}
\label{fig:2hdm_results}
\end{figure}

As shown in Ref.~\cite{HIGG-2022-17}, the addition of the self-coupling operator $c_H$ improves the limits in this region even when using only single-Higgs boson measurements. Adding di-Higgs measurements to the analysis further improves Type-I as shown in Figure~\ref{fig:2hdm_results}, which compare the limits in the $\cos(\beta - \alpha)$, $\tan \beta$ plane for single-Higgs only and for the combination between single and di-Higgs measurements. The other types of 2HDM are already more constrained than Type-I without the inclusion of the Higgs boson self-coupling, and show identical constraints between the two cases.

\subsubsection{Matching to $Z'$ models}
In $Z'$ models, an additional singlet is added to the Lagrangian as:
\begin{equation}
\mathcal{L}_{Z'} = -\frac{1}{4} Z'_{\mu\nu} Z'^{\mu\nu} + \frac{1}{2} m_{Z'}^2 Z'_\mu Z'^\mu  + \frac{\epsilon}{2} B_{\mu\nu} Z'^{\mu\nu} + g_H^2 Z'_\mu Z'^\mu |H|^2 + Z'_\mu \mathcal{J}^\mu,
\end{equation}
where $Z'_\mu$ is the new gauge boson, $g_H$ is its coupling to the Higgs boson doublet, and $\mathcal{J}^\mu$ is the current associated with the new gauge boson.

The models are defined by the hypercharges $Q_i$ of the SM particles under the new $U(1)'$ group and three free parameters: $\epsilon$, the kinetic mixing parameter; $m_{Z'}$, the mass of the new gauge boson; and $g_D$, which enters the expressions for the couplings between the $Z'$ and both the Higgs boson and fermions ($\mathcal{J}^\mu$).
The SMEFT coefficients that can be matched at dimension six are:
\begin{itemize}
\item Four-fermion operators:
$c_{ll}$, $c_{lq}^{(1)}$, $c_{qq}^{(1)}$, $c_{ff}$, $c_{f'f}$, $c_{ud}^{(1)}$, $c_{lf}$, $c_{ql}^{(1)}$ where $q,l$ are the left-handed fermions and $f,f' = u, d, e$ are the right-handed fermions.
\item Higgs boson operators: $c_{H l}^{(1)}$, $\cHQone$, $\ensuremath{c_{Hf}}$, $\cHbox$ and $\cHDD$.
\end{itemize}
The different models are listed in Table~1 of Ref.~\cite{dawson2024smeftmatchingzprimemodels}, and the matching expressions for the dimension-six operators are given in Eq. (19)-(26), summarised here in Table~\ref{tab:Zprime_matching}. The $Z'$ constraints are obtained by placing these expressions  in the likelihood fit, while setting all other Wilson coefficients to zero. In this work, the $B{-}L$ and mirrored hypercharge models are considered.

\begin{table}
\caption{Matching expressions~\cite{dawson2024smeftmatchingzprimemodels} for the dimension-six operators in the $Z'$ models considered in this work. The different models are defined by the hypercharges $Q_i$ of the SM particles under the new U(1)$'$ group.}
\centering
{\tiny
\renewcommand{\arraystretch}{1.6}
\setlength{\tabcolsep}{10pt}
\begin{tabular}{|c l|c l|}
\hline
\multicolumn{4}{|c|}{4-fermion operators} \\
\hline
$C_{ll}[ijkl]/\Lambda^2$ & $-\dfrac{1}{2M_{Z'}^2}(g_{ij}^{lL} + \epsilon g' Y_q \delta_{ij})(g_{kl}^{lL} + \epsilon g' Y_q \delta_{kl})$
& $C_{lq}^{(1)}[ijkl]/\Lambda^2$ & $-\dfrac{1}{M_{Z'}^2}(g_{ij}^{lL} + \epsilon g' Y_q \delta_{ij})(g_{kl}^{qL} + \epsilon g' Y_q \delta_{kl})$ \\

$C_{qq}^{(1)}[ijkl]/\Lambda^2$ & $-\dfrac{1}{2M_{Z'}^2}(g_{ij}^{qL} + \epsilon g' Y_q \delta_{ij})(g_{kl}^{qL} + \epsilon g' Y_q \delta_{kl})$
& $C_{ff}[ijkl]/\Lambda^2$ & $-\dfrac{1}{2M_{Z'}^2}(g_{ij}^{fR} + \epsilon g' Y_f \delta_{ij})(g_{kl}^{fR} + \epsilon g' Y_f \delta_{kl})$ \\

$C_{f'f}[ijkl]/\Lambda^2$ & $-\dfrac{1}{M_{Z'}^2}(g_{ij}^{f'R} + \epsilon g' Y_{f'} \delta_{ij})(g_{kl}^{fR} + \epsilon g' Y_f \delta_{kl}), $
& $C_{ud}^{(1)}[ijkl]/\Lambda^2$ & $-\dfrac{1}{M_{Z'}^2}(g_{ij}^{uR} + \epsilon g' Y_u \delta_{ij})(g_{kl}^{dR} + \epsilon g' Y_d \delta_{kl})$ \\

$C_{lf}[ijkl]/\Lambda^2$ & $-\dfrac{1}{M_{Z'}^2}(g_{ij}^{lL} + \epsilon g' Y_q \delta_{ij})(g_{kl}^{fR} + \epsilon g' Y_f \delta_{kl})$
& $C_{ql}^{(1)}[ijkl]/\Lambda^2$ & $-\dfrac{1}{M_{Z'}^2}(g_{ij}^{qL} + \epsilon g' Y_q \delta_{ij})(g_{kl}^{lR} + \epsilon g' Y_l \delta_{kl})$ \\
\hline\hline
\multicolumn{4}{|c|}{$\psi^2 H^2 D$ and $H^4 D^2$ operators} \\
\hline
$C_{Hl}^{(1)}[ij]/\Lambda^2$ & $-\dfrac{1}{2M_{Z'}^2}(2g_{ij} + \epsilon g')(g_{ij}^{lL} + \epsilon g' Y_l \delta_{ij})$
& $C_{Hq}^{(1)}[ij]/\Lambda^2$ & $-\dfrac{1}{2M_{Z'}^2}(2g_{ij} + \epsilon g')(g_{ij}^{qL} + \epsilon g' Y_q \delta_{ij})$ \\

$C_{Hf}[ij]/\Lambda^2$ & $-\dfrac{1}{2M_{Z'}^2}(2g_{ij} + \epsilon g')(g_{ij}^{fL} + \epsilon g' Y_f \delta_{ij})$
& $C_{H\Box}/\Lambda^2$ & $-\dfrac{1}{8M_{Z'}^2}(2g_H + \epsilon g')^2$ \\

$C_{HDD}/\Lambda^2$ & $-\dfrac{1}{2M_{Z'}^2}(2g_H + \epsilon g')^2$
& & \\
\hline\hline
\multicolumn{4}{|c|}{$g_H = Q_H g_D \quad g_{ij}^{qL} = Q_{q_L}\delta_{ij} \quad g_{ij}^{lL} = Q_{l_L}\delta_{ij} \quad g_{ij}^{fR} = Q_{f_R}\delta_{ij}$} \\
\hline
\end{tabular}
}

\label{tab:Zprime_matching}
\end{table}

\begin{figure}
\centering
\includegraphics[width=0.48\linewidth]{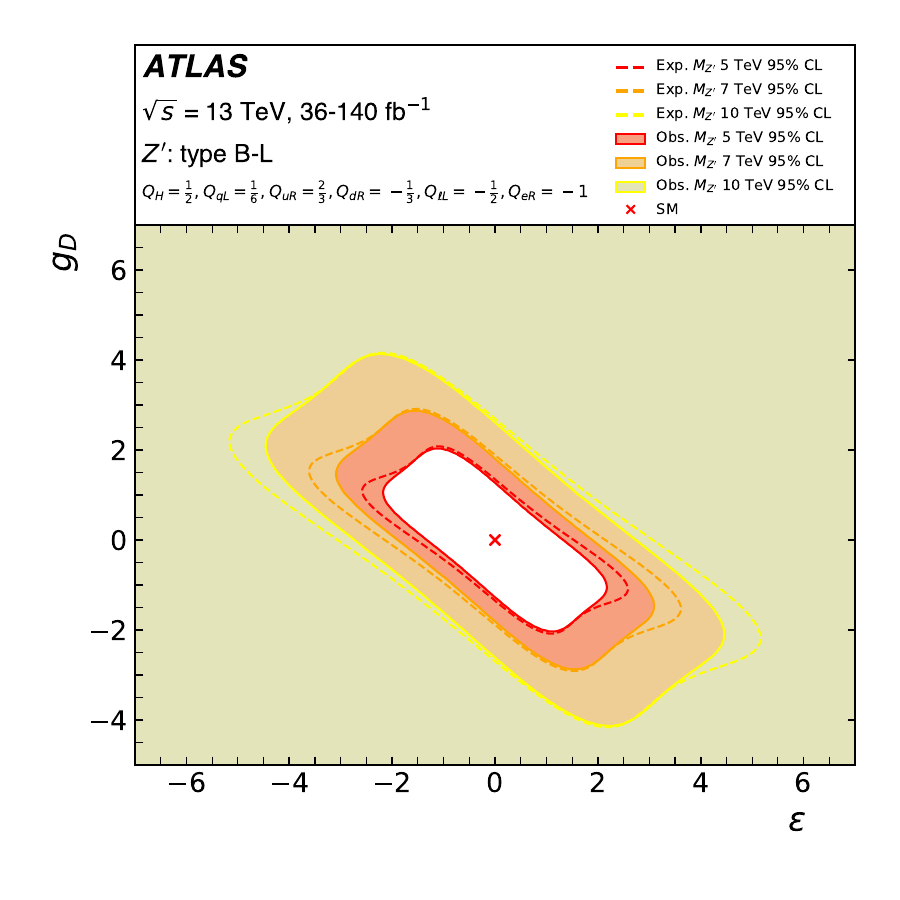}
\includegraphics[width=0.48\linewidth]{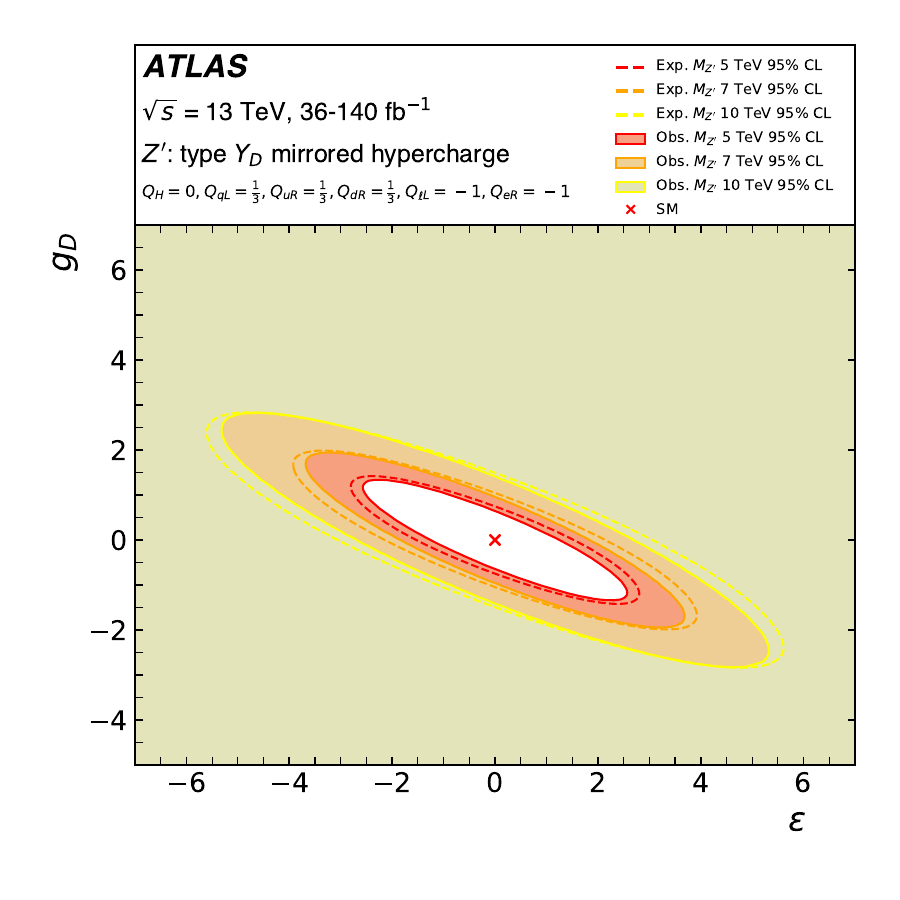}
\caption{Regions of the $Z'$($g_D$, $\epsilon$) parameter plane excluded at 95$\%$ CL  in the SMEFT-based approach by the analyses included in the combination  and for the $B{-}L$ (left) and $Y_D$ mirrored hypercharge (right) models for different hypotheses of the new boson mass $m_{Z'}$ (5, 7, 10~TeV). The dashed lines show the borders of the corresponding expected exclusion regions for the SM hypothesis.}
\label{fig:Zprime_results}
\end{figure}

The results for these two models are shown in Figure~\ref{fig:Zprime_results}. The limits are displayed in the $g_D$ vs. $\epsilon$ plane for different values of $m_{Z'}$ (5, 7, 10~TeV). The constraints are driven in both models by the high-mass tails of the neutral-current Drell--Yan $\tau\tau$ measurement and by the electroweak precision observables. The remaining measurements are found to have a negligible impact on the results. The results are found to be dependent on the mass $m_{Z'}$ of the new boson and the given set of hypercharges, with stronger constraints for lower masses.

%
%
%
%

%
%
%

%
\section{Conclusion}
\label{sec:conclusion}

This work presented the results of the Standard Model effective field theory interpretation of a wide range of input measurements from different groups of processes: Higgs-boson production and decay in the simplified template cross-sections framework, differential cross-section measurements of $WW$, $WZ$ and EW $Zjj$ production, high-mass Drell--Yan analyses, $t\bar{t}$ production, and two di-Higgs channels, $b\bar{b}\gamma\gamma$ and $b\bar{b}\tau\tau$, all measured by the ATLAS Collaboration using Run 2 data, together with precision electroweak observables from LEP, SLD and ATLAS. Constraints on individual Wilson coefficients in the Warsaw basis and simultaneous constraints on linear combinations of Wilson coefficients are reported, extending previous ATLAS effective field theory interpretations in both the number of input measurements and the number of simultaneously constrained parameters.

The main set of results, where di-Higgs measurements are not included, is accompanied by the corresponding simplified likelihood model results to facilitate the reinterpretation of the results outside the ATLAS Collaboration. A total of 47 operators and linear combinations of operators are simultaneously constrained. A variant of these results including di-Higgs measurements is also provided, where the Higgs-boson self-coupling coefficient $c_H$ is added to the fit basis, and constrained simultaneously with the other coefficients. No significant deviations from the SM predictions are observed.

A comparison of results interpreted with a linearised Standard Model effective field theory model that only considers terms suppressed by factors up to $\Lambda^{-2}$, and a variant including quadratic terms that considers all available terms, including those with suppression factor $\Lambda^{-4}$, shows that the effect of operators suppressed by $\Lambda^{-4}$ can significantly affect constraints on Wilson coefficients for a mass scale of $\Lambda$=1 TeV.

The results discussed also include the re-interpretation of the measurements in terms of the two-Higgs-doublet models and heavy-vector-boson models, establishing effective field theory as an interface between the measurements and beyond-Standard Model physics.


%

%
%
%
%
%
\section*{Acknowledgements}
%

%

%
%

%
%

We thank CERN for the very successful operation of the LHC and its injectors, as well as the support staff at
CERN and at our institutions worldwide without whom ATLAS could not be operated efficiently.

The crucial computing support from all WLCG partners is acknowledged gratefully, in particular from CERN, the ATLAS Tier-1 facilities at TRIUMF/SFU (Canada), NDGF (Denmark, Norway, Sweden), CC-IN2P3 (France), KIT/GridKA (Germany), INFN-CNAF (Italy), NL-T1 (Netherlands), PIC (Spain), RAL (UK) and BNL (USA), the Tier-2 facilities worldwide and large non-WLCG resource providers. Major contributors of computing resources are listed in Ref.~\cite{ATL-SOFT-PUB-2026-001}.

We gratefully acknowledge the support of ANPCyT, Argentina; YerPhI, Armenia; ARC, Australia; BMWFW and FWF, Austria; ANAS, Azerbaijan; CNPq and FAPESP, Brazil; NSERC, NRC and CFI, Canada; CERN; ANID, Chile; CAS, MOST and NSFC, China; Minciencias, Colombia; MEYS CR, Czech Republic; DNRF and DNSRC, Denmark; IN2P3-CNRS and CEA-DRF/IRFU, France; SRNSFG, Georgia; BMFTR, HGF and MPG, Germany; GSRI, Greece; RGC and Hong Kong SAR, China; ICHEP and Academy of Sciences and Humanities, Israel; INFN, Italy; MEXT and JSPS, Japan; CNRST, Morocco; NWO, Netherlands; RCN, Norway; MNiSW, Poland; FCT, Portugal; MNE/IFA, Romania; MSTDI, Serbia; MSSR, Slovakia; ARIS and MVZI, Slovenia; DSI/NRF, South Africa; MICIU/AEI, Spain; SRC and Wallenberg Foundation, Sweden; SERI, SNSF and Cantons of Bern and Geneva, Switzerland; NSTC, Taipei; TENMAK, T\"urkiye; STFC/UKRI, United Kingdom; DOE and NSF, United States of America.

Individual groups and members have received support from BCKDF, CANARIE, CRC and DRAC, Canada; CERN-CZ, FORTE and PRIMUS, Czech Republic; COST, ERC, ERDF, Horizon 2020 and Marie Sk{\l}odowska-Curie Actions, European Union; Investissements d'Avenir Labex, Investissements d'Avenir Idex and ANR, France; DFG and AvH Foundation, Germany; Herakleitos, Thales and Aristeia programmes co-financed by EU-ESF and the Greek NSRF, Greece; BSF-NSF and MINERVA, Israel; NCN and NAWA, Poland; La Caixa Banking Foundation, CERCA and AGAUR programs from Generalitat de Catalunya and PROMETEO and GenT Programmes Generalitat Valenciana, Spain; G\"{o}ran Gustafssons Stiftelse, Sweden; The Royal Society and Leverhulme Trust, United Kingdom; Eric and Wendy Schmidt Fund for Strategic Innovation, United States of America.

In addition, individual members wish to acknowledge support from Chile: Agencia Nacional de Investigaci\'on y Desarrollo (ANID FONDECYT reg. 1230987, FONDECYT 1230812, FONDECYT 1240864, Fondecyt 3240661, Fondecyt Regular 1240721); China: Chinese Ministry of Science and Technology (MOST-2023YFA1605700, MOST-2023YFA1609300), National Natural Science Foundation of China (NSFC 12275265, NSFC-W2543005); Czech Republic: Czech Science Foundation (GACR - 24-11373S), Ministry of Education Youth and Sports (ERC-CZ-LL2327, FORTE CZ.02.01.01/00/22\_008/0004632), PRIMUS Research Programme (PRIMUS/21/SCI/017); EU: H2020 European Research Council (ERC - 101002463); European Union: European Research Council (BARD No. 101116429, ERC - 101219398, ERC - 948254, ERC 101089007), European Regional Development Fund (HE COFUND GA No.101081355, ERDF), Marie Sklodowska-Curie Actions (GAP-101168829); France: Agence Nationale de la Recherche (ANR-21-CE31-0013, ANR-22-EDIR-0002, ANR-24-CE31-0504-01); Germany: Deutsche Forschungsgemeinschaft (DFG - 469666862); China: Research Grants Council (GRF); Italy: Istituto Nazionale di Fisica Nucleare (LHC-MIUR - 28003/2025), Ministero dell'Università e della Ricerca (NextGenEU 153D23001490006 M4C2.1.1, NextGenEU I53D23000820006 M4C2.1.1, NextGenEU I53D23001490006 M4C2.1.1, SOE2024\_0000023); Japan: Japan Society for the Promotion of Science (JSPS KAKENHI  JP25H0063, JSPS KAKENHI JP22H01227, JSPS KAKENHI JP22H04944, JSPS KAKENHI JP22KK0227, JSPS KAKENHI JP24K23939, JSPS KAKENHI JP24KK0251, JSPS KAKENHI JP25H00650, JSPS KAKENHI JP25H01291, JSPS KAKENHI JP25K01011, JSPS KAKENHI JP25K01023); Poland: Polish National Science Centre (NCN 2021/42/E/ST2/00350, NCN OPUS 2023/51/B/ST2/02507, NCN OPUS nr 2022/47/B/ST2/03059, NCN UMO-2019/34/E/ST2/00393, UMO-2022/47/O/ST2/00148, UMO-2023/49/B/ST2/04085, UMO-2023/51/B/ST2/00920, UMO-2024/53/N/ST2/00869); Spain: Agència de Gestió d'Ajuts Universitaris i de Recerca. (AGAUR - 2023 BP 00141), Ministry of Science and Innovation (RYC2019-028510-I, RYC2020-030254-I, RYC2021-031273-I, RYC2022-038164-I), Ministerio de Ciencia, Innovación y Universidades/Agencia Estatal de Investigaci\'on (EU NextGenerationEU (PRTR-C17.I1), PID2022-142604OB-C22); Sweden: Carl Trygger Foundation (Carl Trygger Foundation CTS 22:2312), Swedish Research Council (Swedish Research Council 2023-04654, VR 2021-03651, VR 2022-03845, VR 2022-04683, VR 2023-03403, VR 2024-05451, VR 2025-05940), Knut and Alice Wallenberg Foundation (KAW 2023.0366); Switzerland: Swiss National Science Foundation (SNSF - PCEFP2\_194658); United Kingdom: The Binks Trust, Royal Society (NIF-R1-231091); United States of America: U.S. Department of Energy (ECA DE-AC02-76SF00515), John Templeton Foundation (John Templeton Foundation 63206), Neubauer Family Foundation.

%
%


%
\clearpage
\appendix
\part*{Appendix}
\addcontentsline{toc}{part}{Appendix}
\label{sec:appendix}

\label{ch:appendixB}
The validity of the confidence intervals derived using Wilks' theorem for the results shown in Section~\ref{sec:quad} including quadratic terms is discussed in this appendix, following from the results presented in Ref. \cite{wilkspaper}.
Wilks' theorem is commonly used to derive confidence intervals, but it is often not valid in EFT fits that include quadratic terms. The quadratic parameterisation can violate the regularity conditions of the theorem by introducing boundaries in the parameter space. For instance, for a simple measurement parameterised as $\mu = 1 + c^2$, no value of $c$ would lead to the measurement of $\mu < 1$, creating a boundary at $c=0$. Near such boundaries, the profile likelihood ratio distribution does not follow the $\chi^2$ distribution assumed by Wilks theorem, so confidence intervals derived from it may not accurately reflect the true 68$\%$ and 95$\%$ quantiles. Depending on the relative sizes of the linear and quadratic terms, these intervals can underestimate the true uncertainties \cite{wilkspaper}.

\begin{figure}[htbp]
\centering
\includegraphics[width=0.9\textwidth]{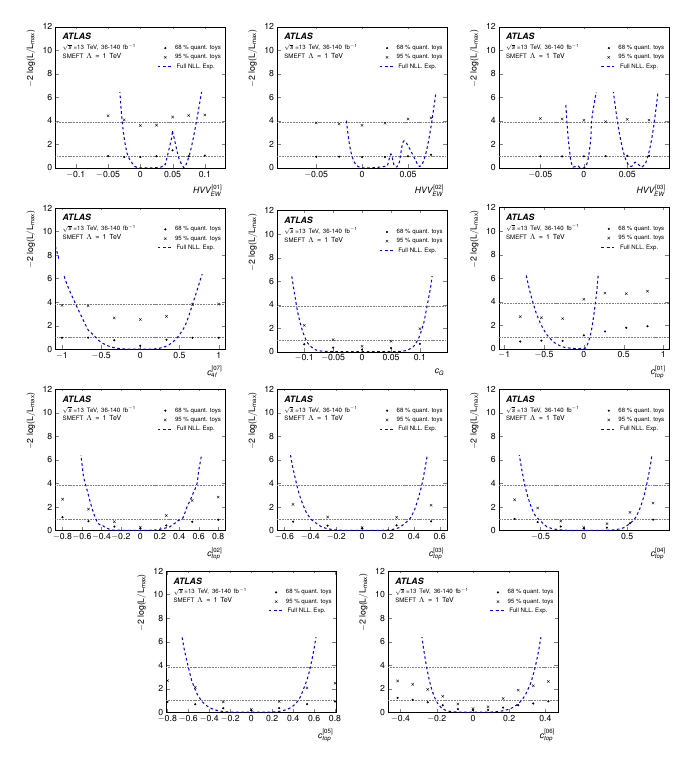}
\caption{Profile likelihood ratio scan of the coefficients of the fit basis that are dominated by quadratic terms using the linear-plus-quadratic model, confidence intervals calculated with the Wilks theorem (intersections with the dashed lines), and  68$\%$ CL (black dots) and 95$\%$ CL (black crosses) quantiles obtained with toys.}
\label{fig:appendixC:toy_vs_quad}
\end{figure}

To study this effect, the simplified Gaussian likelihood results are compared with results obtained from toy data. Figure \ref{fig:appendixC:toy_vs_quad} shows the profile likelihood ratio scans for the coefficients dominated by quadratic terms. The intersections with the toy-derived quantiles (black dots and crosses) indicate the true 68$\%$ and 95$\%$ confidence intervals, which can differ significantly from the Wilks-based intervals, especially near the boundaries. Coefficients dominated by linear terms, in contrast, remain well-described by Wilks theorem.

The constraints of operators that are strongly dominated by quadratic terms ($c_{4f}^{[07]}$, $c_G$, operators of the \textbf{$c_{top}$} group, except $c_{top}^{[01]}$) are stronger than the ones obtained with the Wilks theorem construction (with the partial exception of the right side of the interval for $c_G$), while operators whose sensitivity is degraded with the inclusion of quadratic terms (e.g. $c_{top}^{[01]}$, $c_{HVV,Vff}^{[01]}$, $c_{HVV,Vff}^{[02]}$, $c_{HVV,Vff}^{[03]}$) behave like linear-dominated operators. Overall, this study shows that in this analysis the Wilks theorem assumption is generally conservative.

These results show however that standard Wilks-based intervals may not always be reliable for SMEFT fits including quadratic contributions. While toy-based quantiles could be used to define corrected intervals, this approach is computationally prohibitive for the full likelihood. This strongly motivates the development of analytical approximations for the profile likelihood ratio in quadratic models.


\clearpage
\printbibliography

@Article{Butterworth:2015oua,
      author         = "Butterworth, Jon and others",
      title          = "{PDF4LHC recommendations for LHC Run II}",
      journal        = "J. Phys. G",
      volume         = "43",
      year           = "2016",
      pages          = "023001",
      doi            = "10.1088/0954-3899/43/2/023001",
      eprint         = "1510.03865",
      archivePrefix  = "arXiv",
      primaryClass   = "hep-ph",
      reportNumber   = "OUTP-15-17P, SMU-HEP-15-12, TIF-UNIMI-2015-14,
                        LCTS-2015-27, CERN-PH-TH-2015-249",
      SLACcitation   = "%%CITATION = ARXIV:1510.03865;%%"
}

@article{Lai:2010vv,
      author         = "Lai, H.-L. and others",
      title          = "{New parton distributions for collider physics}",
      journal        = "Phys. Rev. D",
      volume         = "82",
      pages          = "074024",
      doi            = "10.1103/PhysRevD.82.074024",
      year           = "2010",
      eprint         = "1007.2241",
      archivePrefix  = "arXiv",
      primaryClass   = "hep-ph",
      reportNumber   = "MSUHEP-100707, SMU-HEP-10-10",
      SLACcitation   = "%%CITATION = ARXIV:1007.2241;%%",
}

@article{Dulat:2015mca,
      author         = "Dulat, Sayipjamal and Hou, Tie-Jiun and Gao, Jun and
                        Guzzi, Marco and Huston, Joey and Nadolsky, Pavel and
                        Pumplin, Jon and Schmidt, Carl and Stump, Daniel and Yuan,
                        C. P.",
      title          = "{New parton distribution functions from a global analysis
                        of quantum chromodynamics}",
      journal        = "Phys. Rev. D",
      volume         = "93",
      year           = "2016",
      number         = "3",
      pages          = "033006",
      doi            = "10.1103/PhysRevD.93.033006",
      eprint         = "1506.07443",
      archivePrefix  = "arXiv",
      primaryClass   = "hep-ph",
      SLACcitation   = "%%CITATION = ARXIV:1506.07443;%%"
}

@article{Harland-Lang:2014zoa,
      author         = "Harland-Lang, L. A. and Martin, A. D. and Motylinski, P.
                        and Thorne, R. S.",
      title          = "{Parton distributions in the LHC era: MMHT 2014 PDFs}",
      journal        = "Eur. Phys. J. C",
      number         = "5",
      volume         = "75",
      pages          = "204",
      doi            = "10.1140/epjc/s10052-015-3397-6",
      year           = "2015",
      eprint         = "1412.3989",
      archivePrefix  = "arXiv",
      primaryClass   = "hep-ph",
      reportNumber   = "LCTS-2014-47, IPPP-14-97, DCPT-14-194",
      SLACcitation   = "%%CITATION = ARXIV:1412.3989;%%",
}

@article{Ball:2012cx,
      author         = "{NNPDF Collaboration} and Ball, Richard D. and others",
      title          = "{Parton distributions with LHC data}",
      journal        = "Nucl. Phys. B",
      volume         = "867",
      year           = "2013",
      pages          = "244",
      doi            = "10.1016/j.nuclphysb.2012.10.003",
      eprint         = "1207.1303",
      archivePrefix  = "arXiv",
      primaryClass   = "hep-ph",
      reportNumber   = "EDINBURGH-2012-08, IFUM-FT-997, FR-PHENO-2012-014,
                        RWTH-TTK-12-25, CERN-PH-TH-2012-037, SFB-CPP-12-47\,
                        --CERN-PH-TH-2012-037",
      SLACcitation   = "%%CITATION = ARXIV:1207.1303;%%"
}

@article{Ball:2014uwa,
      author         = "{NNPDF Collaboration} and Ball, Richard D. and others",
      title          = "{Parton distributions for the LHC run II}",
      journal        = "JHEP",
      volume         = "04",
      year           = "2015",
      pages          = "040",
      doi            = "10.1007/JHEP04(2015)040",
      eprint         = "1410.8849",
      archivePrefix  = "arXiv",
      primaryClass   = "hep-ph",
      reportNumber   = "EDINBURGH-2014-15, IFUM-1034-FT, CERN-PH-TH-2013-253,
                        OUTP-14-11P, CAVENDISH-HEP-14-11",
      SLACcitation   = "%%CITATION = ARXIV:1410.8849;%%"
}

@article{Sjostrand:2014zea,
      author         = "Sj{\"o}strand, Torbj{\"o}rn and Ask, Stefan and Christiansen,
                        Jesper R. and Corke, Richard and Desai, Nishita and Ilten,
                        Philip and Mrenna, Stephen and Prestel, Stefan and
                        Rasmussen, Christine O. and Skands, Peter Z.",
      title          = "{An introduction to PYTHIA 8.2}",
      journal        = "Comput. Phys. Commun.",
      volume         = "191",
      year           = "2015",
      pages          = "159",
      doi            = "10.1016/j.cpc.2015.01.024",
      eprint         = "1410.3012",
      archivePrefix  = "arXiv",
      primaryClass   = "hep-ph",
      reportNumber   = "LU-TP-14-36, MCNET-14-22, CERN-PH-TH-2014-190,
                        FERMILAB-PUB-14-316-CD, DESY-14-178, SLAC-PUB-16122,
                        --FERMILAB-PUB-14-316-CD",
      SLACcitation   = "%%CITATION = ARXIV:1410.3012;%%"
}

@Article{Alwall:2011uj,
      author         = "Alwall, Johan and Herquet, Michel and Maltoni, Fabio and
                        Mattelaer, Olivier and Stelzer, Tim",
      title          = "{MadGraph 5: going beyond}",
      journal        = "JHEP",
      volume         = "06",
      pages          = "128",
      doi            = "10.1007/JHEP06(2011)128",
      year           = "2011",
      eprint         = "1106.0522",
      archivePrefix  = "arXiv",
      primaryClass   = "hep-ph"
}

@Article{Bahr:2008pv,
      author         = "B{\"a}hr, M. and others",
      title          = "{Herwig++ physics and manual}",
      journal        = "Eur. Phys. J. C",
      volume         = "58",
      year           = "2008",
      pages          = "639",
      doi            = "10.1140/epjc/s10052-008-0798-9",
      eprint         = "0803.0883",
      archivePrefix  = "arXiv",
      primaryClass   = "hep-ph",
      reportNumber   = "CERN-PH-TH-2008-038, CAVENDISH-HEP-08-03, KA-TP-05-2008,
                        DCPT-08-22, IPPP-08-11, CP3-08-05",
      SLACcitation   = "%%CITATION = ARXIV:0803.0883;%%"
}

@Article{Bellm:2015jjp,
      author         = "Bellm, Johannes and others",
      title          = "{Herwig 7.0/Herwig++ 3.0 release note}",
      journal        = "Eur. Phys. J. C",
      volume         = "76",
      year           = "2016",
      number         = "4",
      pages          = "196",
      doi            = "10.1140/epjc/s10052-016-4018-8",
      eprint         = "1512.01178",
      archivePrefix  = "arXiv",
      primaryClass   = "hep-ph",
      reportNumber   = "CERN-PH-TH-2015-289, MAN-HEP-2015-15, IFJPAN-IV-2015-13,
                        HERWIG-2015-01, KA-TP-18-2015, DCPT-15-142, MCNET-15-28,
                        IPPP-15-71, --HERWIG-2015-01",
      SLACcitation   = "%%CITATION = ARXIV:1512.01178;%%"
}

@Article{Nason:2004rx,
      author         = "Nason, Paolo",
      title          = "{A new method for combining NLO QCD with shower Monte Carlo algorithms}",
      journal        = "JHEP",
      volume         = "11",
      pages          = "040",
      doi            = "10.1088/1126-6708/2004/11/040",
      year           = "2004",
      eprint         = "hep-ph/0409146",
      archivePrefix  = "arXiv",
}

@Article{Frixione:2007vw,
      author         = "Frixione, Stefano and Nason, Paolo and Oleari, Carlo",
      title          = "{Matching NLO QCD computations with parton shower
                        simulations: the POWHEG method}",
      journal        = "JHEP",
      volume         = "11",
      pages          = "070",
      doi            = "10.1088/1126-6708/2007/11/070",
      year           = "2007",
      eprint         = "0709.2092",
      archivePrefix  = "arXiv",
      primaryClass   = "hep-ph",
}

@Article{Alioli:2010xd,
      author         = "Alioli, Simone and Nason, Paolo and Oleari, Carlo and Re,
                        Emanuele",
      title          = "{A general framework for implementing NLO calculations in
                        shower Monte Carlo programs: the POWHEG BOX}",
      journal        = "JHEP",
      volume         = "06",
      pages          = "043",
      doi            = "10.1007/JHEP06(2010)043",
      year           = "2010",
      eprint         = "1002.2581",
      archivePrefix  = "arXiv",
      primaryClass   = "hep-ph",
}

@article{Hartanto:2015uka,
      author         = "Hartanto, Heribertus B. and J{\"a}ger, Barbara and Reina,
                        Laura and Wackeroth, Doreen",
      title          = "{Higgs boson production in association with top quarks in
                        the POWHEG BOX}",
      journal        = "Phys. Rev. D",
      volume         = "91",
      year           = "2015",
      number         = "9",
      pages          = "094003",
      doi            = "10.1103/PhysRevD.91.094003",
      eprint         = "1501.04498",
      archivePrefix  = "arXiv",
      primaryClass   = "hep-ph",
      SLACcitation   = "%%CITATION = ARXIV:1501.04498;%%"
}

@article{Alioli:2008gx,
      author         = "Alioli, Simone and Nason, Paolo and Oleari, Carlo and Re,
                        Emanuele",
      title          = "{NLO vector-boson production matched with shower in
                        POWHEG}",
      journal        = "JHEP",
      volume         = "07",
      year           = "2008",
      pages          = "060",
      doi            = "10.1088/1126-6708/2008/07/060",
      eprint         = "0805.4802",
      archivePrefix  = "arXiv",
      primaryClass   = "hep-ph",
      reportNumber   = "NSF-KITP-08-74",
      SLACcitation   = "%%CITATION = ARXIV:0805.4802;%%"
}

@article{Anastasiou:2003ds,
      author         = "Anastasiou, Charalampos and Dixon, Lance and Melnikov,
                        Kirill and Petriello, Frank",
      title          = "{High-precision QCD at hadron colliders: Electroweak
                        gauge boson rapidity distributions at next-to-next-to leading order}",
      journal        = "Phys. Rev. D",
      volume         = "69",
      year           = "2004",
      pages          = "094008",
      doi            = "10.1103/PhysRevD.69.094008",
      eprint         = "hep-ph/0312266",
      archivePrefix  = "arXiv",
      reportNumber   = "SLAC-PUB-10288, UH-511-1042-03",
      SLACcitation   = "%%CITATION = HEP-PH/0312266;%%"
}

@article{Frixione:2007nw,
      author         = "Frixione, Stefano and Ridolfi, Giovanni and Nason, Paolo",
      title          = "{A positive-weight next-to-leading-order Monte Carlo for
                        heavy flavour hadroproduction}",
      journal        = "JHEP",
      volume         = "09",
      pages          = "126",
      doi            = "10.1088/1126-6708/2007/09/126",
      year           = "2007",
      eprint         = "0707.3088",
      archivePrefix  = "arXiv",
      primaryClass   = "hep-ph",
      SLACcitation   = "%%CITATION = ARXIV:0707.3088;%%",
}

@article{deFlorian:2016spz,
      author         = "de Florian, D. and others",
      title          = "{Handbook of LHC Higgs Cross Sections: 4. Deciphering the
                        Nature of the Higgs Sector}",
      collaboration  = "LHC Higgs Cross Section Working Group",
      doi            = "10.23731/CYRM-2017-002",
      year           = "2017",
      eprint         = "1610.07922",
      archivePrefix  = "arXiv",
      primaryClass   = "hep-ph",
      reportNumber   = "FERMILAB-FN-1025-T, CERN-2017-002-M",
      SLACcitation   = "%%CITATION = ARXIV:1610.07922;%%"
}

@Article{Hoeche:2009rj,
      author         = "H{\"o}che, Stefan and Krauss, Frank and Schumann, Steffen and Siegert, Frank",
      title          = "{QCD matrix elements and truncated showers}",
      journal        = "JHEP",
      volume         = "05",
      pages          = "053",
      doi            = "10.1088/1126-6708/2009/05/053",
      year           = "2009",
      eprint         = "0903.1219",
      archivePrefix  = "arXiv",
      primaryClass   = "hep-ph",
}

@Article{Gleisberg:2008fv,
      author         = "Gleisberg, Tanju and H{\"o}che, Stefan",
      title          = "{Comix, a new matrix element generator}",
      journal        = "JHEP",
      volume         = "12",
      pages          = "039",
      doi            = "10.1088/1126-6708/2008/12/039",
      year           = "2008",
      eprint         = "0808.3674",
      archivePrefix  = "arXiv",
      primaryClass   = "hep-ph",
}

@article{Hoeche:2012yf,
      author         = "H{\"o}che, Stefan and Krauss, Frank and Sch{\"o}nherr, Marek and
                        Siegert, Frank",
      title          = "{QCD matrix elements + parton showers. The NLO case}",
      journal        = "JHEP",
      volume         = "04",
      year           = "2013",
      pages          = "027",
      doi            = "10.1007/JHEP04(2013)027",
      eprint         = "1207.5030",
      archivePrefix  = "arXiv",
      primaryClass   = "hep-ph",
      reportNumber   = "SLAC-PUB-15191, IPPP-12-52, DCPT-12-104, LPN12-081,
                        FR-PHENO-2012-017, MCNET-12-09, --FR-PHENO-2012-017",
      SLACcitation   = "%%CITATION = ARXIV:1207.5030;%%"
}

@article{Bothmann:2019yzt,
      author = "Bothmann, Enrico and others",
      title = "{Event generation with Sherpa 2.2}",
      journal = "SciPost Phys.",
      volume = "7",
      year = "2019",
      number = "3",
      pages = "034",
      doi = "10.21468/SciPostPhys.7.3.034",
      reportNumber = "FERMILAB-PUB-19-218-T, SLAC-PUB-17433, IPPP/19/42, MCNET-19-11",
      eprint = "1905.09127",
      archivePrefix = "arXiv",
      primaryClass = "hep-ph",
}

@article{Hoeche:2011fd,
      author         = "H{\"o}che, Stefan and Krauss, Frank and Sch{\"o}nherr, Marek and
                        Siegert, Frank",
      title          = "{A critical appraisal of NLO+PS matching methods}",
      journal        = "JHEP",
      volume         = "09",
      year           = "2012",
      pages          = "049",
      doi            = "10.1007/JHEP09(2012)049",
      eprint         = "1111.1220",
      archivePrefix  = "arXiv",
      primaryClass   = "hep-ph",
      reportNumber   = "SLAC-PUB-14661, IPPP-11-67, DCPT-11-134, LPN11-58,
                        FR-PHENO-2011-019, MCNET-11-24",
      SLACcitation   = "%%CITATION = ARXIV:1111.1220;%%"
}

@article{Catani:2001cc,
      author         = "Catani, S. and Krauss, F. and Webber, B. R. and Kuhn, R.",
      title          = "{QCD Matrix Elements + Parton Showers}",
      journal        = "JHEP",
      volume         = "11",
      year           = "2001",
      pages          = "063",
      doi            = "10.1088/1126-6708/2001/11/063",
      eprint         = "hep-ph/0109231",
      archivePrefix  = "arXiv",
      reportNumber   = "CERN-TH-2000-367, CAVENDISH-HEP-00-03",
      SLACcitation   = "%%CITATION = HEP-PH/0109231;%%"
}

@article{Cascioli:2011va,
      author         = "Cascioli, Fabio and Maierh{\"o}fer, Philipp and Pozzorini, Stefano",
      title          = "{Scattering Amplitudes with Open Loops}",
      journal        = "Phys. Rev. Lett.",
      volume         = "108",
      year           = "2012",
      pages          = "111601",
      doi            = "10.1103/PhysRevLett.108.111601",
      eprint         = "1111.5206",
      archivePrefix  = "arXiv",
      primaryClass   = "hep-ph",
      reportNumber   = "ZU-TH-23-11, LPN11-66",
      SLACcitation   = "%%CITATION = ARXIV:1111.5206;%%"
}

@article{Buccioni:2019sur,
    author = "Buccioni, Federico and Lang, Jean-Nicolas and Lindert, Jonas M. and Maierh{\"o}fer, Philipp and Pozzorini, Stefano and Zhang, Hantian and Zoller, Max F.",
    title = "{OpenLoops 2}",
    eprint = "1907.13071",
    archivePrefix = "arXiv",
    primaryClass = "hep-ph",
    reportNumber = "IPPP/19/62, FR-PHENO-2019-12, PSI-PR-19-15, ZU-TH 37/19",
    doi = "10.1140/epjc/s10052-019-7306-2",
    journal = "Eur. Phys. J. C",
    volume = "79",
    number = "10",
    pages = "866",
    year = "2019"
}

@article{Denner:2016kdg,
      author         = "Denner, Ansgar and Dittmaier, Stefan and Hofer, Lars",
      title          = "{\textsc{Collier}: A fortran-based complex one-loop library in
                        extended regularizations}",
      journal        = "Comput. Phys. Commun.",
      volume         = "212",
      year           = "2017",
      pages          = "220-238",
      doi            = "10.1016/j.cpc.2016.10.013",
      eprint         = "1604.06792",
      archivePrefix  = "arXiv",
      primaryClass   = "hep-ph",
      reportNumber   = "FR-PHENO-2016-003, ICCUB-16-016",
      SLACcitation   = "%%CITATION = ARXIV:1604.06792;%%"
}

@article{Hamilton:2013fea,
      author         = "Hamilton, Keith and Nason, Paolo and Re, Emanuele and
                        Zanderighi, Giulia",
      title          = "{NNLOPS simulation of Higgs boson production}",
      journal        = "JHEP",
      volume         = "10",
      year           = "2013",
      pages          = "222",
      doi            = "10.1007/JHEP10(2013)222",
      eprint         = "1309.0017",
      archivePrefix  = "arXiv",
      primaryClass   = "hep-ph",
      reportNumber   = "MCNET-13-11, CERN-PH-TH-2013-205, OUTP-13-18P",
      SLACcitation   = "%%CITATION = ARXIV:1309.0017;%%"
}

@article{Hamilton:2015nsa,
      author         = "Hamilton, Keith and Nason, Paolo and Zanderighi, Giulia",
      title          = "{Finite quark-mass effects in the NNLOPS POWHEG+MiNLO
                        Higgs generator}",
      journal        = "JHEP",
      volume         = "05",
      year           = "2015",
      pages          = "140",
      doi            = "10.1007/JHEP05(2015)140",
      eprint         = "1501.04637",
      archivePrefix  = "arXiv",
      primaryClass   = "hep-ph",
      reportNumber   = "CERN-PH-TH-2015-006",
      SLACcitation   = "%%CITATION = ARXIV:1501.04637;%%"
}

@techreport{ATL-PHYS-PUB-2014-021,
      title         = "{ATLAS Pythia 8 tunes to 7 TeV data}",
      institution   = "CERN",
      reportNumber  = "ATL-PHYS-PUB-2014-021",
      address       = "Geneva",
      year          = "2014",
      url           = "https://cds.cern.ch/record/1966419",
}

@Booklet{ATL-PHYS-PUB-2021-011,
    author         = "{ATLAS Collaboration}",
    title          = "{Evaluating statistical uncertainties and correlations using the bootstrap method}",
    howpublished   = "{ATL-PHYS-PUB-2021-011}",
    url            = "https://cds.cern.ch/record/2759945",
    year           = "2021",
}

@article{Nason:2009ai,
      author         = "Nason, Paolo and Oleari, Carlo",
      title          = "{NLO Higgs boson production via vector-boson fusion
                        matched with shower in POWHEG}",
      journal        = "JHEP",
      volume         = "02",
      year           = "2010",
      pages          = "037",
      doi            = "10.1007/JHEP02(2010)037",
      eprint         = "0911.5299",
      archivePrefix  = "arXiv",
      primaryClass   = "hep-ph",
      SLACcitation   = "%%CITATION = ARXIV:0911.5299;%%"
}

@article{Anastasiou:2016cez,
      author         = "Anastasiou, Charalampos and Duhr, Claude and Dulat, Falko
                        and Furlan, Elisabetta and Gehrmann, Thomas and Herzog,
                        Franz and Lazopoulos, Achilleas and Mistlberger, Bernhard",
      title          = "{High precision determination of the gluon fusion Higgs
                        boson cross-section at the LHC}",
      journal        = "JHEP",
      volume         = "05",
      year           = "2016",
      pages          = "058",
      doi            = "10.1007/JHEP05(2016)058",
      eprint         = "1602.00695",
      archivePrefix  = "arXiv",
      primaryClass   = "hep-ph",
      reportNumber   = "CP3-16-01, ZU-TH-27-15, NIKHEF-2016-004,
                        CERN-TH-2016-006",
      SLACcitation   = "%%CITATION = ARXIV:1602.00695;%%"
}

@article{Anastasiou:2015ema,
      author         = "Anastasiou, Charalampos and Duhr, Claude and Dulat, Falko
                        and Herzog, Franz and Mistlberger, Bernhard",
      title          = "{Higgs Boson Gluon-Fusion Production in QCD at Three
                        Loops}",
      journal        = "Phys. Rev. Lett.",
      volume         = "114",
      year           = "2015",
      pages          = "212001",
      doi            = "10.1103/PhysRevLett.114.212001",
      eprint         = "1503.06056",
      archivePrefix  = "arXiv",
      primaryClass   = "hep-ph",
      reportNumber   = "CERN-PH-TH-2015-055, CP3-15-07",
      SLACcitation   = "%%CITATION = ARXIV:1503.06056;%%"
}

@article{Actis:2008ug,
      author         = "Actis, Stefano and Passarino, Giampiero and Sturm,
                        Christian and Uccirati, Sandro",
      title          = "{NLO electroweak corrections to Higgs boson production at
                        hadron colliders}",
      journal        = "Phys. Lett. B",
      volume         = "670",
      year           = "2008",
      pages          = "12-17",
      doi            = "10.1016/j.physletb.2008.10.018",
      eprint         = "0809.1301",
      archivePrefix  = "arXiv",
      primaryClass   = "hep-ph",
      reportNumber   = "PITHA-08-20, SFB-CPP-08-62, TTP08-38",
      SLACcitation   = "%%CITATION = ARXIV:0809.1301;%%"
}

@article{Luisoni:2013cuh,
    author = "Luisoni, Gionata and Nason, Paolo and Oleari, Carlo and Tramontano, Francesco",
    title = "{\(HW^{\pm}/HZ\) + 0 and 1 jet at NLO with the POWHEG BOX interfaced to GoSam and their merging within MiNLO}",
    eprint = "1306.2542",
    archivePrefix = "arXiv",
    primaryClass = "hep-ph",
    doi = "10.1007/JHEP10(2013)083",
    journal = "JHEP",
    volume = "10",
    pages = "083",
    year = "2013"
}

@article{Bredenstein:2006nk,
      author         = "Bredenstein, A. and Denner, Ansgar and Dittmaier, S. and Weber, M. M.",
      title          = "{Precision calculations for the Higgs decays \(H \to ZZ/WW \to 4\) leptons}",
      booktitle      = "{Proceedings, 8th DESY Workshop on Elementary Particle
                        Theory: Loops and Legs in Quantum Field Theory}",
      venue          = "Eisenach, Germany",
      eventdate      = "2006-04-23/2006-04-28",
      journal = {{Nucl.\ Phys.\ Proc.\ Suppl.}},
      volume         = "160",
      year           = "2006",
      pages          = "131-135",
      doi            = "10.1016/j.nuclphysbps.2006.09.104",
      eprint         = "hep-ph/0607060",
      archivePrefix  = "arXiv",
      primaryClass   = "hep-ph",
      reportNumber   = "PSI-PR-06-09",

}

@article{wilkspaper,
      title={Cover Your Bases: Asymptotic Distributions of the Profile Likelihood Ratio When Constraining Effective Field Theories in High-Energy Physics}, 
      author={Florian U. Bernlochner and Daniel C. Fry and Stephen B. Menary and Eric Persson},
      year={2022},
      archivePrefix={arXiv},
      url={https://arxiv.org/abs/2207.01350}, 
}

@article{newlumi,
    author         = "{ATLAS Collaboration}",
    title          = "{Luminosity determination in \(pp\) collisions at \(\sqrt{s} = 13\,\text{TeV}\) using the ATLAS detector at the LHC}",
    journal        = "Eur. Phys. J. C",
    volume         = "83",
    year           = "2023",
    pages          = "982",
    doi            = "10.1140/epjc/s10052-023-11747-w",
    reportNumber   = "CERN-EP-2022-281",
    eprint         = "2212.09379",
    archivePrefix  = "arXiv",
    primaryClass   = "hep-ex"
    }

@article{Brivio_2021, 
    author = "Brivio, Ilaria",
    title = "{SMEFTsim 3.0 \textemdash{} a practical guide}",
    eprint = "2012.11343",
    archivePrefix = "arXiv",
    primaryClass = "hep-ph",
    doi = "10.1007/JHEP04(2021)073",
    journal = "JHEP",
    volume = "04",
    pages = "073",
    year = "2021"
}

@article{PhysRevD.102.055012,
  title = {Putting standard model EFT fits to work},
  author = {Dawson, Sally and Homiller, Samuel and Lane, Samuel D.},
  journal = {Phys. Rev. D},
  issue = {5},
  pages = {055012},
  numpages = {19},
  year = {2020},
  month = {Sep},
  publisher = {American Physical Society},
  doi = {10.1103/PhysRevD.102.055012},
  eprint         = "2007.01296",
  archivePrefix  = "arXiv",
  primaryClass   = "hep-ph"
}

@article{Dawson_2022,
	doi = {10.1103/physrevd.106.055012},
	year = 2022,
	month = 9,
	publisher = {American Physical Society ({APS})},
	volume = {106},
	number = {5},
	author = {Sally Dawson and Duarte Fontes and Samuel Homiller and Matthew Sullivan},
	title = {Role of dimension-eight operators in an {EFT} for the 2HDM},
	journal = {Phys. Rev. D},
      pages = {055012},
      eprint         = "2205.01561",
    archivePrefix  = "arXiv",
    primaryClass   = "hep-ph"
}

@article{Heinrich_2019,
   title={Probing the trilinear Higgs boson coupling in di-Higgs production at NLO QCD including parton shower effects},
   volume={2019},
   ISSN={1029-8479},
   DOI={10.1007/jhep06(2019)066},
   number={06},
   journal={JHEP},
   volume = "066",
   publisher={Springer Science and Business Media LLC},
   author={Heinrich, G. and Jones, S. P. and Kerner, M. and Luisoni, G. and Scyboz, L.},
   year={2019},
   month=june ,
      eprint         = "1903.081371",
    archivePrefix  = "arXiv",
    primaryClass   = "hep-ph"
}

@article{Heinrich_2017,
   title={NLO predictions for Higgs boson pair production with full top quark mass dependence matched to parton showers},
   volume={2017},
   ISSN={1029-8479},
   DOI={10.1007/jhep08(2017)088},
   number={8},
   journal={JHEP},
      volume = "088",
   publisher={Springer Science and Business Media LLC},
   author={Heinrich, G. and Jones, S.P. and Kerner, M. and Luisoni, G. and Vryonidou, E.},
   year={2017},
   month=aug ,
      eprint         = "1703.09252",
    archivePrefix  = "arXiv",
    primaryClass   = "hep-ph"


   }

@article{STDM-2018-41,
    author         = "{ATLAS Collaboration}",
    title          = "{Measurement of double-differential charged-current Drell-Yan cross-sections at high transverse masses in \(pp\) collisions at \(\sqrt{s} = 13\,\text{TeV}\) with the ATLAS detector}",
    eprint         = "2502.21088",
    year= "2025",
    doi = {10.1007/JHEP07(2025)026},
    journal        = "JHEP",
    volume         = "07",
    year           = "2025",
    pages          = "026",
    archivePrefix  = "arXiv",
    primaryClass   = "hep-ex",
}

@article{EXOT-2022-42,
    author         = "{ATLAS Collaboration}",
    title          = "{A measurement of the high-mass \(\tau\bar{\tau}\) production cross-section at \(\sqrt{s} = 13\,\text{TeV}\) with the ATLAS detector and constraints on new particles and couplings}",
    doi = {10.1007/JHEP10(2025)05},
    eprint         = "2503.19836",
        journal        = "JHEP",
    volume         = "10",
    year           = "2025",
    pages          = "05",
    archivePrefix  = "arXiv",
    primaryClass   = "hep-ex",
}

@Article{STDM-2012-20,
    author         = "{ATLAS Collaboration}",
    title          = "{Precision measurement and interpretation of inclusive \(W^+\), \(W^-\) and \(Z/\gamma^*\) production cross sections with the ATLAS detector}",
    journal        = "Eur. Phys. J. C",
    volume         = "77",
    year           = "2017",
    pages          = "367",
    doi            = "10.1140/epjc/s10052-017-4911-9",
    reportNumber   = "CERN-EP-2016-272",
    eprint         = "1612.03016",
    archivePrefix  = "arXiv",
    primaryClass   = "hep-ex",
}

@Article{STDM-2017-24,
    author         = "{ATLAS Collaboration}",
    title          = "{Measurement of fiducial and differential \(W^+W^-\) production cross-sections at \(\sqrt{s} = 13\,\text{TeV}\) with the ATLAS detector}",
    journal        = "Eur. Phys. J. C",
    volume         = "79",
    year           = "2019",
    pages          = "884",
    doi            = "10.1140/epjc/s10052-019-7371-6",
    reportNumber   = "CERN-EP-2019-055",
    eprint         = "1905.04242",
    archivePrefix  = "arXiv",
    primaryClass   = "hep-ex",
}

@Article{STDM-2017-27,
    author         = "{ATLAS Collaboration}",
    title          = "{Differential cross-section measurements for the electroweak production of dijets in association with a \(Z\) boson in proton--proton collisions at ATLAS}",
    journal        = "Eur. Phys. J. C",
    volume         = "81",
    year           = "2021",
    pages          = "163",
    doi            = "10.1140/epjc/s10052-020-08734-w",
    reportNumber   = "CERN-EP-2020-045",
    eprint         = "2006.15458",
    archivePrefix  = "arXiv",
    primaryClass   = "hep-ex",
}

@Article{HIGG-2018-28,
    author         = "{ATLAS Collaboration}",
    title          = "{Higgs boson production cross-section measurements and their EFT interpretation in the \(4\ell\) decay channel at \(\sqrt{s} = 13\,\text{TeV}\) with the ATLAS detector}",
    journal        = "Eur. Phys. J. C",
    volume         = "80",
    year           = "2020",
    pages          = "957",
    doi            = "10.1140/epjc/s10052-020-8227-9",
    reportNumber   = "CERN-EP-2020-034",
    eprint         = "2004.03447",
    archivePrefix  = "arXiv",
    primaryClass   = "hep-ex",
    related        = "HIGG-2018-28-err",
    relatedstring  = "Erratum:",
}

@Article{HIGG-2018-42,
    author         = "{ATLAS Collaboration}",
    title          = "{A search for the \(Z\gamma\) decay mode of the Higgs boson in \(pp\) collisions at \(\sqrt{s} = 13\,\text{TeV}\) with the ATLAS detector}",
    journal        = "Phys. Lett. B",
    volume         = "809",
    year           = "2020",
    pages          = "135754",
    doi            = "10.1016/j.physletb.2020.135754",
    reportNumber   = "CERN-EP-2020-052",
    eprint         = "2005.05382",
    archivePrefix  = "arXiv",
    primaryClass   = "hep-ex",
}

@Article{STDM-2018-03,
    author         = "{ATLAS Collaboration}",
    title          = "{Measurement of \(W^{\pm}Z\) production cross sections and gauge boson polarisation in \(pp\) collisions at \(\sqrt{s} = 13\,\text{TeV}\) with the ATLAS detector}",
    journal        = "Eur. Phys. J. C",
    volume         = "79",
    year           = "2019",
    pages          = "535",
    doi            = "10.1140/epjc/s10052-019-7027-6",
    reportNumber   = "CERN-EP-2018-327",
    eprint         = "1902.05759",
    archivePrefix  = "arXiv",
    primaryClass   = "hep-ex",
}

@Article{TOPQ-2018-26,
    author         = "{ATLAS Collaboration}",
    title          = "{Inclusive and differential cross-sections for dilepton \(t\bar{t}\) production measured in \(\sqrt{s} = 13\,\text{TeV}\) \(pp\) collisions with the ATLAS detector}",
    journal        = "JHEP",
    volume         = "07",
    year           = "2023",
    pages          = "141",
    doi            = "10.1007/JHEP07(2023)141",
    reportNumber   = "CERN-EP-2023-016",
    eprint         = "2303.15340",
    archivePrefix  = "arXiv",
    primaryClass   = "hep-ex",
}

@Article{TOPQ-2018-29,
    author         = "{ATLAS Collaboration}",
    title          = "{Test of the universality of \(\tau\) and \(\mu\) lepton couplings in \(W\)-boson decays with the ATLAS detector}",
    journal        = "Nature Physics",
    volume         = "17",
    year           = "2021",
    pages          = "813--818",
    doi            = "10.1038/s41567-021-01236-w",
    reportNumber   = "CERN-EP-2020-139",
    eprint         = "2007.14040",
    archivePrefix  = "arXiv",
    primaryClass   = "hep-ex",
}

@Article{HDBS-2019-27,
    author         = "{ATLAS Collaboration}",
    title          = "{Search for the nonresonant production of Higgs boson pairs via gluon fusion and vector-boson fusion in the \(b\bar{b} \tau^+\tau^-\) final state in proton--proton collisions at \(\sqrt{s} = 13\,\text{TeV}\) with the ATLAS detector}",
    journal        = "Phys. Rev. D",
    volume         = "110",
    year           = "2024",
    pages          = "032012",
    doi            = "10.1103/PhysRevD.110.032012",
    reportNumber   = "CERN-EP-2024-091",
    eprint         = "2404.12660",
    archivePrefix  = "arXiv",
    primaryClass   = "hep-ex",
}

@Article{HIGG-2019-04,
    author         = "{ATLAS Collaboration}",
    title          = "{Measurements of Higgs bosons decaying to bottom quarks from vector boson fusion production with the ATLAS experiment at \(\sqrt{s} = 13\,\text{TeV}\)}",
    journal        = "Eur. Phys. J. C",
    volume         = "81",
    year           = "2021",
    pages          = "537",
    doi            = "10.1140/epjc/s10052-021-09192-8",
    reportNumber   = "CERN-EP-2020-195",
    eprint         = "2011.08280",
    archivePrefix  = "arXiv",
    primaryClass   = "hep-ex",
}

@Article{HIGG-2019-09,
    author         = "{ATLAS Collaboration}",
    title          = "{Measurements of Higgs boson production cross-sections in the \(H\to\tau^{+}\tau^{-}\) decay channel in \(pp\) collisions at \(\sqrt{s} = 13\,\text{TeV}\) with the ATLAS detector}",
    journal        = "JHEP",
    volume         = "08",
    year           = "2022",
    pages          = "175",
    doi            = "10.1007/JHEP08(2022)175",
    reportNumber   = "CERN-EP-2021-217",
    eprint         = "2201.08269",
    archivePrefix  = "arXiv",
    primaryClass   = "hep-ex",
}

@Article{HIGG-2019-14,
    author         = "{ATLAS Collaboration}",
    title          = "{A search for the dimuon decay of the Standard Model Higgs boson with the ATLAS detector}",
    journal        = "Phys. Lett. B",
    volume         = "812",
    year           = "2021",
    pages          = "135980",
    doi            = "10.1016/j.physletb.2020.135980",
    reportNumber   = "CERN-EP-2020-117",
    eprint         = "2007.07830",
    archivePrefix  = "arXiv",
    primaryClass   = "hep-ex",
}

@Article{STDM-2019-24,
    author         = "{ATLAS Collaboration}",
    title          = "{Measurement of the \(W\)-boson mass and width with the ATLAS detector using proton--proton collisions at \(\sqrt{s} = 7\,\text{TeV}\)}",
    journal        = "Eur. Phys. J. C",
    volume         = "84",
    year           = "2024",
    pages          = "1309",
    doi            = "10.1140/epjc/s10052-024-13190-x",
    reportNumber   = "CERN-EP-2024-074",
    eprint         = "2403.15085",
    archivePrefix  = "arXiv",
    primaryClass   = "hep-ex",
}

@Article{TOPQ-2019-23,
    author         = "{ATLAS Collaboration}",
    title          = "{Measurements of differential cross-sections in top-quark pair events with a high transverse momentum top quark and limits on beyond the Standard Model contributions to top-quark pair production with the ATLAS detector at \(\sqrt{s} = 13\,\text{TeV}\)}",
    journal        = "JHEP",
    volume         = "06",
    year           = "2022",
    pages          = "063",
    doi            = "10.1007/JHEP06(2022)063",
    reportNumber   = "CERN-EP-2022-003",
    eprint         = "2202.12134",
    archivePrefix  = "arXiv",
    primaryClass   = "hep-ex",
}

@Article{HIGG-2020-16,
    author         = "{ATLAS Collaboration}",
    title          = "{Measurement of the properties of Higgs boson production at \(\sqrt{s} = 13\,\text{TeV}\) in the \(H \to \gamma\gamma\) channel using \(139\,\text{fb}^{-1}\) of \(pp\) collision data with the ATLAS experiment}",
    journal        = "JHEP",
    volume         = "07",
    year           = "2023",
    pages          = "088",
    doi            = "10.1007/JHEP07(2023)088",
    reportNumber   = "CERN-EP-2022-094",
    eprint         = "2207.00348",
    archivePrefix  = "arXiv",
    primaryClass   = "hep-ex",
}

@Article{HIGG-2020-20,
    author         = "{ATLAS Collaboration}",
    title          = "{Measurements of \(WH\) and \(ZH\) production with Higgs boson decays into bottom quarks and direct constraints on the charm Yukawa coupling in \(13\,\text{TeV}\) \(pp\) collisions with the ATLAS detector}",
    year           = "2025",
    eprint         = "2410.19611",
    archivePrefix  = "arXiv",
    primaryClass   = "hep-ex",
    journal        = "JHEP",
    volume         = "04",
    year           = "2025",
    pages          = "075",
    doi = "10.1007/JHEP04(2025)075",
}

@Article{HIGG-2020-23,
    author         = "{ATLAS Collaboration}",
    title          = "{Measurement of Higgs boson decay into \(b\)-quarks in associated production with a top-quark pair in \(pp\) collisions at \(\sqrt{s} = 13\,\text{TeV}\) with the ATLAS detector}",
    journal        = "JHEP",
    volume         = "06",
    year           = "2022",
    pages          = "097",
    doi            = "10.1007/JHEP06(2022)097",
    reportNumber   = "CERN-EP-2021-202",
    eprint         = "2111.06712",
    archivePrefix  = "arXiv",
    primaryClass   = "hep-ex",
}

@Article{HDBS-2021-10,
    author         = "{ATLAS Collaboration}",
    title          = "{Studies of new Higgs boson interactions through nonresonant \(HH\) production in the \(b\bar{b}\gamma\gamma\) final state in \(pp\) collisions at \(\sqrt{s} = 13\,\text{TeV}\) with the ATLAS detector}",
    journal        = "JHEP",
    volume         = "01",
    year           = "2024",
    pages          = "066",
    doi            = "10.1007/JHEP01(2024)066",
    reportNumber   = "CERN-EP-2023-206",
    eprint         = "2310.12301",
    archivePrefix  = "arXiv",
    primaryClass   = "hep-ex",
}

@Article{HIGG-2021-08,
    author         = "{ATLAS Collaboration}",
    title          = "{Constraints on Higgs boson production with large transverse momentum using \(H \to b\bar{b}\) decays in the ATLAS detector}",
    journal        = "Phys. Rev. D",
    volume         = "105",
    year           = "2022",
    pages          = "092003",
    doi            = "10.1103/PhysRevD.105.092003",
    reportNumber   = "CERN-EP-2021-185",
    eprint         = "2111.08340",
    archivePrefix  = "arXiv",
    primaryClass   = "hep-ex",
}

@Article{HIGG-2021-20,
    author         = "{ATLAS Collaboration}",
    title          = "{Measurements of Higgs boson production by gluon--gluon fusion and vector-boson fusion using \(H \to W W^{*} \to e\nu \mu\nu\) decays in \(pp\) collisions at \(\sqrt{s} = 13\,\text{TeV}\) with the ATLAS detector}",
    journal        = "Phys. Rev. D",
    volume         = "108",
    year           = "2023",
    pages          = "032005",
    doi            = "10.1103/PhysRevD.108.032005",
    reportNumber   = "CERN-EP-2022-078",
    eprint         = "2207.00338",
    archivePrefix  = "arXiv",
    primaryClass   = "hep-ex",
}

@Article{HDBS-2022-03,
    author         = "{ATLAS Collaboration}",
    title          = "{Constraints on the Higgs boson self-coupling from single- and double-Higgs production with the ATLAS detector using \(pp\) collisions at \(\sqrt{s} = 13\,\text{TeV}\)}",
    journal        = "Phys. Lett. B",
    volume         = "843",
    year           = "2023",
    pages          = "137745",
    doi            = "10.1016/j.physletb.2023.137745",
    reportNumber   = "CERN-EP-2022-149",
    eprint         = "2211.01216",
    archivePrefix  = "arXiv",
    primaryClass   = "hep-ex",
}

@Article{HIGG-2022-17,
    author         = "{ATLAS Collaboration}",
    title          = "{Interpretations of the ATLAS measurements of Higgs boson production and decay rates and differential cross-sections in \(pp\) collisions at \(\sqrt{s}=13\,\text{TeV}\)}",
    journal        = "JHEP",
    volume         = "11",
    year           = "2024",
    pages          = "097",
    doi            = "10.1007/JHEP11(2024)097",
    reportNumber   = "CERN-EP-2024-017",
    eprint         = "2402.05742",
    archivePrefix  = "arXiv",
    primaryClass   = "hep-ex",
}

@Article{TOPQ-2023-28,
    author         = "{ATLAS Collaboration}",
    title          = "{Precise test of lepton flavour universality in \(W\)-boson decays into muons and electrons in \(pp\) collisions at \(\sqrt{s} = 13\,\text{TeV}\) with the ATLAS detector}",
    journal        = "Eur. Phys. J. C",
    volume         = "84",
    year           = "2024",
    pages          = "993",
    doi            = "10.1140/epjc/s10052-024-13070-4",
    reportNumber   = "CERN-EP-2024-063",
    eprint         = "2403.02133",
    archivePrefix  = "arXiv",
    primaryClass   = "hep-ex",
}

@Booklet{ATL-SOFT-PUB-2026-001,
    author         = "{ATLAS Collaboration}",
    title          = "{ATLAS Computing Acknowledgements}",
    howpublished   = "{ATL-SOFT-PUB-2026-001}",
    url            = "https://cds.cern.ch/record/2952666",
    year           = 2026
}

@article{Giani:2023smefit,
  author         = "Giani, Tommaso and Magni, Giovanni and Rojo, Juan",
  title          = "{SMEFiT: a flexible toolbox for global interpretations of particle physics data with effective field theories}",
  journal        = "Eur. Phys. J. C",
  volume         = "83",
  pages          = "393",
  year           = "2023",
  doi ={10.1140/epjc/s10052-023-11534-7},
  eprint         = "2302.06660",
  archivePrefix  = "arXiv",
    primaryClass  = {hep-ph}
}

@article{Brivio:2201.04974,
  author       = "Brivio, Ilaria and others",
  title        = "{Truncation, validity, uncertainties}",
  eprint       = "2201.04974",
  archivePrefix= "arXiv",
  primaryClass = "hep-ph",
  year         = "2022"
}

@article{Cowan2011,
  author    = {Cowan, Glen and Cranmer, Kyle and Gross, Eilam and Vitells, Ofer},
  title     = {Asymptotic formulae for likelihood-based tests of new physics},
  journal = {Eur. Phys. J. C},
  year      = {2011},
  volume    = {71},
  number    = {2},
  pages     = {1554},
  doi       = {10.1140/epjc/s10052-011-1554-0},
     Eprint = "1007.1727",
   Archiveprefix = {arXiv},
   Primaryclass = {physics.data-an},
}

@article{Henning:2014wua,
  author         = "{B. Henning, X. Lu and H. Murayama}",
  title          = "{How to use the Standard Model effective field theory}",
  journal        = "JHEP",
  volume         = "01",
  pages          = "23",
  year           = "2016",
  doi          = {10.1007/JHEP01(2016)023},
  eprint       = {1412.1837},
  archivePrefix= {arXiv},
  primaryClass = {hep-ph},
}

@article{Branco:2011iw,
title = {Theory and phenomenology of two-Higgs-doublet models},
journal = {Phys.Rept.},
volume = {516},
number = {1},
pages = {1-102},
year = {2012},
note = {Theory and phenomenology of two-Higgs-doublet models},
issn = {0370-1573},
doi = {10.1016/j.physrep.2012.02.002},
author = {G.C. Branco and P.M. Ferreira and L. Lavoura and M.N. Rebelo and Marc Sher and João P. Silva},
  eprint       = {1106.0034},
  archivePrefix= {arXiv},
  primaryClass = {hep-ph},
}

@article{Langacker:2008yv,
  title = {The physics of heavy ${Z}^{\ensuremath{'}}$ gauge bosons},
  author = {Langacker, Paul},
  journal = {Rev. Mod. Phys.},
  volume = {81},
  issue = {3},
  pages = {1199--1228},
  numpages = {0},
  year = {2009},
  month = {Aug},
  publisher = {American Physical Society},
  doi = {10.1103/RevModPhys.81.1199},
    eprint       = {0801.1345},
  archivePrefix= {arXiv},
  primaryClass = {hep-ph},


}

@article{Dawson:2020PuttingSMEFTFitsToWork,
  author       = {S.~Dawson and S.~Homiller and S.~D.~Lane},
  title        = {{Putting standard model EFT fits to work}},
  journal      = {Phys. Rev. D},
  volume       = {102},
  number       = {5},
  pages        = {055012},
  year         = {2020},
  doi = {10.1103/PhysRevD.102.055012},
  eprint       = {2007.01296},
  archivePrefix= {arXiv},
  primaryClass = {hep-ph}
}

@article{Falkowski:2020FlavourfulLikelihood,
  author       = {Adam Falkowski and David Straub},
  title        = {{Flavourful SMEFT likelihood for Higgs and electroweak data}},
  journal       = {JHEP},
  volume       = {04},
  pages        = {066},
  doi          = {10.1007/JHEP04(2020)066},
  year         = {2020},
  eprint       = {1911.07866},
  archivePrefix= {arXiv},
  primaryClass = {hep-ph}
}

@article{Ethier:2021CombinedSMEFT,
  author       = "{SMEFiT Collaboration}",
  title        = {{Combined SMEFT interpretation of Higgs, diboson, and top quark data from the LHC}},
  journal       = {JHEP},
  volume       = {11},
  pages        = {089},
  year         = {2021},
  eprint       = {2105.00006},
doi           = {10.1007/JHEP11(2021)089},
  archivePrefix= {arXiv},
  primaryClass = {hep-ph}
}

@article{Falkowski:2015ModelIndependent,
  author       = {Adam Falkowski and Francesco Riva},
  title        = {{Model-independent precision constraints on dimension-6 operators}},
  journal       = {JHEP},
  volume       = {02},
  pages        = {039},
  year         = {2015},
  eprint       = {1411.0669},
  archivePrefix= {arXiv},
  doi = {10.1007/JHEP02(2015)039},
  %url = {https://doi.org/10.1007/JHEP02(2015)039}
  primaryClass = {hep-ph}
  
}

@article{CMS:2025eft,
  author       = "{CMS Collaboration}",
  title        = "{Combined effective field theory interpretation of Higgs boson, electroweak vector boson, top quark, and multijet measurements}",
  journal = "Eur. Phys. J. C",
  volume  = {86},
  number  = {4},
  pages   = {331},
  year    = {2026},
  doi= "10.1140/epjc/s10052-025-14997-y",
  year         = "2026",
  eprint       = "2504.02958",
  archivePrefix= "arXiv",
  primaryClass = "hep-ex"
}

@article{mcsanc,
	author = {Arbuzov, A. and Bardin, D. and Bondarenko, S. and
                  Christova, P. and Kalinovskaya, L. and Klein, U. and
                  Kolesnikov, V. and Rumyantsev, L. and Sadykov,
                  R. and Sapronov, A.},
	date = {2016/01/01},
	doi = {10.1134/S0021364016020041},
	id = {Arbuzov2016},
	isbn = {1090-6487},
	journal = {JETP Letters},
	number = {2},
	pages = {131--136},
	title = {Update of the MCSANC Monte Carlo integrator, v. 1.20},
	volume = {103},
	year = {2016},
      eprint         = "1509.03052",
      archivePrefix  = "arXiv",
      primaryClass   = "hep-ph",

}

@article{POWHEG_1,
      author         = "Nason, Paolo",
      title          = "{A new method for combining NLO QCD with shower Monte
                        Carlo algorithms}",
      journal        = "JHEP",
      volume         = "11",
      year           = "2004",
      pages          = "040",
      doi            = "10.1088/1126-6708/2004/11/040",
      eprint         = "hep-ph/0409146",
      archivePrefix  = "arXiv",
      primaryClass   = "hep-ph",
      reportNumber   = "BICOCCA-FT-04-11",
      SLACcitation   = "%%CITATION = HEP-PH/0409146;%%"
}

@article{POWHEG_2,
      author         = "Frixione, Stefano and Nason, Paolo and Oleari, Carlo",
      title          = "{Matching NLO QCD computations with Parton Shower
                        simulations: the POWHEG method}",
      journal        = "JHEP",
      volume         = "11",
      year           = "2007",
      pages          = "070",
      doi            = "10.1088/1126-6708/2007/11/070",
      eprint         = "0709.2092",
      archivePrefix  = "arXiv",
      primaryClass   = "hep-ph",
      reportNumber   = "BICOCCA-FT-07-9, GEF-TH-21-2007",
      SLACcitation   = "%%CITATION = ARXIV:0709.2092;%%"
}

@article{POWHEG_3,
      author         = "Alioli, Simone and Nason, Paolo and Oleari, Carlo and Re,
                        Emanuele",
      title          = "{A general framework for implementing NLO calculations in
                        shower Monte Carlo programs: the POWHEG BOX}",
      journal        = "JHEP",
      volume         = "06",
      year           = "2010",
      pages          = "043",
      doi            = "10.1007/JHEP06(2010)043",
      eprint         = "1002.2581",
      archivePrefix  = "arXiv",
      primaryClass   = "hep-ph",
      reportNumber   = "DESY-10-018, SFB-CPP-10-22, IPPP-10-11, DCPT-10-22",
      SLACcitation   = "%%CITATION = ARXIV:1002.2581;%%"
}

@article{Campbell:2014kua,
    author = "Campbell, John M. and Ellis, R. Keith and Nason, Paolo and Re, Emanuele",
    title = "{Top-Pair production and decay at NLO matched with parton Showers}",
    eprint = "1412.1828",
    archivePrefix = "arXiv",
    primaryClass = "hep-ph",
    reportNumber = "FERMILAB-PUB-14-504-T, OUTP-14-17P",
    doi = "10.1007/JHEP04(2015)114",
    journal = "JHEP",
    volume = "04",
    pages = "114",
    year = "2015"
}

@article{Mildner:2024wbl,
    author = "Mildner, Hannes",
    title = "{An EWPD SMEFT likelihood for the LHC {\textemdash} and how to improve it with measurements of W and Z boson properties}",
    eprint = "2412.07651",
    archivePrefix = "arXiv",
    primaryClass = "hep-ph",
    doi = "10.1007/JHEP07(2025)089",
    journal = "JHEP",
    volume = "07",
    pages = "089",
    year = "2025"
}

@article{Arnold:2011wj,
    author = "Baglio, J. and others",
    title = "{VBFNLO: A parton level Monte Carlo for Processes with electroweak bosons -- Manual for Version 2.7.0}",
    eprint = "1107.4038",
    archivePrefix = "arXiv",
    primaryClass = "hep-ph",
    reportNumber = "KA-TP-15-2011, FTUV-11-0730, WUB-11-06, FR-PHENO-2011-012, MZ-TH-11-19, CERN-PH-TH-2011-173, LPN11-40, DESY-11-125, TTK-10-51, MAN-HEP-2014-02",
    month = "7",
    year = "2014"
}

@article{Anastasiou:2008tj,
	title         = {{Mixed QCD-electroweak corrections to Higgs boson production in gluon fusion}},
	author        = {Anastasiou, Charalampos and Boughezal, Radja and Petriello, Frank},
	year          = 2009,
	journal       = {JHEP},
	volume        = {04},
	pages         = {003},
	doi           = {10.1088/1126-6708/2009/04/003},
	eprint        = {0811.3458},
	archiveprefix = {arXiv},
	primaryclass  = {hep-ph},
	slaccitation  = {%%CITATION = ARXIV:0811.3458;%%}
}

@article{Alioli:2008tz,
	title         = {{NLO Higgs boson production via gluon fusion matched with shower in POWHEG}},
	author        = {Alioli, Simone and Nason, Paolo and Oleari, Carlo and Re, Emanuele},
	year          = 2009,
	journal       = {JHEP},
	volume        = {04},
	pages         = {002},
	doi           = {10.1088/1126-6708/2009/04/002},
	eprint        = {0812.0578},
	archiveprefix = {arXiv},
	primaryclass  = {hep-ph},
	slaccitation  = {%%CITATION = ARXIV:0812.0578;%%}
}

@inproceedings{Amoroso:2020lgh,
	title         = {{Les Houches 2019: Physics at TeV Colliders: Standard Model Working Group Report}},
	author        = {Amoroso, S. and others},
	year          = 2020,
	month         = 3,
	booktitle     = {{11th Les Houches Workshop on Physics at TeV Colliders}: {PhysTeV Les Houches}},
	eprint        = {2003.01700},
	archiveprefix = {arXiv},
	primaryclass  = {hep-ph},
}

@article{Berger:2019wnu,
	title         = {{Simplified template cross sections - Stage 1.1 and 1.2}},
	author={Nicolas Berger and Claudia Bertella and Matteo Bonanomi and Nihal Brahimi and Thomas P. Calvet and Milene Calvetti and Valerio Dao and Marco Delmastro and Michael Duehrssen-Debling and Paolo Francavilla and Yacine Haddad and Sarah Heim and Jelena Jovicevic and Oleh Kivernyk and Maria Moreno Llacer and Jonathon M. Langford and Changqiao Li and Giovanni Marchiori and Josh A. McFayden and Johannes K. L. Michel and Predrag Milenovic and Carlo E. Pandini and Edward Scott and Frank J. Tackmann and Kerstin Tackmann and Lorenzo Viliani and Meng Xiao and Hongtao Yang},
	journal={SciPost Phys. Comm. Rep.},
	pages={15},
	year={2026},
	publisher={SciPost},
	doi={10.21468/SciPostPhysCommRep.15},
}

@article{LesHouches,
	title         = {{Les Houches 2015: Physics at TeV Colliders Standard Model Working Group Report}},
	author        = {Badger, S. and others},
	year          = 2016,
	eprint        = {1605.04692},
	archiveprefix = {arXiv},
	primaryclass  = {hep-ph}
}

@Article{ALEPH:2005ab,
    author = "Schael, S. and others",
    collaboration = "ALEPH, DELPHI, L3, OPAL, SLD, LEP Electroweak Working Group, SLD Electroweak Group, SLD Heavy Flavour Group",
    title = "{Precision electroweak measurements on the $Z$ resonance}",
    eprint = "hep-ex/0509008",
    archivePrefix = "arXiv",
    reportNumber = "SLAC-R-774",
    doi = "10.1016/j.physrep.2005.12.006",
    journal = "Phys. Rept.",
    volume = "427",
    pages = "257--454",
    year = "2006"
}

@article{pdg2024,
author = "{Particle~Data~Group~Collaboration}",
    title = "{Review of Particle Physics}",
    journal = "Phys. Rev. D",
    volume = "110",
    pages = "030001",
    year = "2024",
    doi = "10.1103/PhysRevD.110.030001"
}

@article{FlavourLatticeAveragingGroupFLAG:2021npn,
    author = "Aoki, Y. and others",
    collaboration = "Flavour Lattice Averaging Group (FLAG)",
    title = "{FLAG Review 2021}",
    eprint = "2111.09849",
    archivePrefix = "arXiv",
    primaryClass = "hep-lat",
    reportNumber = "CERN-TH-2021-191, JLAB-THY-21-3528, FERMILAB-PUB-21-620-SCD-T",
    doi = "10.1140/epjc/s10052-022-10536-1",
    journal = "Eur. Phys. J. C",
    volume = "82",
    number = "10",
    pages = "869",
    year = "2022"
}

@article{Brivio:2020onw,
    author = "Brivio, Ilaria",
    title = "{SMEFTsim 3.0 \textemdash{} a practical guide}",
    eprint = "2012.11343",
    archivePrefix = "arXiv",
    primaryClass = "hep-ph",
    doi = "10.1007/JHEP04(2021)073",
    journal = "JHEP",
    volume = "04",
    pages = "073",
    year = "2021"
}

@article{Dawson:2019clf,
    author = "Dawson, Sally and Giardino, Pier Paolo",
    title = "{Electroweak and QCD corrections to $Z$ and $W$ pole observables in the standard model EFT}",
    eprint = "1909.02000",
    archivePrefix = "arXiv",
    primaryClass = "hep-ph",
    reportNumber = "IFT-UAM/CSIC-19-118",
    doi = "10.1103/PhysRevD.101.013001",
    journal = "Phys. Rev. D",
    volume = "101",
    number = "1",
    pages = "013001",
    year = "2020"
}

@article{Brivio:2017vri,
    author = "Brivio, Ilaria and Trott, Michael",
    title = "{The standard model as an effective field theory}",
    eprint = "1706.08945",
    archivePrefix = "arXiv",
    primaryClass = "hep-ph",
    doi = "10.1016/j.physrep.2018.11.002",
    journal = "Phys. Rept.",
    volume = "793",
    pages = "1--98",
    year = "2019"
}

@article{ALEPH:2013dgf,
    author = "Schael, S. and others",
    collaboration = "ALEPH, DELPHI, L3, OPAL, LEP Electroweak",
    title = "{Electroweak measurements in electron-positron collisions at W-boson-pair energies at LEP}",
    eprint = "1302.3415",
    archivePrefix = "arXiv",
    primaryClass = "hep-ex",
    reportNumber = "CERN-PH-EP-2013-022",
    doi = "10.1016/j.physrep.2013.07.004",
    journal = "Phys. Rept.",
    volume = "532",
    pages = "119--244",
    year = "2013"
}

@article{Degrande:2020evl,
    author = "Degrande, C\'eline and Durieux, Gauthier and Maltoni, Fabio and Mimasu, Ken and Vryonidou, Eleni and Zhang, Cen",
    title = "{Automated one-loop computations in the standard model effective field theory}",
    eprint = "2008.11743",
    archivePrefix = "arXiv",
    primaryClass = "hep-ph",
    reportNumber = "CERN-TH-2020-140, CP3-20-42",
    doi = "10.1103/PhysRevD.103.096024",
    journal = "Phys. Rev. D",
    volume = "103",
    number = "9",
    pages = "096024",
    year = "2021"
}

@article{Corbett:2021eux,
    author = "Corbett, Tyler and Helset, Andreas and Martin, Adam and Trott, Michael",
    title = "{EWPD in the SMEFT to dimension eight}",
    eprint = "2102.02819",
    archivePrefix = "arXiv",
    primaryClass = "hep-ph",
    doi = "10.1007/JHEP06(2021)076",
    journal = "JHEP",
    volume = "06",
    pages = "076",
    year = "2021"
}

@article{Ellis:2020unq,
    author = "Ellis, John and Madigan, Maeve and Mimasu, Ken and Sanz, Veronica and You, Tevong",
    title = "{Top, Higgs, Diboson and Electroweak Fit to the Standard Model effective field theory}",
    eprint = "2012.02779",
    archivePrefix = "arXiv",
    primaryClass = "hep-ph",
    reportNumber = "KCL-PH-TH/2020-73, CERN-TH-2020-202",
    doi = "10.1007/JHEP04(2021)279",
    journal = "JHEP",
    volume = "04",
    pages = "279",
    year = "2021"
}

@inproceedings{Brooijmans:2018xbu,
    author = "Brooijmans, G. and others",
    title = "{Les Houches 2017: Physics at TeV Colliders New Physics Working Group Report}",
    booktitle = "{10th Les Houches Workshop on Physics at TeV Colliders}",
    eprint = "1803.10379",
    archivePrefix = "arXiv",
    primaryClass = "hep-ph",
    reportNumber = "FERMILAB-CONF-17-664-PPD",
    month = "3",
    year = "2018"
}

@Article{EvtGen,
    author         = "{D. J. Lange}",
    title          = "{The EvtGen particle decay simulation package}",
    journal        = "Nucl. Instrum. Meth. A",
    volume         = "462",
    year           = "2001",
    pages          = "152",
    doi            = "10.1016/S0168-9002(01)00089-4",
    primaryClass   = "hep-ex",
}

@article{Brivio:2017btx,
    author = "Brivio, Ilaria and Jiang, Yun and Trott, Michael",
    title = "{The SMEFTsim package, theory and tools}",
    eprint = "1709.06492",
    archivePrefix = "arXiv",
    primaryClass = "hep-ph",
    doi = "10.1007/JHEP12(2017)070",
    journal = "JHEP",
    volume = "12",
    pages = "070",
    year = "2017"
}

@misc{dawson2024smeftmatchingzprimemodels,
      title={SMEFT Matching to $Z^\prime$ Models at Dimension-8}, 
      author={Sally Dawson and Matthew Forslund and Marvin Schnubel},
      year={2024},
      eprint={2404.01375},
      archivePrefix={arXiv},
      primaryClass={hep-ph},
}

@article{Ellis:2018UpdatedGlobalSMEFT,
  author       = {John Ellis and Christopher W. Murphy and Ver\'onica Sanz and Tevong You},
  title        = {{Updated global SMEFT fit to Higgs, diboson and electroweak data}},
  journal       = {JHEP},
  volume       = {06},
  pages        = {146},
  year         = {2018},
  doi= {10.1007/JHEP06(2018)146},
  eprint       = {1803.03252},
  archivePrefix= {arXiv},
  primaryClass = {hep-ph}
}

@article{PhysRevD.71.075009,
  title = {Effective theory analysis of precision electroweak data},
  author = {Han, Zhenyu and Skiba, Witold},
  journal = {Phys. Rev. D},
  volume = {71},
  issue = {7},
  pages = {075009},
  numpages = {12},
  year = {2005},
  month = {Apr},
  publisher = {American Physical Society},
  doi = {10.1103/PhysRevD.71.075009},
  eprint = "hep-ph/0412166",
  archivePrefix = "arXiv",

}

@article{JHEP012014151,
  title = {Towards the ultimate SM fit to close in on Higgs physics},
  author = {Pomarol, Alex and Riva, Francesco},
  journal = {JHEP},
  volume = {01},
  issue = {},
  pages = {151},
  numpages = {},
  year = {2014},
  month = {},
  doi = {10.1007/JHEP01(2014)151},
  eprint = "1308.2803",
  archivePrefix = "arXiv",
  primaryClass = "hep-ph",
}

@article{JHEP072015018,
  title = {Electroweak constraints on flavorful effective theories},
  author = {Efrati, Aielet and Falkowski, Adam and Soreq, Yotam},
  journal = {JHEP},
  volume = {07},
  pages = {18},
  year = {2015},
  doi = {10.1007/JHEP07(2015)018},
  eprint = "1503.07872",
  archivePrefix = "arXiv",
  primaryClass = "hep-ph",
}

@article{JHEP022016069,
  title = {Consistent constraints on the Standard Model Effective Field Theory},
  author = {Berthier, Laure and Trott, Michael},
  journal = {JHEP},
  volume = {02},
  issue = {},
  pages = {69},
  numpages = {},
  year = {2016},
  month = {},
  doi = {10.1007/JHEP02(2016)069},
    eprint = "1508.05060",
  archivePrefix = "arXiv",
  primaryClass = "hep-ph",
}

@article{deBlas:2018Wq,
  author = "de Blas, Jorge  and  Ciuchini, Marco  and  Franco, Enrico  and  Mishima, Satoshi  and  Pierini, Maurizio  and  Reina, Laura  and  Silvestrini, Luca",
  title = "{The Global Electroweak and Higgs Fits in the LHC era}",
  doi = "10.22323/1.314.0467",
  journal = "Proceedings of Science",
  year = 2018,
  volume = "EPS-HEP2017",
  pages = "467"

}

@article{PhysRevD.99.033001,
  title = {Electroweak sector under scrutiny: A combined analysis of LHC and electroweak precision data},
  author = {Almeida, Eduardo da Silva and Alves, Alexandre and Rosa-Agostinho, N. and \'Eboli, Oscar J. P. and Gonzalez--Garcia, M. C.},
  journal = {Phys. Rev. D},
  volume = {99},
  issue = {3},
  pages = {033001},
  numpages = {16},
  year = {2019},
  month = {Feb},
  publisher = {American Physical Society},
  doi = {10.1103/PhysRevD.99.033001},
  eprint = "1812.01009",
  archivePrefix = "arXiv",
  primaryClass = "hep-ph",
}

@Article{SciPostPhys.6.6.064,
	title={{The Gauge-Higgs legacy of the LHC run II}},
	author={Anke Biekoetter and Tyler Corbett and Tilman Plehn},
	journal={SciPost Phys.},
	volume={6},
	pages={064},
	year={2019},
	publisher={SciPost},
	doi={10.21468/SciPostPhys.6.6.064},
      eprint = "1812.07587",
  archivePrefix = "arXiv",
  primaryClass = "hep-ph",
}

@article{s10052-019-6977-z,
  title = {A global likelihood for precision constraints and flavour anomalies},
  author = {Aebischer, Jason and Kumar, Jacky and Stangl, Peter and Straub, David M.},
  journal = {Eur.Phys.J.C},
  volume = {79},
  issue = {},
  pages = {509},
  numpages = {},
  year = {2019},
  month = {},
  doi = {10.1140/epjc/s10052-019-6977-z},
    eprint = "1810.07698",
  archivePrefix = "arXiv",
  primaryClass = "hep-ph",
}

@article{JHEP052023208,
  title = {The importance of flavor in SMEFT Electroweak Precision Fits},
  author = {Bellafronte, Luigi and Dawson, Sally and Giardino, Pier Paolo},
  journal = {JHEP},
  volume = {2023},
  issue = {},
  pages = {208},
  numpages = {},
  year = {2023},
  month = {},
  doi = {10.1007/JHEP05(2023)208},
      eprint = "2304.00029",
  archivePrefix = "arXiv",
  primaryClass = "hep-ph",
}

@article{JHEP092024091,
  title = {Mapping the SMEFT at high-energy colliders: from LEP and the (HL-)LHC to the FCC-ee},
  author = {Celada, Eugenia and Giani, Tommaso and ter Hoeve, Jaco and Mantani, Luca and Rojo, Juan and Rossia, Alejo N. and Thomas, Marion O. A. and Vryonidou, Eleni},
  journal = {JHEP},
  volume = {2024},
  issue = {},
  pages = {91},
  numpages = {},
  year = {2024},
  month = {},
  doi = {10.1007/JHEP09(2024)091},
      eprint = "2404.12809",
  archivePrefix = "arXiv",
  primaryClass = "hep-ph",



}
\clearpage
 
\begin{flushleft}
\hypersetup{urlcolor=black}
{\Large The ATLAS Collaboration}

\bigskip

\AtlasOrcid[0000-0002-6665-4934]{G.~Aad}$^\textrm{\scriptsize 102}$,
\AtlasOrcid[0000-0001-7616-1554]{E.~Aakvaag}$^\textrm{\scriptsize 17}$,
\AtlasOrcid[0000-0002-5888-2734]{B.~Abbott}$^\textrm{\scriptsize 121}$,
\AtlasOrcid[0000-0002-0287-5869]{S.~Abdelhameed}$^\textrm{\scriptsize 83b}$,
\AtlasOrcid[0000-0002-1002-1652]{K.~Abeling}$^\textrm{\scriptsize 54}$,
\AtlasOrcid[0000-0001-5763-2760]{N.J.~Abicht}$^\textrm{\scriptsize 48}$,
\AtlasOrcid[0000-0002-8496-9294]{S.H.~Abidi}$^\textrm{\scriptsize 30}$,
\AtlasOrcid[0009-0003-6578-220X]{M.~Aboelela}$^\textrm{\scriptsize 44}$,
\AtlasOrcid[0000-0002-9987-2292]{A.~Aboulhorma}$^\textrm{\scriptsize 36e}$,
\AtlasOrcid[0000-0001-5329-6640]{H.~Abramowicz}$^\textrm{\scriptsize 154}$
% ...
\bigskip
\\

% ...
$^{17}$Department for Physics and Technology, University of Bergen, Bergen; Norway.\\
% ...
$^{30}$Physics Department, Brookhaven National Laboratory, Upton NY; United States of America.\\
% ...
$^{36}$$^{(a)}$Facult\'e des Sciences Ain Chock, Universit\'e Hassan II de Casablanca;$^{(b)}$Facult\'{e} des Sciences, Universit\'{e} Ibn-Tofail, K\'{e}nitra;$^{(c)}$Facult\'e des Sciences Semlalia, Universit\'e Cadi Ayyad, LPHEA-Marrakech;$^{(d)}$LPMR, Facult\'e des Sciences, Universit\'e Mohamed Premier, Oujda;$^{(e)}$Facult\'e des sciences, Universit\'e Mohammed V, Rabat;$^{(f)}$Institute of Applied Physics, Mohammed VI Polytechnic University, Ben Guerir; Morocco.\\
% ...
$^{44}$Physics Department, Southern Methodist University, Dallas TX; United States of America.\\
% ...
$^{48}$Fakult\"{a}t Physik, Technische Universit{\"a}t Dortmund, Dortmund; Germany.\\
% ...
$^{54}$II. Physikalisches Institut, Georg-August-Universit\"{a}t G\"ottingen, G\"ottingen; Germany.\\
% ...
$^{83}$$^{(a)}$Khalifa University of Science and Technology, Abu Dhabi;$^{(b)}$New York University Abu Dhabi, Abu Dhabi;$^{(c)}$United Arab Emirates University, Al Ain;$^{(d)}$University of Sharjah, Sharjah; United Arab Emirates.\\
% ...
$^{102}$CPPM, Aix-Marseille Universit\'e, CNRS/IN2P3, Marseille; France.\\
% ...
$^{121}$Homer L. Dodge Department of Physics and Astronomy, University of Oklahoma, Norman OK; United States of America.\\
% ...
$^{154}$Raymond and Beverly Sackler School of Physics and Astronomy, Tel Aviv University, Tel Aviv; Israel.\\
% ...
$^{a}$ Also at Affiliated with an institute formerly covered by a cooperation agreement with CERN.\\
$^{b}$ Also at An-Najah National University, Nablus; Palestine.\\
$^{c}$ Also at Borough of Manhattan Community College, City University of New York, New York NY; United States of America.\\
$^{d}$ Also at Center for Interdisciplinary Research and Innovation (CIRI-AUTH), Thessaloniki; Greece.\\
$^{e}$ Also at Centre of Physics of the Universities of Minho and Porto (CF-UM-UP); Portugal.\\
$^{f}$ Also at CERN, Geneva; Switzerland.\\
% ...

\end{flushleft}


%

%
%

%

%
%
%
%

\end{document}